\documentclass[aps, prd, nofootinbib, showkeys, superscriptaddress]{revtex4}

\usepackage{amsmath}%,amssymb}
\allowdisplaybreaks

\usepackage{makeidx}
\usepackage{amsfonts}
\usepackage[usenames,dvipsnames]{pstricks}
\usepackage{subfigure}
\usepackage{epsfig}
\usepackage{pst-grad} % For gradients
\usepackage{mathtools}
\usepackage[colorlinks,hyperindex]{hyperref}
\usepackage{array}
\hypersetup
{	colorlinks,%
	citecolor=black,%
	linkcolor=black,%
	urlcolor=black,%
}

\usepackage{allpurpose}

\newcommand\nn{{\nonumber}}

\begin{document}

\title{Perturbative deflection angles of timelike rays}

\author{Yujie Duan}
\thanks{These authors contributed equally to this work.}
\affiliation{School of Physics and Technology, Wuhan University, Wuhan, 430072, China}

\author{Weiyu Hu}
\thanks{These authors contributed equally to this work.}
\affiliation{School of Physics and Technology, Wuhan University, Wuhan, 430072, China}

\author{Ke Huang}
\thanks{These authors contributed equally to this work.}
\affiliation{School of Physics and Technology, Wuhan University, Wuhan, 430072, China}

\author{Junji Jia}
\affiliation{School of Physics and Technology, Wuhan University, Wuhan, 430072, China}
\affiliation{MOE Key Laboratory of Artificial Micro- and Nano-structures, School of Physics and Technology, Wuhan University, Wuhan, 430072, China}
\email{Corresponding Author: junjijia@whu.edu.cn}

\date{\today}

\begin{abstract}
Geodesics of both lightrays and timelike particles with nonzero mass are deflected in a gravitational field. In this work we apply the perturbative method developed in Ref. \cite{Jia:2020dap} to compute the deflection angle of both null and timelike rays in the weak field limit for four spacetimes. We obtained the deflection angles for the Bardeen spacetime to the eleventh order of $m/b$ where $m$ is the ADM mass and $b$ is the impact parameter, and for the Hayward, Janis-Newman-Winicour and Einstein-Born-Infeld spacetimes to the ninth, seventh and eleventh order respectively. The effect of the impact parameter $b$, velocity $v$ and spacetime parameters on the deflection angle are analyzed in each of the four spacetimes. It is found that in general, the perturbative deflection angle depends on and only on the asymptotic behavior of the metric functions, and in an order-correlated way. Moreover, it is shown that although these deflection angles are calculated in the large $b/m$ limit, their minimal valid $b$ can be as small as a few $m$'s as long as the order is high enough. At these impact parameters, the deflection angle itself is also found large. As velocity decreases, the deflection angle in all spacetime studied increases. For a given $b$, if the spacetime parameters allows a critical velocity $v_c$, then the perturbative deflection angle will deviate from its true value as $v$ decreases to $v_c$. It is also found that if the variation of spacetime parameters can only change the spacetime qualitatively at small but not large radius, then these spacetime parameter will not cause a qualitative change of the deflection angle, although its value is still quantitatively affected. The application and possible extension of the work are discussed.

\end{abstract}

\keywords{deflection angle, gravitational lensing, weak lensing, strong lensing, neutrino mass, massive gravitational wave}
\pacs{95.30.Sf, 04.20.-q, 97.60.Lf}

\maketitle

\section{Introduction}

The deflection of the lightray near a massive object played an important role in both the theoretical development of gravitational theories and the experimental observations in astrophysics. It helped establishing General Relativity as a correct description of gravity at its very early age. It is also the foundation of gravitational lensing (GL) effect, which is an important tool to deduce properties of the signal source, the signal itself and the lens it passes by.  After the first observation of GL in 1979 \cite{Walsh:1979nx}, many features, such as luminous arcs \cite{lynds}, Einstein cross \cite{Huchra:1985zz} and rings \cite{hewitt}, CMB GL \cite{Smith:2007rg,Das:2011ak,vanEngelen:2012va} and GL of supernovas \cite{Quimby:2013lfa,Nordin:2013cfa} have been observed. These GLs can be used to study the coevolution of supermassive
black holes (BHs) and galaxies \cite{Peng:2006ew} and cosmological parameters (for a review see \cite{Lewis:2006fu}), properties of the supernova \cite{Sharon:2014ija}, as well as dark matter substructures \cite{Metcalf:2001ap, Metcalf:2001es}.

Traditionally, electromagnetic wave was the main kind of signal in relevant theoretical or observational studies. However, with the discovery of SN1987A neutrinos and blazer TXS 0506+056 \cite{IceCube:2018dnn,IceCube:2018cha}, and the more recent discovery of gravitational waves (GWs) \cite{Abbott:2016blz,Abbott:2016nmj,Abbott:2017oio,TheLIGOScientific:2017qsa}, more kinds of messengers become possible. It is known now that neutrinos have non-zero mass \cite{Tanabashi:2018oca} and GW in some gravitational theories beyond GR can also be massive. To study the GL of these particles, including the apparent angles of source images and the time delay effect \cite{Jia:2019hih}, in principle one should start from the deflection angles calculated for corresponding \emph{timelike} rays. Recently some of us have examined the deflection angles of massive particles in the simple Schwarzschild and Reissner-Nordstr\"{o}m (RN) spacetimes using exact formulas \cite{Jia:2015zon,Pang:2018jpm} which are valid in both the weak and strong field limits. These results were then used to correlate the GL observations to the properties of the messengers, namely the absolute mass, mass ordering of neutrinos and velocity of GWs.
For the deflection angles in more complicated spacetimes, we proposed a perturbative method which were shown to work to very high orders in the weak field limit for both null and timelike particles \cite{Jia:2020dap}. The results in Schwarzschild and RN spacetimes were shown to work even when the gravitational field is not weak, and when the charge in the RN spacetime passes the extremal value \cite{Jia:2020dap}.

In this work, we extend this perturbative method to the calculation of deflection angles of signals with general velocity in four other spacetimes, namely the Bardeen, Hayward, Janis-Newman-Winicour (JNW) and Einstein-Born-Infeld (EBI) spacetimes. These spacetimes all carry \emph{charges} of their own kind, and allow transition from BH spacetime to non-BH (possibly naked singularity) spacetime in their respective parameter space. The motivation is to see how different kinds of charges will influence the deflection angle of rays with general velocity \cite{Virbhadra:1998dy,Perlick:2010zh}. In particular, we will determine the validity of the calculated deflection angle near and beyond the critical value of the parameters.

To our best knowledge, the state of the art for the computation of the deflection angles is to the fifth order in Bardeen spacetime \cite{Virbhadra:1998dy, Amore:2006pi,Ghaffarnejad:2014zva}, sixth order in Hayward spacetime \cite{Wei:2015qca, Chiba:2017nml}, second order in JNW spacetime\cite{Amore:2006pi,Virbhadra:1998dy,Virbhadra:2007kw} and third order in EBI BH spacetimes \cite{Amore:2006pi}, all for only lightrays but not timelike particles.
In this work, we extend these orders dramatically, to the eleventh, ninth, seventh and eleventh orders for the Bardeen, Hayward, JNW and EBI spacetimes respectively, for both lightlike and timelike rays.

The paper is organized as the following. We first recap the general procedure to find the deflection angle in general static, spherically symmetric and asymptotically flat spacetimes for signal with arbitrary velocity in Sec. \ref{sec:method}. Then this method is applied to the above mentioned four spacetimes in subsections of the section \ref{sec:metrics}. For each spacetime, after the deflection angles is found, the effect of the impact parameter, signal velocity and the spacetime parameters are analyzed. Finally, in Sec. \ref{sec:discussion}, we compare these effects among different spacetimes. Throughout the paper we use the geometric unit $G=c=1$.

\section{Perturbative method for the deflection angles \label{sec:method}}

Because the computation of the deflection angle in this work are based on the perturbative method developed in Ref. \cite{Jia:2020dap}, here we briefly recap the procedure of the computation in weak field limit for signals with arbitrary velocity.

For any static and spherically symmetric spacetimes, we can start from a general metric
\begin{equation}
\dd s^{2}=-A(r) \dd t^{2}+B(r) \dd r^{2}+C(r)\left(\dd \theta^{2}+\sin ^{2} \theta \dd \varphi^{2}\right).
\end{equation}
To be asymptotically flat, the metric functions satisfy \cite{bozza:2002}
\begin{align}
& A(r\to\infty)=1-\frac{2m}{r}+\mathcal{O}(r^{-2}),\\
& B(r\to\infty)=1+\frac{2m}{r}+\mathcal{O}(r^{-2}),\\
& C(r\to\infty)=r^2+\mathcal{O}(r^1),
\end{align}
where $m$ is the ADM mass parameter of the system.

Using this metric, it is easy to find the geodesic equations, from which two first integrals can be carried out. These first integrals define the energy $E$ and angular momentum $L$ at infinity for unit mass. $E$ can be related to the velocity at infinity and $L$ to the minimal radial coordinate $r_0$ of the trajectory by \cite{Jia:2020dap}
\begin{align}
E=\frac{1}{\sqrt{1-v^2}},~L=\frac{(E-\kappa A(r_0))C(r_0)}{A(r_0)}, \label{eq:lx0rel}
\end{align}
where $\kappa=0,~1$ for null and timelike geodesics respectively. Then it is not difficult to show that the change of the angular coordinate for a particle coming from and going back to spacial infinity is given by \cite{Jia:2020dap}
\begin{align}
I(r_0,~E,~L)=&2\int_{r_0}^\infty \frac{L}{C(r)\sqrt{B(r)}} \lb \kappa -\frac{E^2}{A(r)}+\frac{L^2}{C(r)}\rb^{-\frac{1}{2}}\dd r\label{eq:iintinr}\\
=& \int_0^1 y(u,r_0,v,p)\dd u, \label{eqdxdl}
\end{align}
where in the second step a change of variable $u=\frac{r_0}{r}$ was used and we have changed $L$ and $E$ to the signal velocity $v$ and $r_0$ according to Eq. \eqref{eq:lx0rel}. The $y(u,r_0,v,p)$ denotes the integrand and $p$  collectively stands for all parameters in the metric functions.
And furthermore, $r_0$ could be related to the impact parameter $b$ through \cite{Jia:2020dap}
\bea
\frac{1}{b}&=&\frac{\sqrt{E^2-1}}{L}\label{blerel}\\
&=&\frac{\sqrt{E^2-1}}{\sqrt{E^2-\kappa A(r_0)}}\sqrt{\frac{A(r_0)	}{C(r_0)}}\equiv l\lb \frac{1}{r_0}\rb. \label{br0rel}
\eea

In the weak field limit, $r_0\gg m$, then the integrand in Eq. \eqref{eqdxdl} can be expanded in the powers of $m/r_0$, i.e.,
\begin{align}
    y(u,r_0,v,p)=\sum_{n=1}^\infty y_n(u,v,p) \lb \frac{m}{r_0}\rb^n.
\end{align}
We emphasis that the integration for this series expansion is always feasible because of the particular form that $y_i(u,v,p)$ take (see Ref. \cite{Jia:2020dap}) and therefore in principle calculation to arbitrarily desired order is possible, especially if one has enough time and memory space for a symbolic computation tool to carry out these integrals.
After integration, then the total change of the angular coordinate takes the form
\be
I(r_0,v,p)=\sum_{n=0}^\infty I_n(v,p)\lb \frac{m}{r_0}\rb^n. \label{eq:iinx0}
\ee
Here $I_n(v,p)$ should in general depend on the particle velocity $v$ and parameters $p$ of the spacetime. In GL however, the minimal radial coordinate $r_0$ is still not as conveniently connected to observables, such as lens distance and apparent angles of the images, as the impact parameter $b$. Therefore, it is desirable to replace $r_0$ in Eq. \eqref{eq:iinx0} by $b$. For this purpose, we  can inverse the  function $l\lb\frac{1}{r_0}\rb $ in Eq. \eqref{br0rel} to express $\frac{1}{r_0}$ as a function of $\frac{1}{b}$, also in series form,
\begin{align}
\frac{1}{r_0}=\sum_{n=1}^\infty C_n\lb \frac{1}{b}\rb^n. \label{eq:xinb}
\end{align}
Substituting this into Eq. \eqref{eq:iinx0} and sorting the series as powers of $m/b$, one obtains the change of the angular coordinate as
\be
I(b,v,p)=\sum_{n=0}^\infty I_n^\prime(v,p)\lb \frac{m}{b}\rb^n, \label{eq:iinb}
\ee
where the $I_n^\prime(v,p)$ are the new coefficients.

The deflection angle in the weak field limit is then $\alpha=I(b,v,p)-\pi$.
This deflection angle can be directly used in the GL equation to solve the angular positions and magnifications of images of the source, also as series expansions, as was done in Ref. \cite{Jia:2020dap}.
However, since those computation are also lengthy, in this work we will not solve GL equations, but rather restrain ourselves to the perturbative computation and then analysis of the deflection angles in the four spacetimes mentioned in the introduction.

\section{The deflection angles in various spacetimes \label{sec:metrics}}

\subsection{Deflection in the Bardeen spacetime\label{subsec:Bardeen}}

The regular Bardeen spacetime is described by the metric \cite{Bardeen:1968,AyonBeato:2000zs}
\begin{equation}
A(r)=B(r)^{-1}=1-\frac{2 m r^{2}}{\left(r^{2}+g^{2}\right)^{3 / 2}},~C(r)=r^{2}, \label{eq:bardeenmetric}
\end{equation}
where $m$ is the mass of the spacetime and $g$ is the charge parameter for some nonlinear electrodynamics \cite{AyonBeato:2000zs}. When $g=0$, this reduces to the Schwarzschild spacetime. When $g\neq0$, the spacetime is regular everywhere. This metric approaches the Schwarzschild spacetime at large $r$ and the de Sitter spacetime at small $r$. The asymptotic expansion of $A(r)$ indeed is given by
\be
A(r\to\infty)=1-\frac{2 m}{r}+\frac{3 g^2 m}{r^3}+\mathcal{O}\left(\frac{1}{r}\right)^5, \label{eq:bdlarger}
\ee
from which it is clear that the Bardeen spacetime deviates from the Schwarzschild one from the third order of large $r$.
When the dimensionless parameter $\hat{g}\equiv \frac{|g|}{m}<\frac{4}{3\sqrt{3}}\equiv\hat{g}_c$, this spacetime allows two horizons, which become extremal when $\hat{g}=\hat{g}_c$, and then a non-BH spacetime when $\hat{g} >\hat{g}_c$ \cite{Borde:1994ai}. For $\hat{g}_c<\hat{g}<\hat{g}_p\equiv \frac{48}{25\sqrt{5}}$, there can still exist a photon sphere which might diverge the deflection angle if the trajectory is close to it \cite{Ghaffarnejad:2014zva}.

Using metric \eqref{eq:bardeenmetric} and going through the expansion and integration procedure from Eq. \eqref{eqdxdl} to Eq. \eqref{eq:iinb}, the change of the angular coordinate $I_{\mathrm{B}}(r_0,v,\hat{g})$ for a particle ray with the minimal radial coordinate $r_0$ and velocity $v$ in the Bardeen spacetime becomes, to the eleventh order
\begin{equation}
I_{\mathrm{B}}\left(r_0, v,\hat{g}\right)=\sum_{n=0}^{11} B_{n}(v,\hat{g})\left(\frac{m}{r_0}\right)^{n}+\mathcal{O}\left(\frac{m}{r_0}\right)^{12} \label{eq:ibinx0}
\end{equation}
where the coefficients are
\begin{subequations}
\label{eq:bdinr0}
\begin{align}
B_{0}=
&S_0,\\
B_{1}=
&S_1,\\
B_{2}=
&S_2,\\
B_{3}=
&S_3-2\hat{g} ^2 \left(\frac{3}{v^2}+1\right),\\
B_{4}=
&S_4-\frac{3}{16}\hat{g} ^2 \left(9\pi+\frac{72 \pi -48}{ v^2}+\frac{ 24 \pi -112}{ v^4}\right),\\
B_{5}=
&S_5+3\hat{g} ^2 \left(-4+\frac{3 \pi -52}{v^2}+\frac{21 \pi -52}{v^4}+\frac{6 \pi -22}{v^6}\right)+2\hat{g} ^4 \left(1+\frac{5}{v^2}\right)\\
B_{6}=
&S_6+\hat{g} ^2 \left(-\frac{525 \pi }{64}+\frac{45 (24-31 \pi )}{8 v^2}+\frac{3 (343-93 \pi )}{v^4}+\frac{1537-534 \pi }{2 v^6}
+\frac{9 (41-12 \pi )}{2 v^8}\right)\nn\\
&+\hat{g} ^4 \left(\frac{45 \pi }{16}+\frac{135 \pi -91}{4 v^2}+\frac{90 \pi -323}{4 v^4}\right),\\
B_{7}=
&S_{7}+3\hat{g} ^2 \left(-18+\frac{9 (67 \pi -928)}{16 v^2}+\frac{3 (679 \pi -1956)}{4 v^4}+\frac{ 603 \pi -1997}{v^6}+\frac{1272 \pi -3811}{4 v^8}\right)\nn\\
&+\hat{g}^4 \left(\frac{198}{7}+\frac{8224-477 \pi }{16 v^2}+\frac{3424-1179 \pi }{4 v^4}-\frac{2 (81 \pi -263)}{v^6}\right)-2\hat{g} ^6 \left(1+\frac{7}{v^2}\right),\\
B_{8}=
&S_{8}+3\hat{g}^2\left[-\frac{24255 \pi }{2048}+\left(1197-\frac{184527 \pi }{128}\right)\frac{1}{v^2}+\left(18063-\frac{707787 \pi }{128}\right)\frac{1}{v^4}+\left(\frac{70425}{2}-\frac{92277 \pi }{8}\right)\frac{1}{v^6}\right.\nn\\
&\left.+\left(\frac{54843}{2}-\frac{135729 \pi }{16}\right)\frac{1}{v^8}+\left(\frac{73671}{8}-3024 \pi\right)\frac{1}{v^{10}}+\left(\frac{9411}{8}-360 \pi\right)\frac{1}{v^{12}}\right]\nn\\
&+\hat{g}^4\left[\frac{25725 \pi }{1024}+\left(\frac{22773 \pi }{32}-\frac{2215}{4}\right)\frac{1}{v^2}+\left(\frac{56223 \pi }{32}-\frac{22965}{4}\right)\frac{1}{v^4}+\left(\frac{9105 \pi }{4}-\frac{56017}{8}\right)\frac{1}{v^6}\right.\nn\\
&\left.+\left(\frac{6351 \pi }{8}-\frac{20253}{8}\right)\frac{1}{v^8}\right]+3\hat{g}^6\left[-\frac{175 \pi}{128} +\left(\frac{119}{8}-\frac{175 \pi }{8}\right)\frac{1}{v^2}+\left(\frac{563}{8}-\frac{175 \pi }{8}\right)\frac{1}{v^4}\right],\\
B_{9}=
&S_{9}+\hat{g}^2\left[-\frac{1144}{5}+\left(\frac{29691 \pi }{32}-\frac{58392}{5}\right)\frac{1}{v^2}+\left(\frac{1330839 \pi }{64}-\frac{326454}{5}\right)\frac{1}{v^4}+\left(\frac{1771275 \pi }{32}-176768\right)\frac{1}{v^6}\right.\nn\\
&\left.+\left(\frac{260415 \pi }{4}-201168\right)\frac{1}{v^8}+\left(\frac{66465 \pi }{2}-\frac{531849}{5}\right)\frac{1}{v^{10}}+\left(8832 \pi -\frac{542533}{20}\right)\frac{1}{v^{12}}\right.\nn\\
&\left.+\left(864 \pi -\frac{13989}{5}\right)\frac{1}{v^{14}}\right]+\hat{g}^4\left[204+\left(7600-\frac{35223 \pi }{64}\right)\frac{1}{v^2}+\frac{30371-9696 \pi }{v^4}+\left(56307-\frac{141525 \pi }{8}\right)\frac{1}{v^6}\right.\nn\\
&\left.+\frac{40542-13044 \pi }{v^8}+\frac{10086-3183 \pi }{v^{10}}\right]+\hat{g}^6\left[-\frac{1144}{21}+\left(\frac{1185 \pi }{16}-\frac{8896}{7}\right)\frac{1}{v^2}+\left(\frac{7545 \pi }{8}-2968\right)\frac{1}{v^4}\right.\nn\\
&\left.+\left(\frac{3045 \pi }{4}-2410\right)\frac{1}{v^6}\right]+2\hat{g}^8 \left(1+\frac{9}{v^2}\right),\\
B_{10}=
&S_{10}+3\hat{g}^2\left[-\frac{405405 \pi }{8192}+\left(\frac{14121}{5}-\frac{1670733 \pi }{512}\right)\frac{1}{v^2}+\left(\frac{351267}{5}-\frac{5894733 \pi }{256}\right)\frac{1}{v^4}\right.\nn\\
&\left.+\left(\frac{2614779}{10}-\frac{2688321 \pi }{32}\right)\frac{1}{v^6}+\left(\frac{815845}{2}-\frac{4111275 \pi }{32}\right)\frac{1}{v^8}+\left(\frac{2499479}{8}-\frac{402165 \pi }{4}\right)\frac{1}{v^{10}}\right.\nn\\
&\left.+\left(\frac{4911113}{40}-\frac{77127 \pi }{2}\right)\frac{1}{v^{12}}+\left(\frac{2003647}{80}-8112 \pi\right)\frac{1}{v^{14}}+\left(\frac{173047}{80}-672 \pi\right)\frac{1}{v^{16}}\right]\nn\\
&+\hat{g}^4\left[\frac{654885 \pi }{4096}+\left(\frac{8300571 \pi }{1024}-\frac{26995}{4}\right)\frac{1}{v^2}+\left(\frac{10818879 \pi }{256}-\frac{517357}{4}\right)\frac{1}{v^4}+\left(\frac{3678033 \pi }{32}-\frac{2867039}{8}\right)\frac{1}{v^6}\right.\nn\\
&\left.+\left(\frac{3953013 \pi }{32}-\frac{3128781}{8}\right)\frac{1}{v^8}+\left(\frac{488337 \pi }{8}-\frac{6097593}{32}\right)\frac{1}{v^{10}}+\left(11235 \pi -\frac{1135689}{32}\right)\frac{1}{v^{12}}\right]\nn\\
&+\hat{g}^6\left[-\frac{1}{256} (15435 \pi )+\left(\frac{93473}{56}-\frac{68283 \pi }{32}\right)\frac{1}{v^2}+\left(\frac{1221379}{56}-\frac{114111 \pi }{16}\right)\frac{1}{v^4}+\left(\frac{589497}{16}-\frac{23577 \pi }{2}\right)\frac{1}{v^6}\right.\nn\\
&\left.+\left(\frac{286337}{16}-\frac{11343 \pi }{2}\right)\frac{1}{v^8}\right]+9\hat{g}^8\left[\frac{315 \pi }{512}+\left(\frac{1575 \pi }{128}-\frac{539}{64}\right)\frac{1}{v^2}+\left(\frac{525 \pi }{32}-\frac{3179}{64}\right)\frac{1}{v^4}\right]\\
B_{11}=
&S_{11}+3\hat{g}^2\left[-\frac{2210}{7}+\left(\frac{4293165 \pi }{2048}-\frac{174858}{7}\right)\frac{1}{v^2}+\left(\frac{18076749 \pi }{256}-\frac{8181207}{35}\right)\frac{1}{v^4}\right.\nn\\
&\left.+\left(\frac{21350013 \pi }{64}-\frac{5271603}{5}\right)\frac{1}{v^6}+\left(\frac{11113359 \pi }{16}-\frac{43333089}{20}\right)\frac{1}{v^8}+\left(\frac{5680695 \pi }{8}-\frac{8984733}{4}\right)\frac{1}{v^{10}}\right.\nn\\
&\left.+\left(\frac{812145 \pi }{2}-\frac{10128041}{8}\right)\frac{1}{v^{12}}+\left(123486 \pi -\frac{109722421}{280}\right)\frac{1}{v^{14}}+\left(21408 \pi -\frac{148573919}{2240}\right)\frac{1}{v^{16}}\right.\nn\\
&\left.+\left(1536 \pi -\frac{11028373}{2240}\right)\frac{1}{v^{18}}\right]+\hat{g}^4\left[1196+\left(74576-\frac{12156195 \pi }{2048}\right)\frac{1}{v^2}+\left(542955-\frac{84259845 \pi }{512}\right)\frac{1}{v^4}\right.\nn\\
&\left.+\left(1888397-\frac{38284485 \pi }{64}\right)\frac{1}{v^6}+\left(2938605-\frac{3762405 \pi }{4}\right)\frac{1}{v^8}+\left(\frac{4318233}{2}-\frac{10945485 \pi }{16}\right)\frac{1}{v^{10}}\right.\nn\\
&\left.+\left(\frac{6214129}{8}-\frac{993045 \pi }{4}\right)\frac{1}{v^{12}}+\left(\frac{915743}{8}-36300 \pi\right)\frac{1}{v^{14}}\right]+\hat{g}^6\left[-\frac{6460}{11}+\left(\frac{496905 \pi }{256}-\frac{187008}{7}\right)\frac{1}{v^2}\right.\nn\\
&\left.+\left(\frac{1346391 \pi }{32}-\frac{970369}{7}\right)\frac{1}{v^4}+\left(\frac{828813 \pi }{8}-\frac{2286525}{7}\right)\frac{1}{v^6}+\left(\frac{398151 \pi }{4}-311516\right)\frac{1}{v^8}\right.\nn\\
&\left.+\left(\frac{65019 \pi }{2}-102378\right)\frac{1}{v^{10}}\right]+\hat{g}^8\left[\frac{650}{7}+\left(\frac{18510}{7}-\frac{2475 \pi }{16}\right)\frac{1}{v^2}+\left(\frac{55677}{7}-\frac{4815 \pi }{2}\right)\frac{1}{v^4}\right.\nn\\
&\left.+\left(8049-\frac{40905 \pi }{16}\right)\frac{1}{v^6}\right]-2\hat{g}^{10}\left[1+\frac{11}{v^2}\right].
\end{align}
\end{subequations}
Here $S_n$ is the coefficient of order $\lb\frac{m}{r_0}\rb^n$ in the change of the angular coordinate in Schwarzschild spacetime. Their values are given in Ref. \cite{Jia:2020dap} and also listed in Eq. \eqref{eq:angschinx0} of Appendix \ref{appd:snexp}.

Using Eq. \eqref{eq:xinb}, the $\frac{m}{r_0}$ in Eq. \eqref{eq:ibinx0} can be expressed in powers of $\frac{m}{b}$. The result is
\begin{align}
\frac{m}{r_0}=
&\lb\frac{m}{b}\rb+\frac{1}{v^2}\left(\frac{m}{b}\right)^2+\left(\frac{2}{v^2}+\frac{1}{2 v^4}\right)\left(\frac{m}{b}\right)^3+\left(\frac{4}{v^2}+\frac{4}{v^4}-\frac{3 \hat{g}^2}{2 v^2}\right)\left(\frac{m}{b}\right)^4\nn\\
&+\left[\frac{8}{v^2}+\frac{18}{v^4}+\frac{3}{v^6}-\frac{1}{8 v^8}-3\hat{g}^2\left(\frac{2}{v^2}+\frac{3}{2v^4}\right)\right]\left(\frac{m}{b}\right)^5\nn\\
&+\left[16\left(\frac{1}{v^2}+\frac{4}{v^4}+\frac{2}{v^6}\right)-6\hat{g}^2\left(\frac{3}{v^2}+\frac{6}{v^4}+\frac{1}{v^6}\right)+\frac{15\hat{g}^4}{8v^2}\right]\left(\frac{m}{b}\right)^6\nn\\
&+\left[\left(\frac{32}{v^2}+\frac{200}{v^4}+\frac{200}{v^6}+\frac{25}{v^8}-\frac{5}{4v^{10}}+\frac{1}{16v^{12}}\right)-3\hat{g}^2\left(\frac{16}{v^2}+\frac{60}{v^4}+\frac{30}{v^6}+\frac{5}{4v^8}\right)+3\hat{g}^4\left(\frac{4}{v^2}+\frac{5}{v^4}\right)\right]\left(\frac{m}{b}\right)^7\nn\\
&+\left[64\left(\frac{1}{v^2}+\frac{9}{v^4}+\frac{15}{v^6}+\frac{5}{v^8}\right)-120\hat{g}^2\left(\frac{1}{v^2}+\frac{6}{v^4}+\frac{6}{v^6}+\frac{1}{v^8}\right)+\frac{99\hat{g}^4}{2}\left(\frac{1}{v^2}+\frac{3}{v^4}+\frac{1}{v^6}\right)-\frac{35\hat{g}^6}{16v^2}\right]\left(\frac{m}{b}\right)^8\nn\\
&+\left[\left(\frac{128}{v^2}+\frac{1568}{v^4}+\frac{3920}{v^6}+\frac{2450}{v^8}+\frac{245}{v^{10}}-\frac{49}{4v^{12}}+\frac{7}{8v^{14}}-\frac{5}{128v^{16}}\right)-\frac{3\hat{g}^2}{16}\left(\frac{1536}{v^2}+\frac{13440}{v^4}+\frac{22400}{v^6}\right.\right.\nn\\
&\left.\left.+\frac{8400}{v^8}+\frac{420}{v^{10}}-\frac{7}{v^{12}}\right)+\frac{21\hat{g}^4}{8}\left(\frac{64}{v^2}+\frac{336}{v^4}+\frac{280}{v^6}+\frac{35}{v^8}\right)-5\hat{g}^6\left(\frac{4}{v^2}+\frac{7}{v^4}\right)\right]\left(\frac{m}{b}\right)^9\nn\\
&+\left[256\left(\frac{1}{v^2}+\frac{16}{v^4}+\frac{56}{v^6}+\frac{56}{v^8}+\frac{14}{v^{10}}\right)-672\hat{g}^2\left(\frac{1}{v^2}+\frac{12}{v^4}+\frac{30}{v^6}+\frac{20}{v^8}+\frac{3}{v^{10}}\right)\right.\nn\\
&\left.+102\hat{g}^4\left(\frac{5}{v^2}+\frac{40}{v^4}+\frac{60}{v^6}+\frac{20}{v^8}+\frac{1}{v^{10}}\right)-\frac{429\hat{g}^6}{4}\left(\frac{1}{v^2}+\frac{4}{v^4}+\frac{2}{v^6}\right)+\frac{315\hat{g}^8}{218v^2}\right]\left(\frac{m}{b}\right)^{10}\nn\\
&+\left[\left(\frac{512}{v^2}+\frac{10368}{v^4}+\frac{48384}{v^6}+\frac{70560}{v^8}+\frac{31752}{v^{10}}+\frac{2646}{v^{12}}-\frac{126}{v^{14}}+\frac{81}{8 v^{16}}-\frac{45}{64 v^{18}}+\frac{7}{256 v^{20}}\right)\right.\nn\\
&-\frac{3\hat{g}^2}{32}\left(\frac{16384}{v^2}+\frac{258048}{v^4}+\frac{903168}{v^6}+\frac{940800}{v^8}+\frac{282240}{v^{10}}+\frac{14112}{v^{12}}-\frac{336}{v^{14}}+\frac{9}{v^{16}}\right) \nn\\
&+\frac{45\hat{g}^4}{16}\left(\frac{512}{v^2}+\frac{5760}{v^4}+\frac{13440}{v^6}+\frac{8400}{v^8}+\frac{1260}{v^{10}}+\frac{21}{v^{12}}\right)-7\hat{g}^6\left(\frac{64}{v^2}+\frac{432}{v^4}+\frac{504}{v^6}+\frac{105}{v^8}\right)\nn\\
&\left.+\frac{15\hat{g}^8}{2}\left(\frac{4}{v^2}+\frac{9}{v^4}\right)\right]\left(\frac{m}{b}\right)^{11}
+\mathcal{O}\left(\frac{m}{b}\right)^{12}
\end{align}
Putting this into Eq. \eqref{eq:ibinx0}, we obtain the change of the angular coordinate in powers of $\frac{m}{b}$ for general velocity $v$,
\begin{equation}
I_\mathrm{B}(b, v,\hat{g})=\sum_{n=0}^{11} B_{n}^{\prime}(v,\hat{g})\left(\frac{m}{b}\right)^{n}+\mathcal{O}\left(\frac{m}{b}\right)^{12}
\end{equation}
where the coefficients are
\begin{subequations}\label{eq:bardeenanginb}
\begin{align}
 B^\prime_{0}=&S^\prime_0,\\
 B^\prime_{1}=&S^\prime_1,\\
 B^\prime_{2}=&S^\prime_2,\\
 B^\prime_{3}=&S^\prime_3-2\hat{g} ^2 \left(1+\frac{3}{v^2}\right),\label{eq:b3inb}\\
 B^\prime_{4}=&S^\prime_4 -\frac{9\pi}{2}\hat{g} ^2 \left(\frac{3  }{8}+\frac{3  }{ v^2}+\frac{ 1 }{ v^4}\right),\\
 B^\prime_{5}=&S^\prime_5-12\hat{g} ^2 \left(1+\frac{15}{v^2}+\frac{15}{v^4}+\frac{1}{v^6}\right)+2\hat{g} ^4 \left(1+\frac{5}{v^2}\right),\\
 B^\prime_{6}=&S^\prime_6-105\pi\hat{g} ^2 \left(\frac{5 }{64}+\frac{15  }{8 v^2}+\frac{15 }{4 v^4}+\frac{1 }{v^6}\right)+\frac{45\pi}{2}\hat{g} ^4 \left(\frac{1 }{8}+\frac{3 }{2 v^2}+\frac{1 }{ v^4}\right),\\
B^\prime_{7}=&S^\prime_7-6\hat{g} ^2 \left(9+\frac{315}{v^2}+\frac{1050}{v^4}+\frac{630}{v^6}+\frac{45}{v^8}-\frac{1}{v^{10}}\right)+198\hat{g} ^4 \left(\frac{1}{7}+\frac{3}{v^2}+\frac{5}{v^4}+\frac{1}{v^6}\right)-2\hat{g} ^6 \left(1+\frac{7}{v^2}\right),\\
B^\prime_{8}=&S^\prime_8-\frac{10395\pi\hat{g}^2}{2048}\left(7+\frac{336}{v^2}+\frac{1680}{v^4}+\frac{1792}{v^6}+\frac{384}{v^8}\right)+\frac{735\pi\hat{g}^4}{1024}\left(35+\frac{1120}{v^2}+\frac{3360}{v^4}+\frac{1792}{v^6}+\frac{128}{v^8}\right)\nn\\
&-\frac{525\pi\hat{g}^6}{128}\left(1+\frac{16}{v^2}+\frac{16}{v^4}\right),\\
B^\prime_{9}=&S^\prime_9-\frac{8 \hat{g}^2}{5}\left(143+\frac{9009}{v^2}+\frac{63063}{v^4}+\frac{105105}{v^6}+\frac{45045}{v^8}+\frac{3003}{v^{10}}-\frac{91}{v^{12}}+\frac{3}{v^{14}}\right)\nn\\
&+204\hat{g}^4\left(1+\frac{45}{v^2}+\frac{210}{v^4}+\frac{210}{v^6}+\frac{45}{v^8}+\frac{1}{v^{10}}\right)-\frac{1144\hat{g}^6}{21}\left(1+\frac{27}{v^2}+\frac{63}{v^4}+\frac{21}{v^6}\right)+2\hat{g}^8\left(1+\frac{9}{v^2}\right) ,\\
B^\prime_{10}=&S^\prime_{10}-\frac{135135\pi \hat{g}^2}{8192}\left(9+\frac{720}{v^2}+\frac{6720}{v^4}+\frac{16128}{v^6}+\frac{11520}{v^8}+\frac{2048}{v^{10}}\right)+\frac{31185\pi \hat{g}^4}{4096}\left(21+\frac{1260}{v^2}+\frac{8400}{v^4}\right.\nn\\
&\left.+\frac{13440}{v^6}+\frac{5760}{v^8}+\frac{512}{v^{10}}\right)-\frac{2205\pi \hat{g}^6}{256}\left(7+\frac{280}{v^2}+\frac{1120}{v^4}+\frac{896}{v^6}+\frac{128}{v^8}\right)+\frac{945\pi\hat{g}^8}{512}\left(3+\frac{60}{v^2}+\frac{80}{v^4}\right) ,\\
B^\prime_{11}=&S^\prime_{11}-\frac{6\hat{g}^2}{7}\left(1105+\frac{109395}{v^2}+\frac{1312740}{v^4}+\frac{4288284}{v^6}+\frac{4594590}{v^8}+\frac{1531530}{v^{10}}+\frac{92820}{v^{12}}-\frac{3060}{v^{14}}+\frac{153}{v^{16}}-\frac{5}{v^{18}}\right)\nn\\
&+92\hat{g}^4\left(13+\frac{1001}{v^2}+\frac{9009}{v^4}+\frac{21021}{v^6}+\frac{15015}{v^8}+\frac{3003}{v^{10}}+\frac{91}{v^{12}}-\frac{1}{v^{14}}\right)\nn\\
&-\frac{6460\hat{g}^6}{11}\left(1+\frac{55}{v^2}+\frac{330}{v^4}+\frac{462}{v^6}+\frac{165}{v^8}+\frac{11}{v^{10}}\right)+\frac{130\hat{g}^8}{7}\left(5+\frac{165}{v^2}+\frac{495}{v^4}+\frac{231}{v^6}\right)-2\hat{g}^{10}\left(1+\frac{11}{v^2}\right).\\
  \end{align}
\end{subequations}
Here $S_n^\prime$ is the coefficient of order $\lb \frac{m}{b}\rb^n$ in the change of the angular coordinate of the Schwarzschild spacetime given in Eq. \eqref{eq:angschinx0}. Note that if we set $v=c$ for lightray in Eqs. \eqref{eq:bdinr0} and \eqref{eq:bardeenanginb}, their first five orders reduce to the Eqs. (2.18) and (2.20) of Ref. \cite{Ghaffarnejad:2014zva} respectively.

From these equations, it is seen that in $I_\mathrm{B}(b,v,\hat{g})$ the deviation due to parameter $\hat{g}$ from the Schwarzschild case is only present from and above the third order of $m/b$. Since the deflection angle in the weak field limit is expected to be sensitive only to the asymptotic behavior of the spacetime, this implies that the metric functions at large $r$ must only differ from the Schwarzschild spacetime starting from third order too. This is indeed confirmed in the expansion of the lapse function in Eq. \eqref{eq:bdlarger}, where the difference only appears from the third order.

\begin{center}
\begin{figure}[htp!]
\includegraphics[width=0.45\textwidth]{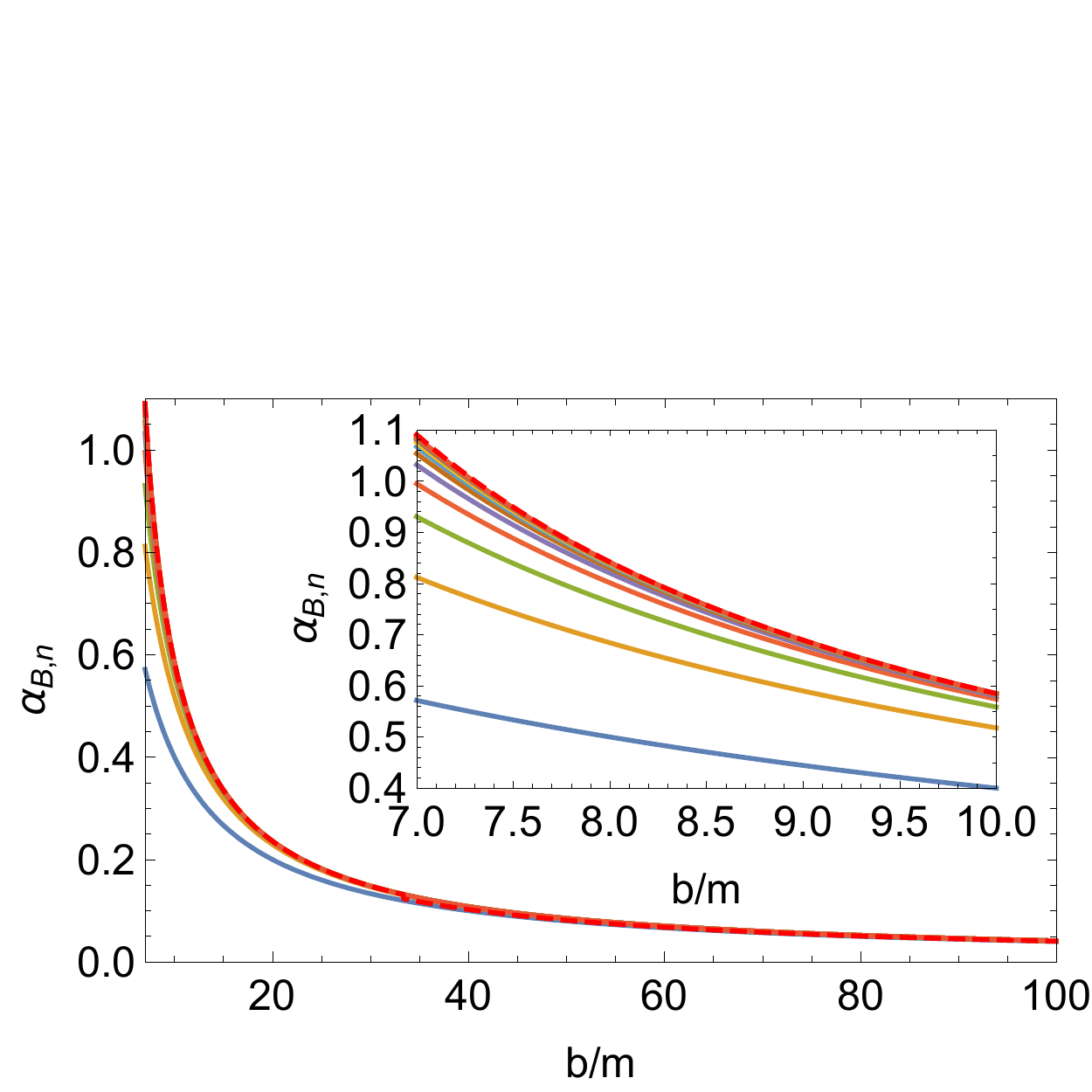}
\hspace{0.5cm}\includegraphics[width=0.45\textwidth]{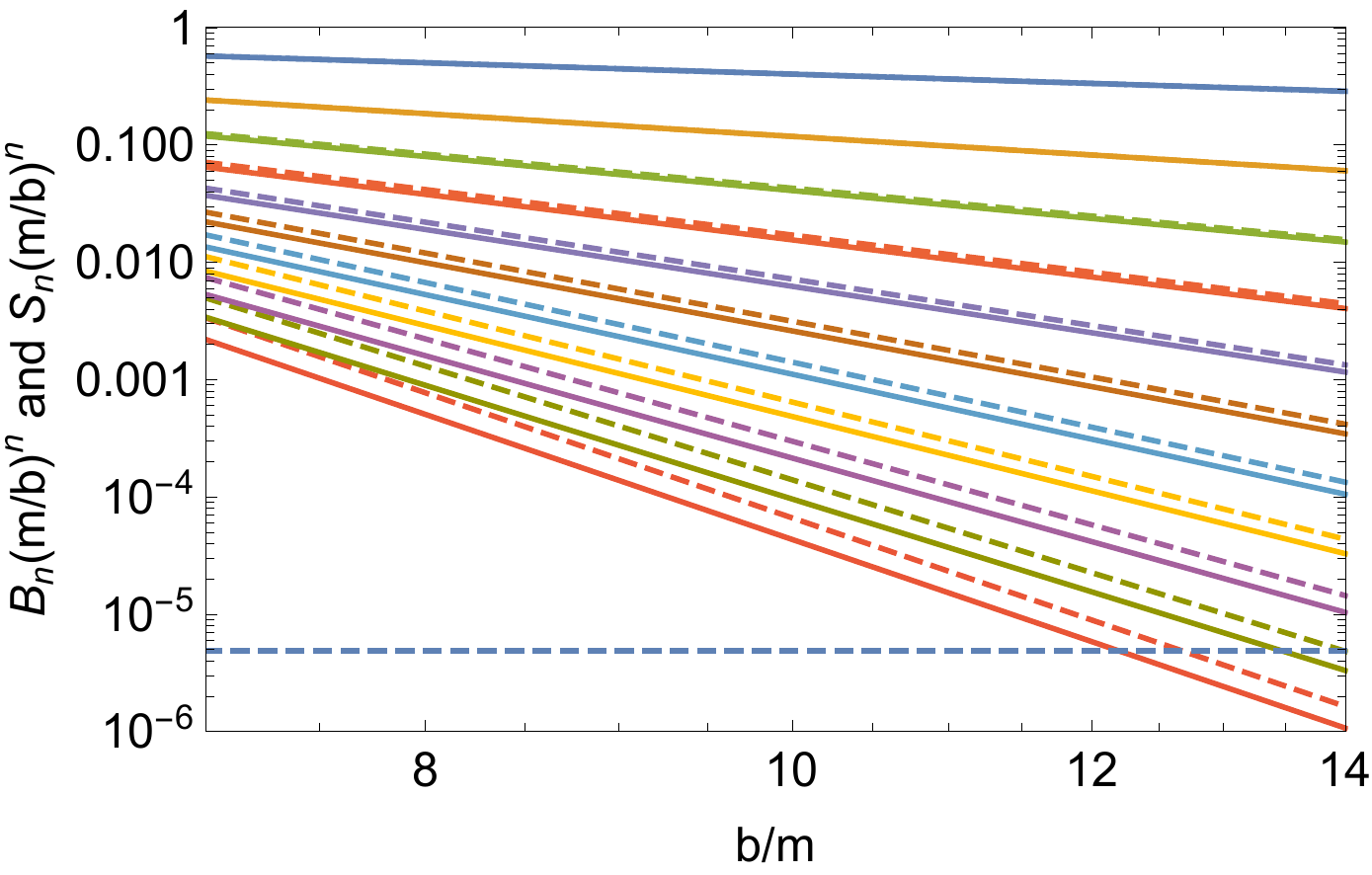}\\
(a)\hspace{8cm}(b)\\
\includegraphics[width=0.45\textwidth]{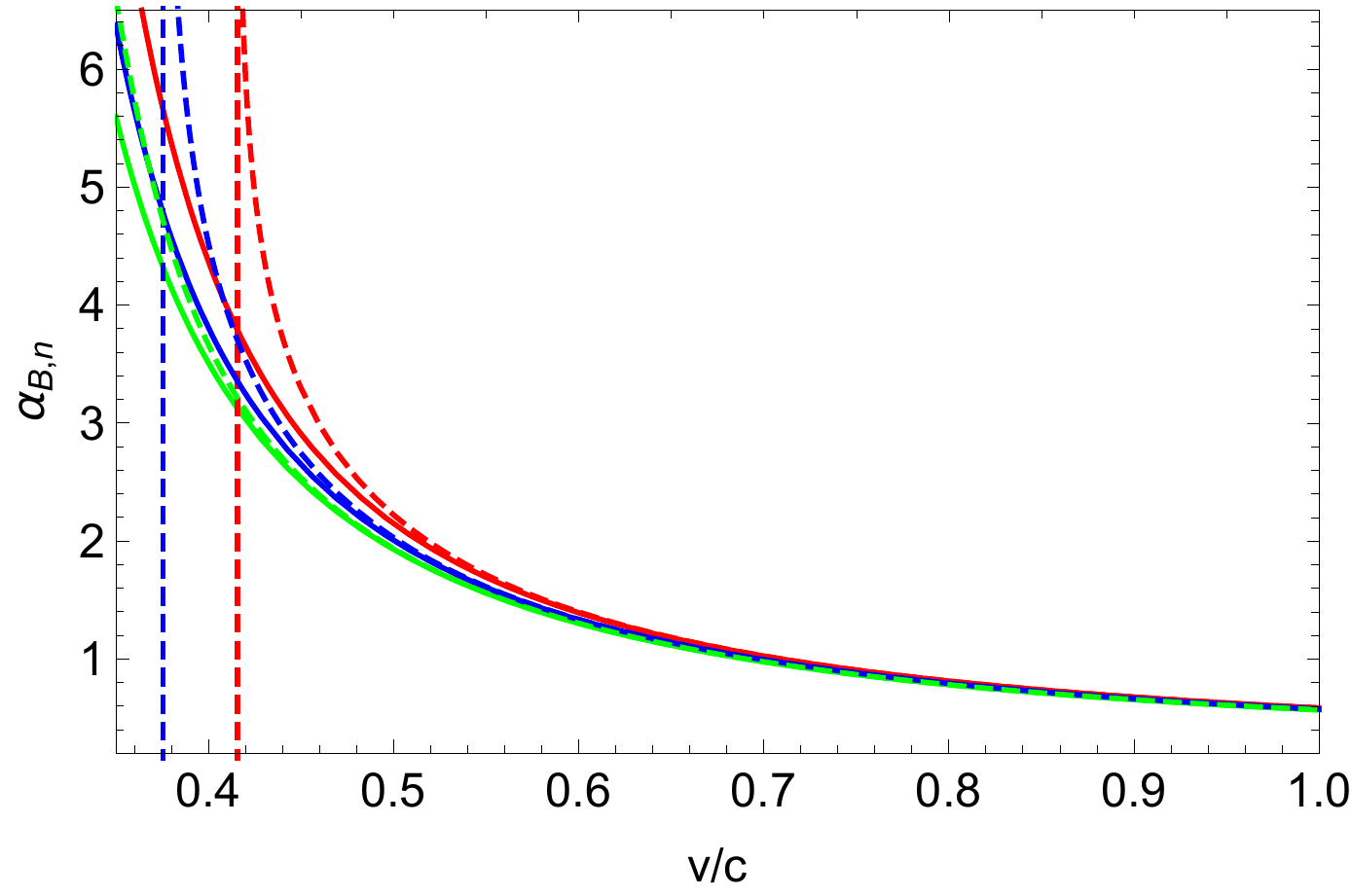}
\hspace{0.5cm}\includegraphics[width=0.45\textwidth]{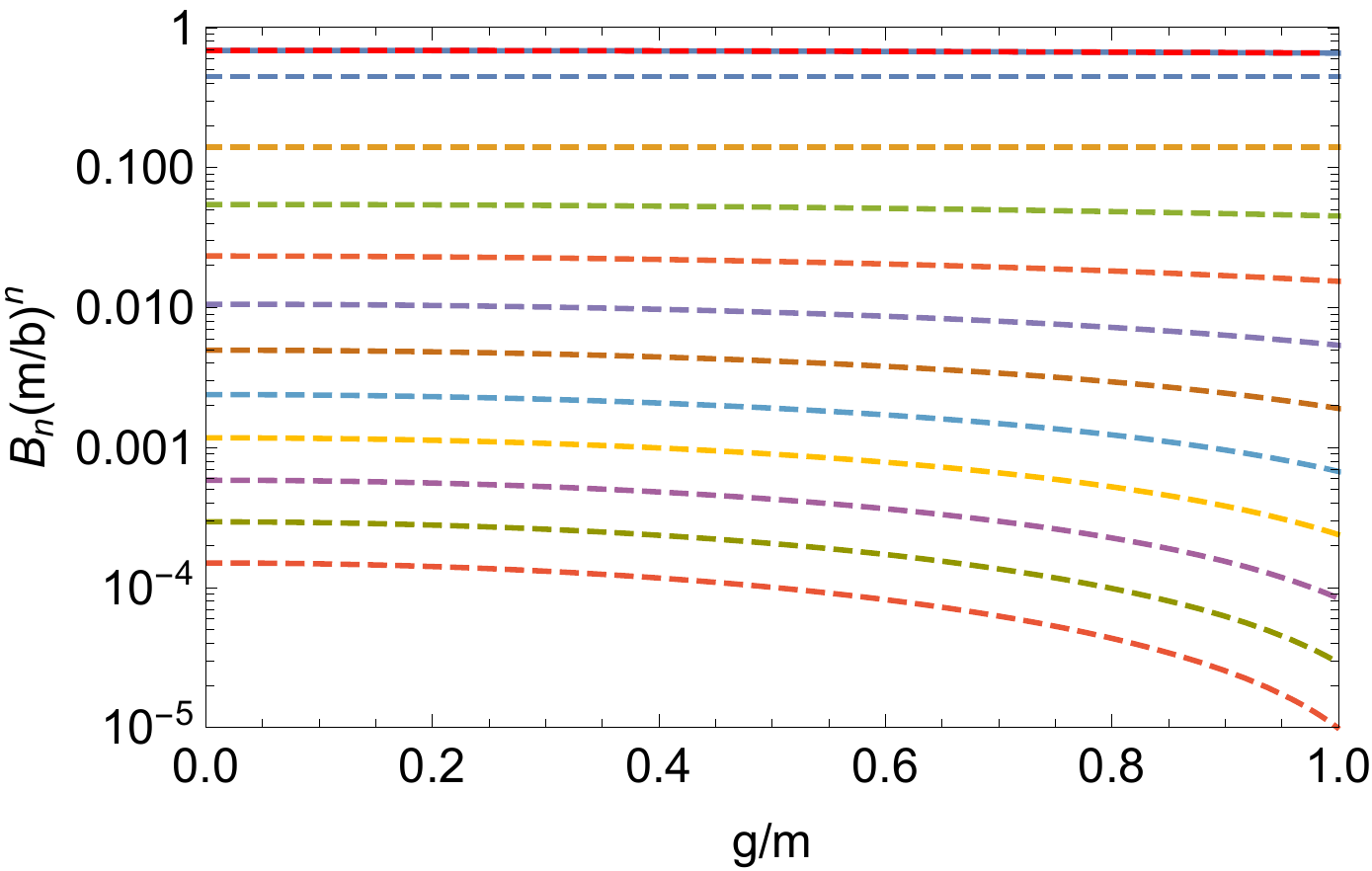}\\
(c)\hspace{8cm}(d)
\caption{The deflection angles in the Bardeen spacetime. (a) Partial sums \eqref{eq:bdpartialsum} (solid curves) and the exact deflection angle (dashed red curve) for $b/m$ from 7 to 100 and from 7 to 14 in the inset and $v=c,~\hat{g}=0.5$. From bottom to top curve in each plot, the maximum index in the partial sum increases from 1 to 7. (b) Contribution from each order of Eq. \eqref{eq:bardeenanginb} (solid lines from top to bottom curve order increases from 1 to 17) for $v=c,~\hat{g}=0.5$ and the corresponding order results in the Schwarzschild case (dashed lines). (c) Partial sums $\alpha_{\mathrm{B},11}$ (solid lines) and exact values (dash lines) of the deflection angle for $b=10m$ and $\hat{g}_1=0.5$ (red lines), $\hat{g}_2=0.8$ (blue lines) and $\hat{g}_3=1.0$ (green lines). The vertical dash lines are two critical velocities. (d) Contributions from each order of Eq. \eqref{eq:bardeenanginb} (dashed curves), partial sum $\alpha_{\mathrm{B},11}$ (top red dashed curve) and exact value (top solid curve) of the deflection angle for $b=10m,~\hat{v}=0.9c$. \label{fig:bardeen1}}
\end{figure}
\end{center}

% no description of the line style and unnecessary other parameters in the text. do this in the caption.

To study how the deflection angle depends on various kinematic and spacetime parameters, we plot the deflection angle $\alpha_\mathrm{B}\equiv I_\mathrm{B}(b,v,\hat{g})-\pi$ in Fig. \ref{fig:bardeen1}. In Fig. \ref{fig:bardeen1} (a), we plot the partial sums defined as the sum of Eqs. \eqref{eq:bardeenanginb} to the $n$-th order as
\begin{align}
    \alpha_{\mathrm{B},n}=\sum_{i=1}^n B^\prime(v,\hat{g})\lb \frac{m}{b}\rb^i,\label{eq:bdpartialsum}
\end{align}
and the exact value of the deflection angle calculated by numerically integration Eq. \eqref{eq:iintinr} for the Bardeen metric. It is seen that the partial sums $\alpha_{\mathrm{B},n}$ approaches the true value as $n$ increases. The highest order partial sum $\alpha_{\mathrm{B},11}$ practically overlaps with the exact value for $b$ greater than $7m$. As $b$ decreases from large value, the deflection angle monotonically increases. The contribution from higher orders in Eqs. \eqref{eq:bardeenanginb} only becomes important when $b$ is small enough. At $b=7m$, the deflection angle reached a value of $\alpha_{\mathrm{B}}=0.27\pi$, which is rather large. Therefore this suggests that our series deflection angle works even when the gravitational field is not weak.

In Fig. \ref{fig:bardeen1} (b), the contribution from each order of Eq. \eqref{eq:bardeenanginb} and the corresponding orders for the Schwarzschild metric are plotted for $v=c$ and $\hat{g}=0.5$. It is seen that for any $b$, as the order increases, the contribution from each order decreases linearly in this log-log plot. This suggests that the summation of these contributions will converge as they should. Moreover, it is also seen that the contributions to the deflection angle in the Bardeen spacetime only deviate from the Schwarzschild ones from the third order. This is a reflection that the asymptotic expansion \eqref{eq:bdlarger} of the lapse function of the Bardeen metric only deviates from the Schwarzschild lapse function from the third order. Furthermore, it is seen that comparing to the Schwarzschild case, a nonzero $\hat{g}$ decreases the deflection angle in all orders.
The horizontal line at 1 [as] here represents the typical precision in the GL by galaxy or galaxy cluster. It shows that the seventh order result can reach this accuracy  for $b$ that is as small as about $12.1m$.

In Fig. \ref{fig:bardeen1} (c), the dependence of the partial sum $\alpha_{\mathrm{B},11}$ and the exact value of the deflection angle are plotted for three representative value of parameter $\hat{g}$, $\hat{g}_1=0.5,~\hat{g}_1=0.8$ and
$\hat{g}_3=1.0$. It is known in the Schwarzschild and RN BH spacetimes that for a given $b$, there exists a critical velocity $v_c$ below which the deflection angle diverges to infinity, i.e., the particle will enter the BH event horizon. Here for $\hat{g}_1$ which is smaller than $\hat{g}_c$ and $\hat{g}_2$ which is between $\hat{g}_c$ and $\hat{g}_p$, it is known that the spacetime contains a photon sphere in both cases and therefore for some fixed $b$ there shall exist a lower limit of the velocity $v_c$ below which the true deflection angle diverges. It is seen then from the plot that this critical value for $\hat{g}_1$ is roughly $0.42c$ and for $\hat{g}_2$ is $0.38c$ when $b=10m$. On the other hand, when $\hat{g}=\hat{g}_3$ which is larger than $\hat{g}_p$, the photon sphere disappears and the partial sum $\alpha_{\mathrm{B},11}$ and the true value of the deflection angles agrees in a wider range of the velocity.

Fig. \ref{fig:bardeen1} (d) shows the effect of the charge $\hat{g}$ on the contributions to deflection angle at various orders, as well as their partial sum $\alpha_{\mathrm{B},11}$ and the exact value of the deflection angle.
As seen from Eq. \eqref{eq:bardeenanginb}, comparing to the Schwarzschild metric, the effect of $\hat{g}$ appears in the third and above orders. Therefore for a finite $\hat{g}$, in general its effect should be small when $b$ is large, and it only becomes apparent when $b$ is reasonably small, i.e., when the gravitational field is not weak anymore. To illustrate its effect, therefore we choose $b=10m$ in the plot.
It is seen that as dictated by Eqs. \eqref{eq:bardeenanginb}, $\hat{g}$ monotonically decreases the deflection angle from the third order and does the same to the partial sum. In the entire range of $\hat{g}$, both below or above the critical $\hat{g}_c$ and $\hat{g}_p$, this partial sum agrees with the true value of the deflection angle.

\subsection{Deflection in the Hayward spacetime\label{subsec:Hayward}}

The Hayward metric is given by \cite{Hayward:2005gi}
\begin{equation}
  \begin{aligned}
  A(r)=B(r)^{-1}=1-\frac{2mr^{2}}{r^{3}+2l^{2}m},~ C(r)=r^2. \label{eq:haywardmetric}
  \end{aligned}
\end{equation}
Here $m$ is the mass of spacetime and $l$ characterize the central energy density $\frac{3}{8\pi l^2}$ \cite{Hayward:2005gi,Chiba:2017nml}. When $l=0$, this reduces to the Schwarzschild spacetime. When $l\neq0$ however, similar to the Bardeen case, the Hayward spacetime is regular everywhere. Moreover, one can also verify that the Hayward spacetime approaches the Schwarzschild one at large $r$ and the de Sitter spacetime at small $r$ by expanding $A(r)$
\begin{align}
&A(r\to\infty)=1-\frac{2 m}{r}+\frac{4 l^2 m^2}{r^4}+\mathcal{O}\left(\frac{1}{r}\right)^7,
\label{eq:haywardarexp}   \\
&A(r\to 0)=1-\frac{r^2}{l^2}+\frac{r^5}{2 l^4 m}+O\left(r\right)^8.
\end{align}
Clearly, the Hayward spacetime deviates from Schwarzschild spacetime at large $r$ only from the fourth order.

Depending on the value of the dimensionless parameter $\hat{l}\equiv \frac{|l|}{m}$, this metric corresponds to a BH or non-BH spacetime. When $\hat{l}$ is smaller than a critical value, i.e., $\hat{l}<\frac{4}{3\sqrt{3}}\equiv \hat{l}_c$, there are two horizons, allowing a normal BH. When  $\hat{l}=\hat{l}_c$, the BH becomes extremal and when $\hat{l}>\hat{l}_c$, the spacetime becomes a non-BH one.
For $\hat{l}_c<\hat{l}<\hat{l}_p\equiv \frac{25}{24}\sqrt{\frac{5}{6}}$, there can still exist a photon sphere which might diverge the deflection angle if the trajectory is close to it \cite{Chiba:2017nml}.
Using metric \eqref{eq:haywardmetric} and going through the expansion and integration procedure from Eq. \eqref{eqdxdl} to Eq. \eqref{eq:iinb}, the change of the angular coordinate takes the form
\be
I_{\mathrm{H}}(r_0,v,\hat{l})=\sum_{n=0}^{13}H_{n}(v,\hat{l})\left(\frac{m}{r_0}\right)^n +\mathcal{O}\left(\frac{m}{r_0}\right)^{14}, \label{eq:hwdix0}
\ee
where
\begin{subequations}
\label{eq:hinx0coeff}
\begin{align}
H_{0}=&S_0,\\
H_{1}=&S_1,\\
H_{2}=&S_2,\\
H_{3}=&S_3,\\
H_{4}=&S_{4}-3 \pi \hat{l}^2 \left(\frac{1}{4}+\frac{1}{v^2}\right),\\
H_{5}=& S_5+\hat{l}^2\left[-\frac{32}{5}+\left(\pi-20\right)\frac{3}{v^2} +\left(3 \pi -7\right)\frac{4}{v^4}\right],\\
H_{6}=&S_6+\hat{l}^2\left[-\frac{75}{16} \pi+\left(12-19 \pi \right) \frac{4}{v^2}  + \left(72-19 \pi \right)\frac{9}{2 v^4}+\left(148-51\pi\right)\frac{1}{v^6}\right],\\
H_{7}=&S_7+\hat{l}^2\left[-32 + \left(417 \pi -6112\right) \frac{1}{8 v^2}+ \left(2217 \pi -6224\right) \frac{1}{4 v^4}+\left(507 \pi -1678\right)\frac{1}{v^6}+\left(186 \pi -566\right)\frac{1}{v^8}\right] \nn\\
& +\frac{128\hat{l}^4}{35} \left(1+\frac{7} {v^2}\right),\\
H_{8}=&S_8+\hat{l}^2 \left[-\frac{11025 \pi
   }{512}+\left(560-\frac{46845\pi}{64}\right)\frac{1}{v^2}+\left(7564-\frac{36405 \pi }{16}\right)\frac{1}{v^4}+\left(11468-\frac{15045\pi}{4}\right)\frac{1}{v^6}\right.\nn\\
   &\left.+\left(7246-\frac{9015 \pi }{4}\right)\frac{1}{v^8}+\left(1854-600\pi \right)\frac{1}{v^{10}}\right] +\frac{3\hat{l}^4}{320}\left[525 \pi +\left(485\pi-224\right)\frac{16}{v^2}+\left(365 \pi -1248\right)\frac{16}{v^4}\right],\\
H_{9}=&S_9+\hat{l}^2\left[-\frac{704}{5}+\left(\frac{30801\pi}{64}-\frac{31276}{5}\right)\frac{1}{v^2}+\left(\frac{73815\pi}{8}-28580\right)\frac{1}{v^4 }+\left(\frac{41355 \pi }{2}-66034\right)\frac{1}{v^6}\right.\nn\\
&\left.+\left(-3437+1110\pi\right)\frac{1}{v^8}+\left(8550 \pi -\frac{272133}{10}\right)\frac{1}{v^{10}}+\left(1776 \pi -\frac{55237}{10}\right)\frac{1}{v^{12}}\right]\nn\\
   &+\hat{l}^4\left[\frac{1024}{21}+\left(\frac{40368}{35}-\frac{555 \pi }{8}\right)\frac{1}{v^2}+\left(\frac{1814}{5}-119 \pi\right)\frac{9 }{v^4}+\left(\frac{3866}{5}-243 \pi \right)\frac{2}{v^6}\right],\\
H_{10}=&S_{10}+\hat{l}^2\left[-\frac{189189 \pi }{2048}+\left(\frac{22096}{5}-\frac{5477631 \pi }{1024}\right)\frac{1}{v^2}+\left(\frac{500404}{5}-\frac{4153713 \pi}{128}\right)\frac{1}{v^4}\right.\nn\\
   &\left.+\left(314612-\frac{6465855 \pi }{64}\right)\frac{1}{v^6}+\left(425270-\frac{2145825 \pi }{16}\right)\frac{1}{v^8}+\left(\frac{1387972}{5}-\frac{712983 \pi }{8}\right)\frac{1}{v^{10}}\right.\nn\\
   &\left.+\left(\frac{923927}{10}-29166 \pi\right)\frac{1}{v^{12}}+\left(\frac{154283}{10}-4944 \pi \right)\frac{1}{v^{14}}\right]\nn\\
   &+\hat{l}^4\left[\frac{11025 \pi }{256}+\left(\frac{47349 \pi}{32}-\frac{36024}{35}\right)\frac{1}{v^2}+\left(4635 \pi -\frac{101368}{7}\right)\frac{1}{v^4}+\left(\frac{13833 \pi}{2}-\frac{107984}{5}\right)\frac{1}{v^6}\right.\nn\\
   &\left.+\left(\frac{6363 \pi }{2}-\frac{50224}{5}\right)\frac{1}{v^8}\right]-\hat{l}^{6}\left[\frac{63 \pi }{32}+\frac{315 \pi }{16 v^2}\right],\\
H_{11}=&S_{11}+\hat{l}^2\left[-\frac{189189 \pi }{2048}+\left(\frac{22096}{5}-\frac{5477631 \pi }{1024}\right)
   \frac{1}{v^2}+\left(\frac{500404}{5}-\frac{4153713 \pi }{128}\right)
   \frac{1}{v^4}\right.\nn\\
   &\left.+\left(314612-\frac{6465855 \pi }{64}\right)
   \frac{1}{v^6}+\left(425270-\frac{2145825 \pi }{16}\right)
   \frac{1}{v^8}+\left(\frac{1387972}{5}-\frac{712983 \pi }{8}\right)
   \frac{1}{v^{10}}\right.\nn\\
   &\left.+\left(\frac{923927}{10}-29166 \pi \right)
   \frac{1}{v^{12}}+\left(\frac{154283}{10}-4944 \pi \right) \frac{1}{v^{14}}\right]\nn\\
   &+\hat{l}^4\left[\frac{11025 \pi }{256}+\left(-\frac{36024}{35}+\frac{47349 \pi }{32}\right)
   \frac{1}{v^2}+\left(-\frac{101368}{7}+4635 \pi \right)
   \frac{1}{v^4}+\left(-\frac{107984}{5}+\frac{13833 \pi }{2}\right)
   \frac{1}{v^6}\right.\nn\\
   &\left.+\left(-\frac{50224}{5}+\frac{6363 \pi }{2}\right) \frac{1}{v^8}\right]-\hat{l}^{6}\left(\frac{63 \pi}{32}+\frac{315 \pi }{16}\frac{1}{v^2}\right),\\
H_{12}=&S_{12}+\hat{l}^2\left[-\frac{4160}{7}+\left(-\frac{1477124}{35}+\frac{3550617 \pi }{1024}\right)
   \frac{1}{v^2}+\left(-\frac{1706276}{5}+\frac{53216523 \pi }{512}\right)
   \frac{1}{v^4}\right.\nn\\
   &\left.+\left(-\frac{6798822}{5}+\frac{27563487 \pi }{64}\right)
   \frac{1}{v^6}+\left(-2432070+\frac{24930075 \pi }{32}\right)
   \frac{1}{v^8}+\left(-\frac{4458653}{2}+\frac{5644605 \pi }{8}\right)
   \frac{1}{v^{10}}\right.\nn\\
   &\left.+\left(-\frac{10966099}{10}+\frac{1404003 \pi }{4}\right)
   \frac{1}{v^{12}}+\left(-\frac{40754733}{140}+92136 \pi \right)
   \frac{1}{v^{14}}+\left(-\frac{5757121}{140}+13152 \pi \right) \frac{1}{v^{16}}\right]\nn\\
   &+\hat{l}^4\left[\frac{1120581 \pi }{4096}+\left(-\frac{451256}{35}+\frac{8388789 \pi }{512}\right)
   \frac{1}{v^2}+\left(-\frac{10870808}{35}+\frac{53537775 \pi }{512}\right)
   \frac{1}{v^4}\right.\nn\\
   &\left.+\left(-\frac{36253528}{35}+\frac{5282055 \pi }{16}\right)
   \frac{1}{v^6}+\left(-\frac{49805888}{35}+\frac{14449935 \pi }{32}\right)
   \frac{1}{v^8}\right.\nn\\
   &\left.+\left(-\frac{4569773}{5}+\frac{1166163 \pi }{4}\right)
   \frac{1}{v^{10}}+\left(-\frac{1251989}{5}+\frac{318411 \pi }{4}\right) \frac{1}{v^{12}}\right]\nn\\
   &+\hat{l}^6\left[-\frac{1}{128} (5775 \pi )+\left(\frac{89344}{105}-\frac{45507 \pi }{32}\right)
   \frac{1}{v^2}+\left(\frac{1141024}{105}-\frac{58407 \pi }{16}\right)
   \frac{1}{v^4}+\left(\frac{356704}{35}-\frac{12963 \pi }{4}\right) \frac{1}{v^6}\right],\\
H_{13}=&S_{13}+\hat{l}^2\left[-\frac{51680}{21}+\left(-\frac{5355932}{21}+\frac{358002735 \pi }{16384}\right)
   \frac{1}{v^2}+\left(-\frac{22483612}{7}+\frac{3857795085 \pi }{4096}\right)
   \frac{1}{v^4}\right.\nn\\
   &+\left(-\frac{137874110}{7}+\frac{3199571385 \pi }{512}\right)
   \frac{1}{v^6}+\left(-57187718+\frac{2338559475 \pi }{128}\right)
   \frac{1}{v^8}\nn\\
   &+\left(-\frac{176216073}{2}+\frac{1789998105 \pi }{64}\right)
   \frac{1}{v^{10}}+\left(-\frac{154959447}{2}+\frac{395556735 \pi }{16}\right)
   \frac{1}{v^{12}}\nn\\
   &+\left(-\frac{1150481683}{28}+\frac{52187835 \pi }{4}\right)
   \frac{1}{v^{14}}+\left(-\frac{184890359}{14}+4215045 \pi \right)
   \frac{1}{v^{16}}\nn\\
   &\left.+\left(-\frac{1656792409}{672}+782400 \pi \right)
   \frac{1}{v^{18}}+\left(-\frac{177870673}{672}+84480 \pi \right) \frac{1}{v^{20}}\right]\nn\\
   &+\hat{l}^4\left[2048+\left(\frac{5311024}{35}-\frac{12435471 \pi }{1024}\right)
   \frac{1}{v^2}+\left(\frac{9003788}{7}-\frac{48647511 \pi }{128}\right)
   \frac{1}{v^4}\right.\nn\\
   &+\left(\frac{37203116}{7}-\frac{216301905 \pi }{128}\right)
   \frac{1}{v^6}+\left(\frac{70329374}{7}-\frac{6413085 \pi }{2}\right)
   \frac{1}{v^8}+\left(\frac{66486484}{7}-\frac{24143805 \pi }{8}\right)
   \frac{1}{v^{10}}\nn\\
   &\left.+\left(\frac{46206759}{10}-1472517 \pi \right)
   \frac{1}{v^{12}}
  +\left(\frac{2082727}{2}-331245 \pi \right) \frac{1}{v^{14}}\right]\nn\\
   &+\hat{l}^6\left[-\frac{204800}{429}+\left(-\frac{8109616}{385}+\frac{45813 \pi }{32}\right)
   \frac{1}{v^2}+\left(-\frac{1888624}{21}+\frac{213273 \pi }{8}\right)
   \frac{1}{v^4}\right.\nn\\
   &\left.+\left(-\frac{2479808}{15}+52671 \pi \right)
   \frac{1}{v^6}+\left(-\frac{3317952}{35}+\frac{60453 \pi }{2}\right) \frac{1}{v^8}\right]+\hat{l}^8\left(\frac{32768}{3003}+\frac{32768}{231}\frac{1}{v^2}\right).
\end{align}
\end{subequations}
Here $S_n$ are the change of the angular coordinate at corresponding orders in the Schwarzschild spacetime given in  Eq. \eqref{eq:angschinx0}.

To express the deflection angle in terms of the impact parameter $b$, we express $r_0$ using $b$ using Eq. \eqref{eq:xinb}
\begin{align}
\frac{m}{r_0}=&\frac{m}{b}+\left(\frac{1}{v^2}\right)\left(\frac{m}{b}\right)^2 +\left(\frac{2}{v^2}+\frac{1}{2 v^4}\right)\left(\frac{m}{b}\right)^3 +4\left(\frac{1}{v^2}+ \frac{1}{v^4}\right)\left(\frac{m}{b}\right)^4\nn\\
&+\left[\left(\frac{8}{v^2}+\frac{18}{v^4}+\frac{3}{v^6}-\frac{1}{8 v^8}\right)-2 \hat{l}^2 \frac{1}{v^2}\right]\left(\frac{m}{b}\right)^5+\left[16\left(\frac{1}{v^2}+\frac{4}{v^4}+\frac{2}{v^6}\right)-8 \hat{l}^2 \left(\frac{1}{v^2}+\frac{1}{v^4}\right)\right]\left(\frac{m}{b}\right)^6\nn\\
&+\left[\left(\frac {32} {v^2} + \frac {200} {v^4}  + \frac {200} {v^6} + \frac {25} {v^8} - \frac {5} {4 v^{10}}  +\frac {1} {16 v^{12}} \right)-\hat{l}^2 \left(\frac{24}{v^2}+\frac{60}{v^4}+\frac{15}{v^6}\right)\right]\left(\frac{m}{b}\right)^7\nn\\
&+\left[\left(\frac{320}{v^8}+\frac{960}{v^6}+\frac{576}{v^4}+\frac{64}{v^2}\right)-16 \hat{l}^2 \left(\frac{4}{v^2}+\frac{18}{v^4}+\frac{12}{v^6}+\frac{1}{v^8}\right)\right]\left(\frac{m}{b}\right)^8 \nn\\
&+\left[\frac{128}{v^2}+\frac{1568}{v^4}+\frac{3920}{v^6}+\frac{2450}{v^8}+\frac{245}{v^{10}}-\frac{49}{4 v^{12}}+\frac{7}{8 v^{14}}-\frac{5}{128 v^{16}}\right.\nn\\
&\left.-\hat{l}^2\left( \frac{160}{v^2}+\frac{1120}{v^4}+\frac{1400}{v^6}+\frac{350}{v^8}+\frac{35}{4 v^{10}}\right)+\hat{l}^4 \left(\frac{24}{v^2}+\frac{42}{v^4}\right)\right]\left(\frac{m}{b}\right)^9\nn\\
&+\left[256 \left(\frac{1}{v^2}+\frac{16 }{v^4}+\frac{56 }{v^6}+\frac{56 }{v^8}+\frac{14 }{v^{10}}\right)-384\hat{l}^2\left(\frac{1}{v^2}+\frac{10 }{v^4}+\frac{20 }{v^6}+\frac{10 }{v^8}+\frac{1}{v^{10}}\right)\right.\nn\\
&\left.+96\hat{l}^4 \left(\frac{1}{v^2}+\frac{4 }{v^4}+\frac{2 }{v^6}\right)\right]\left(\frac{m}{b}\right)^{10}\nn\\
&+\left[\frac{1}{256} \left(\frac{131072 }{v^2}+\frac{2654208 }{v^4}+\frac{12386304 }{v^6}+\frac{18063360 }{v^8}+\frac{8128512 }{v^{10}}+\frac{677376 }{v^{12}}\right.\right.\nn\\
&\left.\left.-\frac{32256 }{v^{14}}+\frac{2592 }{v^{16}}-\frac{180 }{v^{18}}+\frac{7}{v^{20}}\right)+\hat{l}^2 \left(\frac{-896 }{v^2}-\frac{12096 }{v^4}-\frac{35280 }{v^6}-\frac{29400 }{v^8}-\frac{6615 }{v^{10}}\right.\right.\nn\\
&\left.\left.-\frac{441 }{2 v^{12}}+\frac{21 }{8 v^{14}}\right)+5\hat{l}^4 \left(64 \frac{1}{v^2}+\frac{432 }{v^4}+\frac{504 }{v^6}+\frac{105 }{v^8}\right)-8\hat{l}^6 \frac{1}{v^2}\right]\left(\frac{m}{b}\right)^{11}\nn\\
&+\left[1024 \left(\frac{1}{v^2}+\frac{25 }{v^4}+\frac{150 }{v^6}+\frac{300 }{v^8}+\frac{210 }{v^{10}}+\frac{42 }{v^{12}}\right)-1024\hat{l}^2\left(\frac{2 }{v^2}+\frac{35 }{v^4}+\frac{140 }{v^6}+\frac{175 }{v^8}+\frac{70 }{v^{10}}\right.\right.\nn\\
&\left.\left.+\frac{7 }{v^{12}}\right)+960\hat{l}^4\left(\frac{1}{v^2}+\frac{10 }{v^4}+\frac{20 }{v^6}+\frac{10 }{v^8}+\frac{1}{v^{10}}\right)-32\hat{l}^6\left(\frac{2 }{v^2}+\frac{5 }{v^4}\right)\right]\left(\frac{m}{b}\right)^{12}\nn\\
&+\left[\frac{2048 }{v^2}+\frac{61952 }{v^4}+\frac{464640 }{v^6}+\frac{1219680 }{v^8}+\frac{1219680 }{v^{10}}+\frac{426888 }{v^{12}}+\frac{30492 }{v^{14}}-\frac{5445 }{4v^{16}}\right.\nn\\
&\left.+\frac{1815 }{16v^{18}}-\frac{605
}{64v^{20}}+\frac{77 }{128v^{22}}-\frac{21 }{1024v^{24}}-\frac{9\hat{l}^2}{64}\left(\frac{32768 }{v^2}+\frac{720896 }{v^4}+\frac{3784704 }{v^6}\right.\right.\nn\\
&\left.\left.+\frac{6623232 }{v^8}+\frac{4139520 }{v^{10}}+\frac{827904 }{v^{12}}+\frac{29568 }{v^{14}}-\frac{528 }{v^{16}}+\frac{11 }{v^{18}}\right)+\frac{21\hat{l}^4}{4}\left(\frac{512 }{v^2}+\frac{7040 }{v^4}\right.\right.\nn\\
&\left.\left.+\frac{21120 }{v^6}+\frac{18480 }{v^8}+\frac{4620 }{v^{10}}+\frac{231 }{v^{12}}\right)-40\hat{l}^6\left(\frac{8 }{v^2}+\frac{44 }{v^4}+\frac{33 }{v^6}\right)\right]\left(\frac{m}{b}\right)^{13}+\mathcal{O}\lb \frac{m}{b}\rb^{14}.
\end{align}
Using this in Eq. \eqref{eq:hwdix0}, the change of the angular coordinate in terms of the impact parameter is obtained as
\be
I_{\mathrm{H}}(b,v,\hat{l}))=\sum_{n=0}^{13}H_{n}^\prime (v,\hat{l})\left(\frac{m}{b}\right)^n +\mathcal{O}\left(\frac{m}{b}\right)^{14}, \label{eq:hwdib}
\ee
where the coefficients are
\begin{subequations} \label{eq:hinbcoeffs}
\begin{align}
H^\prime_{0}=&S^\prime_0,\\
  H^\prime_{1}=&S^\prime_1,\label{eq:halphaone}\\
  H^\prime_{2}=&S^\prime_2,\\
  H^\prime_{3}=&S^\prime_3,\\
  H^\prime_{4}=&S^\prime_4 -\hat{l}^2 \left(\frac{3 \pi }{4}+\frac{3 \pi }{v^2}\right),\label{eq:halphafour}\\
  H^\prime_{5}=&S^\prime_5-32\hat{l}^2 \left(\frac{1}{5}+\frac{2}{v^2}+\frac{1}{v^4}\right),\\
  H^\prime_{6}=&S^\prime_6-\frac{15\pi \hat{l}^2 }{16}\left(5+\frac{90}{v^2}+\frac{120}{v^4}+\frac{16}{v^6}\right)
  ,\\
  H^\prime_{7}=&S^\prime_7-32\hat{l}^2\left(1+\frac{28}{v^2}+\frac{70}{v^4}+\frac{28}{v^6}+\frac{1}{v^8}\right)+\frac{128\hat{l}^4 }{5}\left(\frac{1}{7}+\frac{1}{v^2}\right),\\
  H^\prime_{9}=&S^\prime_9-\frac{64\hat{l}^2 }{5}\left(11+\frac{594}{v^2}+\frac{3465}{v^4}+\frac{4620}{v^6}+\frac{1485}{v^8}+\frac{66}{v^{10}}-\frac{1}{v^{12}}\right)+\hat{l}^{4}\left(\frac{1024}{21}+\frac{9216}{7 v^2}+\frac{3072}{v^4}+\frac{1024}{v^6}\right),\\
  H^\prime_{10}=&S^\prime_{10}-\frac{63063\pi}{2048}\hat{l}^2\left(3+ \frac{210}{v^2}+ \frac{1680}{v^4}+ \frac{3360}{v^6}+ \frac{1920}{v^8}+\frac{256}{v^{10}}\right)\nn\\
  &+\frac{1575\pi}{256}\hat{l}^4\left(7+\frac{280 }{v^2}+\frac{1120 }{v^4}+\frac{896 }{v^6}+\frac{128 }{v^8}\right)-\frac{63\pi}{32}\hat{l}^6\left(1+\frac{10}{v^2}\right)\\
  H^\prime_{11}=&S^\prime_{11}-\frac{64}{7} \hat{l}^2 \left(65+\frac{5720 }{v^2}+\frac{60060
   }{v^4}+\frac{168168 }{v^6}+\frac{150150 }{v^8}+\frac{40040 }{v^{10}}+\frac{1820
   }{v^{12}}-\frac{40 }{v^{14}}+\frac{1}{v^{16}}\right)\nn\\
   &+\frac{3840}{11} \hat{l}^4 \left(1+\frac{55 }{v^2}+\frac{330
   }{v^4}+\frac{462 }{v^6}+\frac{165 }{v^8}+\frac{11
   }{v^{10}}\right)-\frac{8192}{231} \hat{l}^6 \left(1+\frac{22 }{v^2}+\frac{33}{v^4}\right)\\
  H^\prime_{12}=&S^\prime_{12}
  -\frac{2297295\pi }{65536}\hat{l}^2 \left(11+\frac{1188
   }{v^2}+\frac{15840 }{v^4}+\frac{59136 }{v^6}+\frac{76032 }{v^8}+\frac{33792
   }{v^{10}}+\frac{4096 }{v^{12}}\right)\nn\\
   &+\frac{4851\pi }{4096}\hat{l}^4 \left(231+\frac{16632 }{v^2}+\frac{138600
   }{v^4}+\frac{295680 }{v^6}+\frac{190080 }{v^8}+\frac{33792 }{v^{10}}+\frac{1024
   }{v^{12}}\right)\nn\\
   &-\frac{5775\pi}{128} \hat{l}^6 \left(1+\frac{36 }{v^2}+\frac{120 }{v^4}+\frac{64
   }{v^6}\right)\\
  H^\prime_{13}=&S^\prime_{13}
  +\frac{160}{21} \hat{l}^2 \left(-323-\frac{41990
   }{v^2}-\frac{692835 }{v^4}-\frac{3325608 }{v^6}-\frac{5819814
   }{v^8}-\frac{3879876 }{v^{10}}-\frac{881790 }{v^{12}}-\frac{38760
   }{v^{14}}+\frac{969 }{v^{16}}\right.\nn\\
   &\left.-\frac{38 }{v^{18}}+\frac{1}{v^{20}}\right) +2048 \hat{l}^4 \left(1+\frac{91 }{v^2}+\frac{1001 }{v^4}+\frac{3003
   }{v^6}+\frac{3003 }{v^8}+\frac{1001 }{v^{10}}+\frac{91
   }{v^{12}}+\frac{1}{v^{14}}\right)\nn\\
   &-\frac{40960}{429} \hat{l}^6
   \left(5+\frac{260 }{v^2}+\frac{1430 }{v^4}+\frac{1716 }{v^6}+\frac{429
   }{v^8}\right)
   +\frac{32768 }{3003}\hat{l}^8 \left(1+\frac{13 }{v^2}\right).
  \end{align}
\end{subequations}
Here the coefficients $S_n^\prime$ are the coefficients in the change of the angular coordinate in the Schwarzschild spacetime given in \eqref{eq:angschinb}.
Note that if we set $v=c$ for lightray in Eqs. \eqref{eq:hinx0coeff} and \eqref{eq:hinbcoeffs}, the first sixth orders agree with Eq. (18) and (19) of Ref. \cite{Chiba:2017nml} respectively.

From these equations, it is seen that the modification comparing the Schwarzschild case due to parameter $\hat{l}$ is only present from and above the fourth order of $m/b$. Since the deflection angle in the weak field limit is expected to be sensitive only to the asymptotic behavior of the spacetime, this implies that the metric functions at large $r$ must only differ from the Schwarzschild spacetime starting from high orders too. This is indeed confirmed in the expansion of the lapse function in Eq. \eqref{eq:haywardarexp}, where the difference only appears starting from the fourth order.

\begin{center}
\begin{figure}[htp!]
\includegraphics[width=0.45\textwidth]{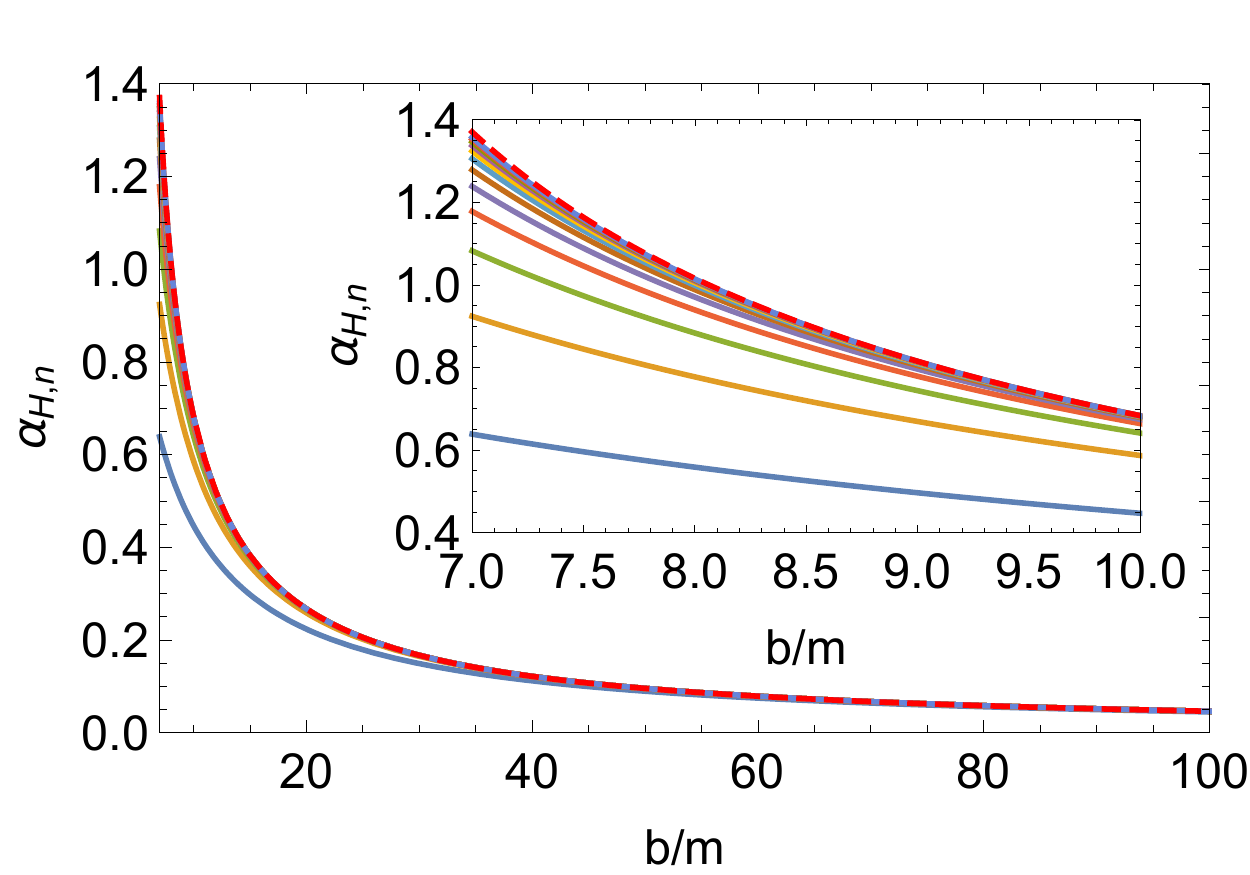}
\hspace{0.5cm}\includegraphics[width=0.45\textwidth]{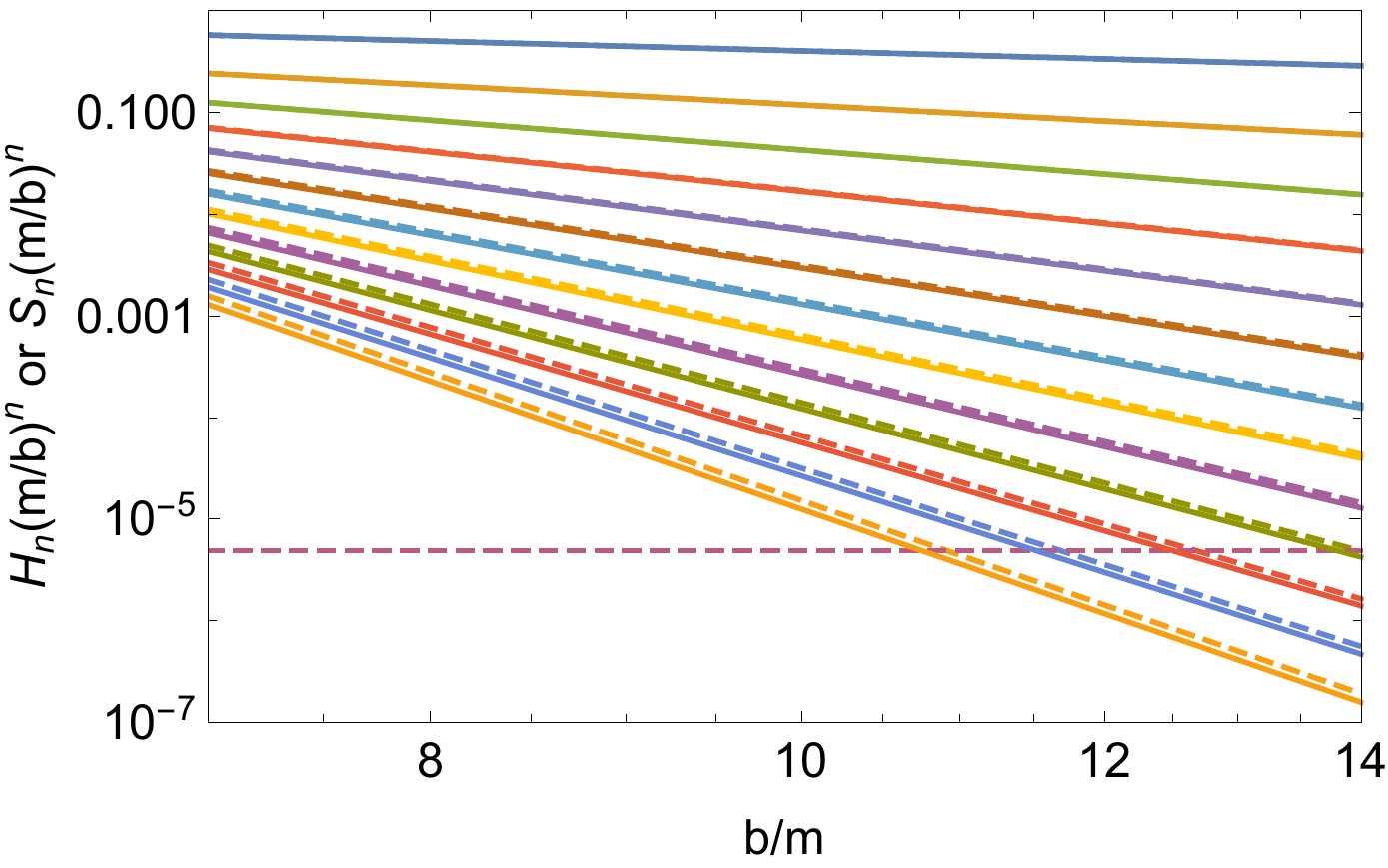}\\
(a)\hspace{8cm}(b)\\
\includegraphics[width=0.45\textwidth]{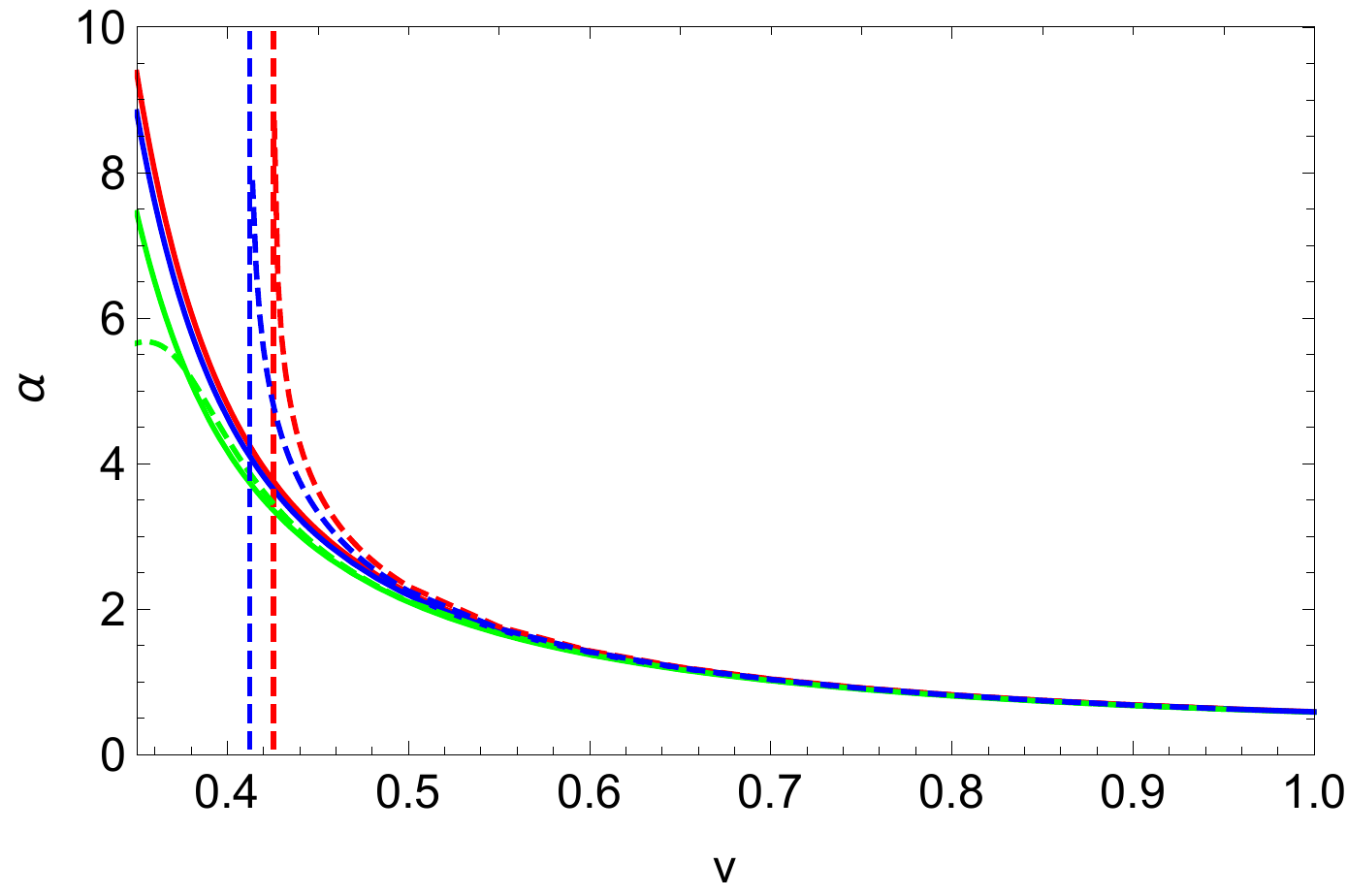}
\hspace{0.5cm}\includegraphics[width=0.45\textwidth]{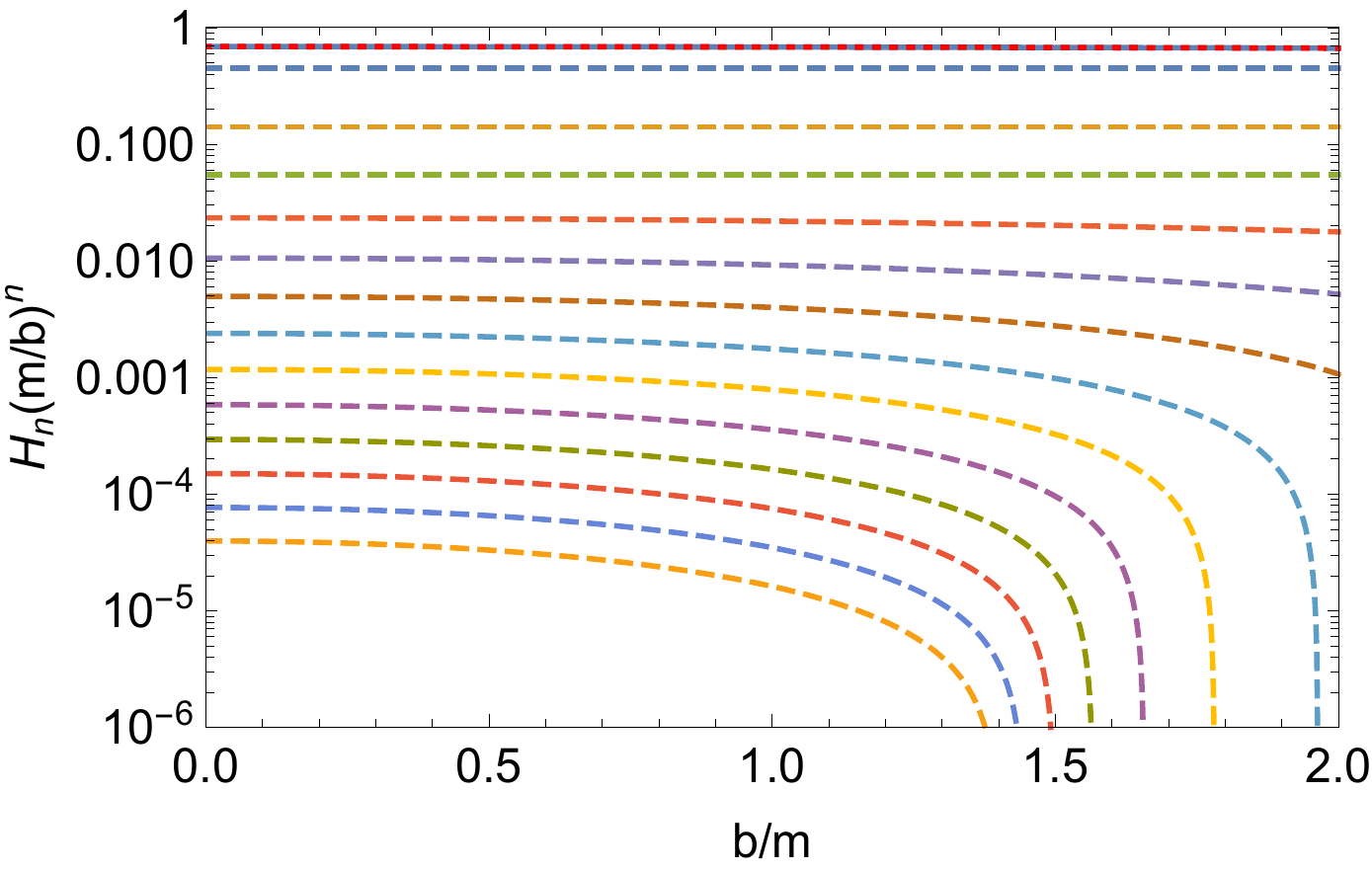}\\
(c)\hspace{8cm}(d)
\caption{Deflection angles in the Hayward spacetime. (a) Partial sums \eqref{eq:hangpsum} (solid curves) and the exact value (red dashed curve) of the exact deflection angle for $b/m$ from 7 to 100 and from 7 to 10 in the inset and $v=c,~\hat{l}=0.5$. From bottom to top curve in each plot, the maximum index in the partial sum increases from 1 to 13. (b) Contribution from each order in Eq. \eqref{eq:hinbcoeffs} (from top to bottom curve the order increases from 1 to 13) for $v=c,~\hat{l}=0.5$ and corresponding results in the Schwarzschild case (dashed lines).
(c) Partial sum $\alpha_{\mathrm{H},13}$ (solid curves) and the exact deflection angle (dashed curves) for $b=10m$ and $\hat{l}_1=0.5$ (red lines), $\hat{l}_2=0.85$ (blue lines) and $\hat{l}_3=1.4$ (green lines).  (d) Contribution from each order (dashed curves), partial sum $\alpha_{\mathrm{H},13}$ (top solid curve) and exact deflection angle (top red dashed curve) for $b=10m,~v=0.9$.
\label{fig:hayward1}}
\end{figure}
\end{center}

To study how the deflection angle depends on various kinematic and spacetime parameters, we plot the deflection angle $\alpha_\mathrm{H}\equiv I_H(b,v,\hat{l})-\pi$ in Fig. \ref{fig:hayward1}. In Fig. \ref{fig:hayward1} (a), the partial sum of the deflection angle defined as the sum of Eqs. \eqref{eq:hinbcoeffs} to the $n$-th order
\begin{align}
    \alpha_{\mathrm{H},n}=\sum_{i=1}^n H_{n}^\prime (v,\hat{l})\left(\frac{m}{b}\right)^i, \label{eq:hangpsum}
\end{align}
as well as the exact value of the deflection angle obtained using numerical integration are shown. It is seen that in general, the deflection angle decreases monotonically as $b$ increases, becoming completely determined by the leading order formula \eqref{eq:halphaone} and overlaps with the exact value when $b$ is large. As $b$ decreases, the contribution from higher orders start to manifest (see the insert for a zoom-in at small $b$). However, the partial sum to the thirteenth order still overlaps with the true value of the deflection angle, even when $b$ is as small as $7m$, at which point $\alpha_\mathrm{H}\approx 0.51\pi$. This shows that the deflection angle we calculated perturbatively can even work when the gravitational field is not weak or when deflection angle is not small anymore.

In Fig. \ref{fig:hayward1} (b), the contribution from each order of Eqs. \eqref{eq:hinbcoeffs} are plotted (solid lines), together with the values at the corresponding orders for the Schwarzschild spacetime for comparison (dashed lines). It is seen that the contribution from each order decreases roughly by the same factor as the order increases, suggesting that the summation \eqref{eq:hwdib} is convergent for the given range of $b$. It is also clear that as dictated by the asymptotic expansion \eqref{eq:haywardarexp} and Eq. \eqref{eq:halphafour}, the contribution starts to deviate from that Schwarzschild spacetime result from and above the fourth order. Moreover, from these deviations, one sees that an increase of $\hat{l}$ indeed decreases the deflection angle. Furthermore, the horizontal line at 1 [as] shows that the thirteenth order result can reach this resolution for the impact parameter $b$ as small as $10.7m$.

%we can compute the relation between the critical values of $b,~v$ and $\hat{l}$ from the existence of minimal radial coordinate $r_0$ in the denominator of Eq. \eqref{eq:iintinr}. For

In Fig. \ref{fig:hayward1} (c), the deflection angle is plotted against the velocity $v$ for some fixed $b$ and three representative values of $\hat{l}$, i.e., $\hat{l}_1=0.5<\hat{l}_c$ and $\hat{l}_c<\hat{l}_2=0.85>\hat{l}_p$ and $\hat{l}_3=1.4>\hat{l}_p$. Similar to the Bardeen BH spacetime case, for  Hayward spacetime with $b=10m$, their exist a critical velocity around $0.43c$ for $\hat{l}=\hat{l}_1$ and $0.41c$ for $\hat{l}=\hat{l}_2$. From Fig. \ref{fig:hayward1} (c) it is seen that for $\hat{l}=\hat{l}_1$ or $\hat{l}_2$, as velocity decreases from $c$, the deflection angle calculated both from the partial sum $\alpha_{\mathrm{H},13}$ and numerical integration increase monotonically. As $v$ approaches $v_c$, the deflection angle $\alpha_{\mathrm{H},13}$ calculated using the perturbative method starts to deviate from the true deflection angle around $0.50c$ for $\hat{l}_1$ and $0.48c$ for $\hat{l}_2$. Below these, the true deflection angles in both cases diverge as $v\to v_c$ from above and the perturbative deflection angle becomes invalid. On the other hand, when $\hat{l}=\hat{l}_3$, the photon sphere disappears and the partial sum $\alpha_{\mathrm{H},13}$ still roughly agree with the true value even when the velocity is as small as $0.38c$.

In Fig. \ref{fig:hayward1} (d), we examine the effect of parameter $\hat{l}$ on the deflection angles. It is seen that the partial sum $\alpha_{\mathrm{H},13}$ and the numerical value overlaps for the entire range of $\hat{l}$ from 0 to 2, which shows that our result also works after the parameter $\hat{l}$ passes through the critical value. The deflection angle monotonically decreases as $\hat{l}$ increases in the whole range. In particular, when $\hat{l}$ passes $\hat{l}_c$ and $\hat{l}_p$, the deflection angle does not experience any qualitative change, which is a reflection that the perturbative expansion depends only on the asymptotic expansions of the metric functions.

\subsection{Deflection in the JNW spacetime\label{subsec:JNW}}
% Hu Weiyu is responsible for this subsection

We consider now the most general spherically symmetric asymptotically flat exact solution to the Einstein massless scalar equations, i.e., JNW metric \cite{Janis:1968zz}
\begin{equation}
A(r)=B(r)^{-1}=\left(1-\dfrac{\beta}{r}\right)^{\lambda},~C(r)=r^{2} \left(1-\dfrac{\beta}{r}\right)^{1-\lambda},
\label{eq:jnwmetric}
\end{equation}
where $\beta$ is the location of the naked curvature singularity and  $\beta<r<\infty$.  $\beta, \lambda$ are related to the ADM mass $m$ and scalar charge $q$ by
\be
\beta=2\sqrt{m^2+q^2},~\lambda=\frac{2m}{\beta}. \label{eq:jnwpara}
\ee
Clearly, when $q=0$, we will have $\lambda=1$ and the JNW spacetime reduces to the Schwarzschild spacetime.
In order for $m\geq 0$, it is required that $0\leq \lambda\leq 1$ \cite{Claudel:2000yi,Perlick:2010zh}. Moreover, it was known that in the strong field limit, the deflection angle as well as the GL features of the JNW metric with $\frac{1}{2}<\lambda<1$ are quite similar to the case of Schwarzschild metric, because of the existence of a photon sphere at
\begin{align}
    r_{\mathrm{J,p}}=\beta\frac{1+2\lambda}{2} \label{eq:jnwps}
\end{align} in this case \cite{Claudel:2000yi}. While if $0<\lambda<\frac{1}{2}$, the above photon sphere is within the naked singularity at $r=\beta$ and therefore not accessible. Consequently, the GL will only have finite number of images \cite{Perlick:2010zh}. We would like to see whether the critical value $\lambda_c=\frac{1}{2}$, i.e. $q=m$ will play any role in the deflection angle in the weak field limit.

Using the metric functions \eqref{eq:jnwmetric}, and going through the procedure from Eq. \eqref{eqdxdl} to Eq. \eqref{eq:iinb}, the change of the angular coordinate for a particle ray with velocity $v$ and impact parameter $b$ in the JNW spacetime becomes
\begin{equation}
I_\mathrm{J}(r_0, v,\lambda)=\sum_{n=0}^{7} J_{n}(v,\lambda)\left(\frac{\beta}{r_0}\right)^{n}+\mathcal{O}\left(\frac{\beta}{r_0}\right)^{8}, \label{eq:ijnwinx0}
\end{equation}
where the coefficients are
\begin{subequations}
\label{eq:jninx0}
\begin{align}
J_{0}=
&\pi,\\
J_{1}=
&\lambda  \left(\frac{1}{v^2}+1\right),\\
J_{2}=
&-\frac{\pi }{16}+\lambda  \left(\frac{1}{2}+\frac{1}{2 v^2}\right)+\lambda ^2 \left[\frac{\pi }{4}-\frac{1}{2}+\left(\frac{3 \pi }{4}-1\right)\frac{1}{v^2}-\frac{1}{2 v^4}\right],\\
J_{3}=
&-\frac{\pi }{16}+\lambda\left[\frac{1}{24}+\frac{\pi }{16}+\left(\frac{1}{24}+\frac{\pi }{16}\right)\frac{1}{v^2}\right]+\lambda^2\left[\frac{\pi }{4}-\frac{1}{2}+\left(\frac{3 \pi }{4}-1\right)\frac{1}{v^2}-\frac{1}{2 v^4}\right]\nn\\
&+\lambda^3\left[\frac{7}{8}-\frac{\pi }{4}+\left(\frac{101}{24}-\pi\right)\frac{1}{v^2}+\left(\frac{13}{8}-\frac{3 \pi }{4}\right)\frac{1}{v^4}+\frac{7}{24 v^6}\right],\\
J_{4}=
&-\frac{55 \pi }{1024}+\lambda\left[\frac{3 \pi }{32}-\frac{3}{16}+\left(\frac{3 \pi }{32}-\frac{3}{16}\right)\frac{1}{v^2}\right] +\lambda^2\left[\frac{1}{48}+\frac{\pi }{16}+\left(\frac{1}{24}+\frac{21 \pi }{64}\right)\frac{1}{v^2}+\left(\frac{1}{48}-\frac{9 \pi }{64}\right)\frac{1}{v^4}\right]\nn\\
&+\lambda^3\left[\frac{21}{16}-\frac{3 \pi }{8}+\left(\frac{101}{16}-\frac{3 \pi }{2}\right)\frac{1}{v^2}+\left(\frac{39}{16}-\frac{9 \pi }{8}\right)\frac{1}{v^4}+\frac{7}{16 v^6}\right]\nn\\
&+\lambda^4\left[\frac{3 \pi }{8}-\frac{55}{48}+\left(\frac{81 \pi }{32}-\frac{175}{24}\right)\frac{1}{v^2}+\left(\frac{75 \pi }{32}-\frac{47}{6}\right)\frac{1}{v^4}+\left(\frac{3 \pi }{4}-\frac{15}{8}\right)\frac{1}{v^6}-\frac{3}{16 v^8}\right],\\
J_{5}=
&-\frac{23 \pi}{512}+\lambda\left[\frac{149 \pi }{1536}-\frac{1513}{5760}+\left(\frac{149 \pi }{1536}-\frac{1513}{5760}\right)\frac{1}{v^2}\right]+\lambda^2\left[\frac{13}{24}-\frac{\pi }{8}
+\left(\frac{13}{12}-\frac{3 \pi }{32}\right)\frac{1}{v^2}
+\left(\frac{13}{24}-\frac{9 \pi }{32}\right)\frac{1}{v^4}\right]\nn\\
&+\lambda^3\left[\frac{323}{576}-\frac{13 \pi }{96}+\left(\frac{2291}{576}-\frac{35 \pi }{48}\right)\frac{1}{v^2}
+\left(\frac{9}{64}-\frac{7 \pi }{16}\right)\frac{1}{v^4}
+\left(\frac{7 \pi }{32}-\frac{37}{192}\right)\frac{1}{v^6}\right]\nn\\
&+\lambda^4\left[\frac{3 \pi }{4}-\frac{55}{24}+\left(\frac{81 \pi }{16}-\frac{175}{12}\right)\frac{1}{v^2}
+\left(\frac{75 \pi }{16}-\frac{47}{3}\right)\frac{1}{v^4}
+\left(\frac{3 \pi }{2}-\frac{15}{4}\right)\frac{1}{v^6}-\frac{3}{8 v^8}\right]\nn\\
&+\lambda^5\left[\frac{1975}{1152}-\frac{13 \pi }{24}+\left(\frac{87683}{5760}
-\frac{227 \pi }{48}\right)\frac{1}{v^2}
+\left(\frac{2287}{192}-\frac{59 \pi }{16}\right)\frac{1}{v^6} +\left(\frac{255}{128}-\frac{3 \pi }{4}\right)\frac{1}{v^8}+\frac{83}{640 v^{10}}\right],\\
J_{6}=
&-\frac{617 \pi }{16384}+\lambda\left[\frac{265 \pi }{3072}-\frac{601}{2304}+\left(\frac{265 \pi }{3072}-\frac{601}{2304}\right)\frac{1}{v^2}\right]\nn\\
&+\lambda^2\left[\frac{9413}{11520}-\frac{2815 \pi }{12288}
+\left(\frac{9413}{5760}-\frac{1389 \pi }{4096}\right)\frac{1}{v^2}+\left(\frac{9413}{11520}-\frac{713 \pi }{2048}\right)\frac{1}{v^4}\right]\nn\\
&+\lambda^3\left[\frac{55 \pi }{192}-\frac{905}{1152}
+\left(\frac{65 \pi }{96}-\frac{665}{1152}\right)\frac{1}{v^2}+\left(\frac{25 \pi }{32}-\frac{475}{128}\right)\frac{1}{v^4}+\left(\frac{35 \pi }{64}-\frac{155}{128}\right)\frac{1}{v^6}\right]\nn\\
&
+\lambda^4\left[\frac{1511 \pi }{3072}-\frac{1687}{1152}+\left(\frac{4409 \pi }{1024}-\frac{6799}{576}\right)\frac{1}{v^2}+\left(\frac{1519 \pi }{512}-\frac{1535}{144}\right)\frac{1}{v^4}
+\left(\frac{17}{192}+\frac{103 \pi }{256}\right)\frac{1}{v^6}+\left(\frac{157}{384}-\frac{19 \pi }{64}\right)\frac{1}{v^8}\right]\nn\\
&+\lambda^5\left[\frac{9875}{2304}-\frac{65 \pi }{48}+\left(\frac{87683}{2304}-\frac{1135 \pi }{96}\right)\frac{1}{v^2}
+\left(\frac{7645}{128}-\frac{305 \pi }{16}\right)\frac{1}{v^4}+\left(\frac{11435}{384}-\frac{295 \pi }{32}\right)\frac{1}{v^6}+\left(\frac{1275}{256}-\frac{15 \pi }{8}\right)\frac{1}{v^8}\right.\nn\\
&\left.+\frac{83}{256 v^{10}}\right]+\lambda^6\left[\frac{635 \pi }{768}-\frac{29863}{11520}+\left(\frac{2365 \pi }{256}-\frac{166063}{5760}\right)\frac{1}{v^2}+\left(\frac{2683 \pi }{128}-\frac{756193}{11520}\right)\frac{1}{v^4}\right.\nn\\
&\left.+\left(\frac{1031 \pi }{64}-\frac{9709}{192}\right)\frac{1}{v^6}+\left(\frac{161 \pi }{32}-\frac{12383}{768}\right)\frac{1}{v^8}+\left(\frac{3 \pi }{4}-\frac{263}{128}\right)\frac{1}{v^{10}}-\frac{73}{768 v^{12}}\right],\\
J_{7}=
&-\frac{523 \pi }{16384}+\lambda\left[\frac{17477 \pi }{245760}-\frac{368839}{1612800}+\left(\frac{17477 \pi }{245760}-\frac{368839}{1612800}\right)\frac{1}{v^2}\right]\nn\\
&+\lambda^2\left[\frac{3253}{3840}-\frac{1023 \pi }{4096}
+\left(\frac{3253}{1920}-\frac{1671 \pi }{4096}\right)\frac{1}{v^2}+\left(\frac{3253}{3840}-\frac{699 \pi }{2048}\right)\frac{1}{v^4}\right]\nn\\
&+\lambda^3\left[\frac{38473 \pi }{61440}-\frac{144349}{76800}
+\left(\frac{56761 \pi }{30720}-\frac{1028999}{230400}\right)\frac{1}{v^2}+\left(\frac{35933 \pi }{20480}-\frac{34481}{5120}\right)\frac{1}{v^4} +\left(\frac{4799 \pi }{6144}-\frac{90137}{46080}\right)\frac{1}{v^6}\right]\nn\\
&+\lambda^4\left[\frac{171}{128}-\frac{409 \pi }{1024}+\left(\frac{67}{64}+\frac{267 \pi }{1024}\right)\frac{1}{v^2}+\left(\frac{115}{16}-\frac{1443 \pi }{512}\right)\frac{1}{v^4}+\left(\frac{617}{64}-\frac{651 \pi }{256}\right)\frac{1}{v^6}
\right.\nn\\
&\left.+\left(\frac{277}{128}-\frac{57 \pi }{64}\right)\frac{1}{v^8}\right]+\lambda^5\left[\frac{296099}{76800}-\frac{18617 \pi }{15360}+\left(\frac{9510937}{230400}-\frac{97451 \pi }{7680}\right)\frac{1}{v^2}
+\left(\frac{90829}{1536}-\frac{19389 \pi }{1024}\right)\frac{1}{v^4}\right.\nn\\
&\left.+\left(\frac{94747}{4608}-\frac{9373 \pi }{1536}\right)\frac{1}{v^6}+\left(-\frac{2501}{3072}-\frac{57 \pi }{256}\right)\frac{1}{v^8}+\left(\frac{3 \pi }{8}
-\frac{9857}{15360}\right)\frac{1}{v^{10}}\right]\nn\\
&+\lambda^6\left[\frac{635 \pi }{256}-\frac{29863}{3840}+\left(\frac{7095 \pi }{256}-\frac{166063}{1920}\right)\frac{1}{v^2}+\left(\frac{8049 \pi }{128}-\frac{756193}{3840}\right)\frac{1}{v^4}\right. \nn\\
&\left.+\left(\frac{3093 \pi }{64}-\frac{9709}{64}\right)\frac{1}{v^6}+\left(\frac{483 \pi }{32}-\frac{12383}{256}\right)\frac{1}{v^8}+\left(\frac{9 \pi }{4}-\frac{789}{128}\right)\frac{1}{v^{10}}-\frac{73}{256 v^{{12}}}\right]\nn\\
&+\lambda^7\left[\frac{930367}{230400}-\frac{4933 \pi }{3840}+\left(\frac{88329593}{1612800}-\frac{11137 \pi }{640}\right)\frac{1}{v^2}+\left(\frac{2553593}{15360}-\frac{67613 \pi }{1280}\right)\frac{1}{v^4}+\left(\frac{8364067}{46080}-\frac{22201 \pi }{384}\right)\frac{1}{v^6}\right.\nn\\
&\left.+\left(\frac{269615}{3072}-\frac{1791 \pi }{64}\right)\frac{1}{v^8}+\left(\frac{312923}{15360}-\frac{51 \pi }{8}\right)\frac{1}{v^{10}}+\left(\frac{10713}{5120}-\frac{3 \pi}{4}\right)\frac{1}{v^{12}}+\frac{523}{7168 v^{14}}\right].
\end{align}
\end{subequations}

The $\beta/r_0$ in Eq. \eqref{eq:jninx0} can be expanded as a power series of $\beta/b$ using Eq. \eqref{eq:xinb}, given by
\begin{align}
\dfrac{\beta}{r_0}&=\frac{\beta}{b}+\left(\frac{\beta}{b}\right)^2\left[-\frac{1}{2}+\lambda  \left(\frac{1}{2 v^2}+\frac{1}{2}\right)\right]+\left(\frac{\beta}{b}\right)^{3}\left[\frac{1}{8}+\lambda  \left(-\frac{1}{2 v^2}-\frac{1}{2}\right)
+\lambda^2\left(\frac{3}{8}+\frac{1}{v^2}+\frac{1}{8v^4}\right)\right]\nn\\
&
+\left(\frac{\beta}{b}\right)^{4}\left[\lambda  \left(\frac{1}{6 v^2}+\frac{1}{6}\right)+\lambda ^2 \left(-\frac{1}{2}-\frac{5}{4 v^2}-\frac{1}{4 v^4}\right)
+\lambda ^3 \left(\frac{1}{3}+\frac{19}{12 v^2}+\frac{3}{4 v^4}\right)\right]\nn\\
&
+\left(\frac{\beta}{b}\right)^{5}\left[-\frac{1}{128}+\lambda  \left(\frac{1}{48 v^2}+\frac{1}{48}\right)
+\lambda ^2 \left(\frac{35}{192}+\frac{7}{16 v^2}+\frac{7}{64 v^4}\right)\right.\nn\\
&\left.+\lambda ^3 \left(-\frac{25}{48}-\frac{109}{48 v^2}-\frac{21}{16 v^4}-\frac{1}{16 v^6}\right)
+\lambda ^4 \left(\frac{125}{384}+\frac{37}{16 v^2}+\frac{149}{64 v^4}+\frac{1}{4 v^6}-\frac{1}{128 v^8}\right)\right]\nn\\
&
+\left(\frac{\beta}{b}\right)^{6}\left[\lambda  \left(-\frac{1}{40 v^2}-\frac{1}{40}\right)
+\lambda ^2 \left(\frac{1}{16}+\frac{7}{48 v^2}+\frac{1}{24 v^4}\right)+\lambda ^3 \left(\frac{3}{16}+\frac{37}{48 v^2}+\frac{1}{2 v^4}+\frac{1}{24 v^6}\right)\right.\nn\\
&\left.
+\lambda ^4 \left(-\frac{9}{16}-\frac{175}{48 v^2}-\frac{97}{24 v^4}-\frac{3}{4 v^6}\right)+\lambda ^5 \left(\frac{27}{80}+\frac{781}{240 v^2}+\frac{11}{2 v^4}+\frac{41}{24 v^6}\right)\right]\nn\\
&
+\left(\frac{\beta}{b}\right)^{7}\left[\frac{1}{1024}+\lambda  \left(-\frac{3}{1280 v^2}-\frac{3}{1280}\right)+\lambda ^2 \left(-\frac{2387}{46080}-\frac{341}{2880 v^2}-\frac{341}{9216 v^4}\right)\right.\nn\\
&\left.+\lambda ^3 \left(\frac{49}{384}+\frac{193}{384 v^2}+\frac{45}{128 v^4}+\frac{5}{128 v^6}\right)+\lambda ^4 \left(\frac{1715}{9216}+\frac{325}{288 v^2}+\frac{6125}{4608 v^4}+\frac{125}{384 v^6}+\frac{25}{3072 v^8}\right)\right.\nn\\
&\left.+\lambda ^5 \left(-\frac{2401}{3840}-\frac{21121}{3840 v^2}-\frac{1245}{128 v^4}-\frac{505}{128 v^6}-\frac{55}{256 v^8}+\frac{1}{256 v^{10}}\right)\right.\nn\\
&\left.+\lambda ^6 \left(\frac{16807}{46080}+\frac{12931}{2880 v^2}+\frac{103291}{9216 v^4}+\frac{2575}{384 v^6}+\frac{1835}{3072 v^8}-\frac{3}{128 v^{10}}+\frac{1}{1024 v^{12}}\right)\right]
+\mathcal{O}\left(\frac{\beta}{b}\right)^{8}
\end{align}
Substituting this into Eq. \eqref{eq:ijnwinx0}, we obtain the change of angular coordinates as a series of  $\beta/b$ for general velocity
\begin{equation}
I_\mathrm{J}(b, v,\lambda)=\sum_{n=0}^{7} J_{n}^{\prime}(v,\lambda)\left(\frac{\beta}{b}\right)^{n}+\mathcal{O}\left(\frac{\beta}{b}\right)^{8} \label{eq:ijnwinb}
\end{equation}
where the coefficients  for the first few orders are
\begin{subequations}
\begin{align}
 J^\prime_{0}=&\pi,\\
 J^\prime_{1}=&\lambda\left(\frac{1}{v^2}+1\right),\label{eq:j1inb}\\
 J^\prime_{2}=&-\frac{\pi}{16}+\lambda ^2 \pi\left(\frac{1}{4}+\frac{3  }{4 v^2}\right),\\
 J^\prime_{3}=&-\lambda  \left(\frac{1}{3}+\frac{1}{3 v^2}\right)+\lambda ^3 \left(\frac{3}{4}+\frac{49}{12 v^2}+\frac{5}{4 v^4}-\frac{1}{12 v^6}\right),\\
 J^\prime_{4}=&\frac{9 \pi }{1024}-\lambda ^2 \pi\left(\frac{5  }{32}+\frac{25  }{64 v^2}+\frac{5  }{64 v^4}\right)+\lambda ^4 \pi\left(\frac{1 }{4}+\frac{65  }{32 v^2}+\frac{55  }{32 v^4}\right),\label{eq:j4inb}\\
 J^\prime_{5}=&\lambda  \left(\frac{4}{45}+\frac{4}{45 v^2}\right)-\lambda ^3 \left(\frac{25}{36}+\frac{109}{36 v^2}+\frac{7}{4 v^4}+\frac{1}{12 v^6}\right)+\lambda ^5 \left(\frac{125}{144}+\frac{6841}{720 v^2}\right.\nn\\
 &\left.+\frac{119}{8 v^4}+\frac{65}{24 v^6}-\frac{3}{16 v^8}+\frac{1}{80 v^{10}}\right),\\
 J^\prime_{6}=&-\frac{25 \pi }{16384}+\lambda ^2\pi \left(\frac{259  }{4096}+\frac{1813  }{12288 v^2}+\frac{259  }{6144 v^4}\right)-\lambda ^4 \pi\left(\frac{315  }{1024}+\frac{6125  }{3072 v^2}+\frac{3395  }{1536 v^4}+\frac{105  }{256 v^6}\right)\nn\\
 &+\lambda ^6 \pi\left(\frac{81  }{256}+\frac{3367  }{768 v^2}+\frac{4081  }{384 v^4}+\frac{315  }{64 v^6}\right)
  ,\\
 J^\prime_{7}=&-\lambda  \left(\frac{4}{175}+\frac{4}{175 v^2}\right)+\lambda ^3 \left(\frac{343}{900}+\frac{1351}{900 v^2}+\frac{21}{20 v^4}+\frac{7}{60 v^6}\right)\nn\\
 &-\lambda ^5 \left(+\frac{2401}{1800}+\frac{21121}{1800 v^2}+\frac{83}{4 v^4}+\frac{101}{12 v^6}+\frac{11}{24 v^8}-\frac{1}{120 v^{10}}\right)\nn\\
 &+\lambda ^7 \left(\frac{16807}{14400}+\frac{1979713}{100800 v^2}+\frac{21319}{320 v^4}+\frac{17671}{320 v^6}+\frac{1375}{192 v^8}-\frac{437}{960 v^{10}}+\frac{13}{320 v^{12}}-\frac{1}{448 v^{14}}\right)
  \end{align}\label{eq:jnwinbcoeffs}
\end{subequations}

\begin{center}
\begin{figure}[htp!]
\includegraphics[width=0.45\textwidth]{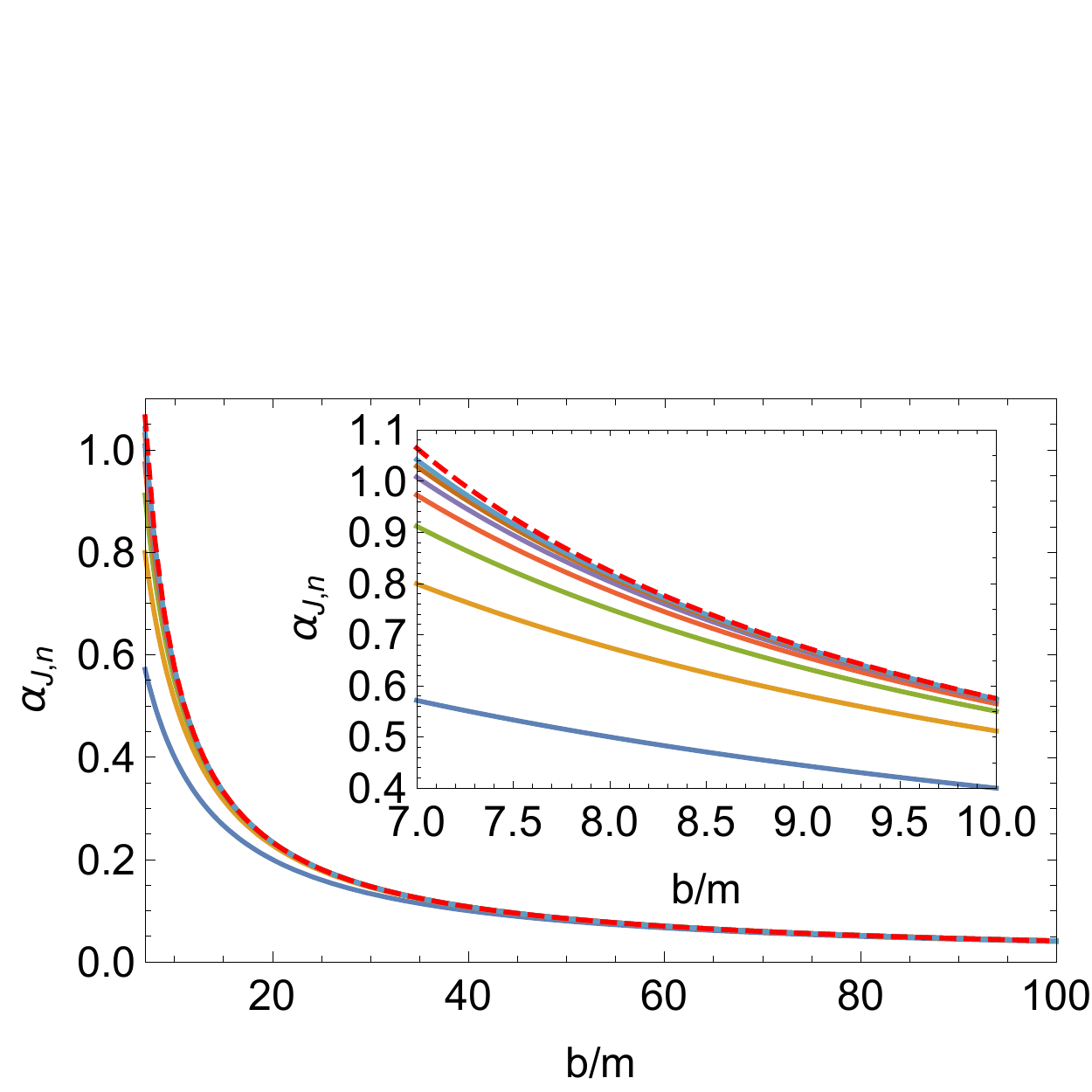}
\hspace{0.5cm}\includegraphics[width=0.45\textwidth]{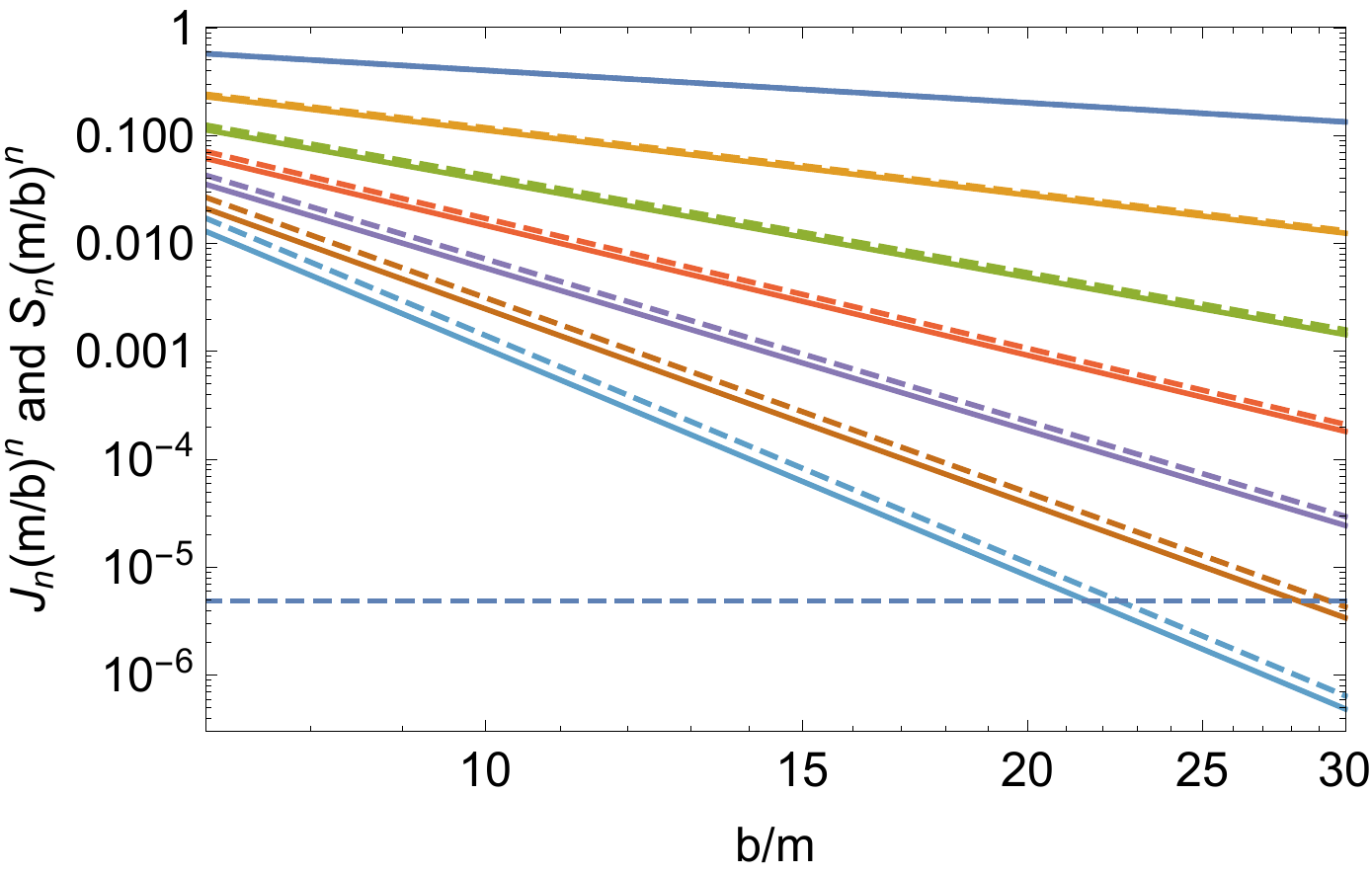}\\
(a)\hspace{8cm}(b)\\
\includegraphics[width=0.45\textwidth]{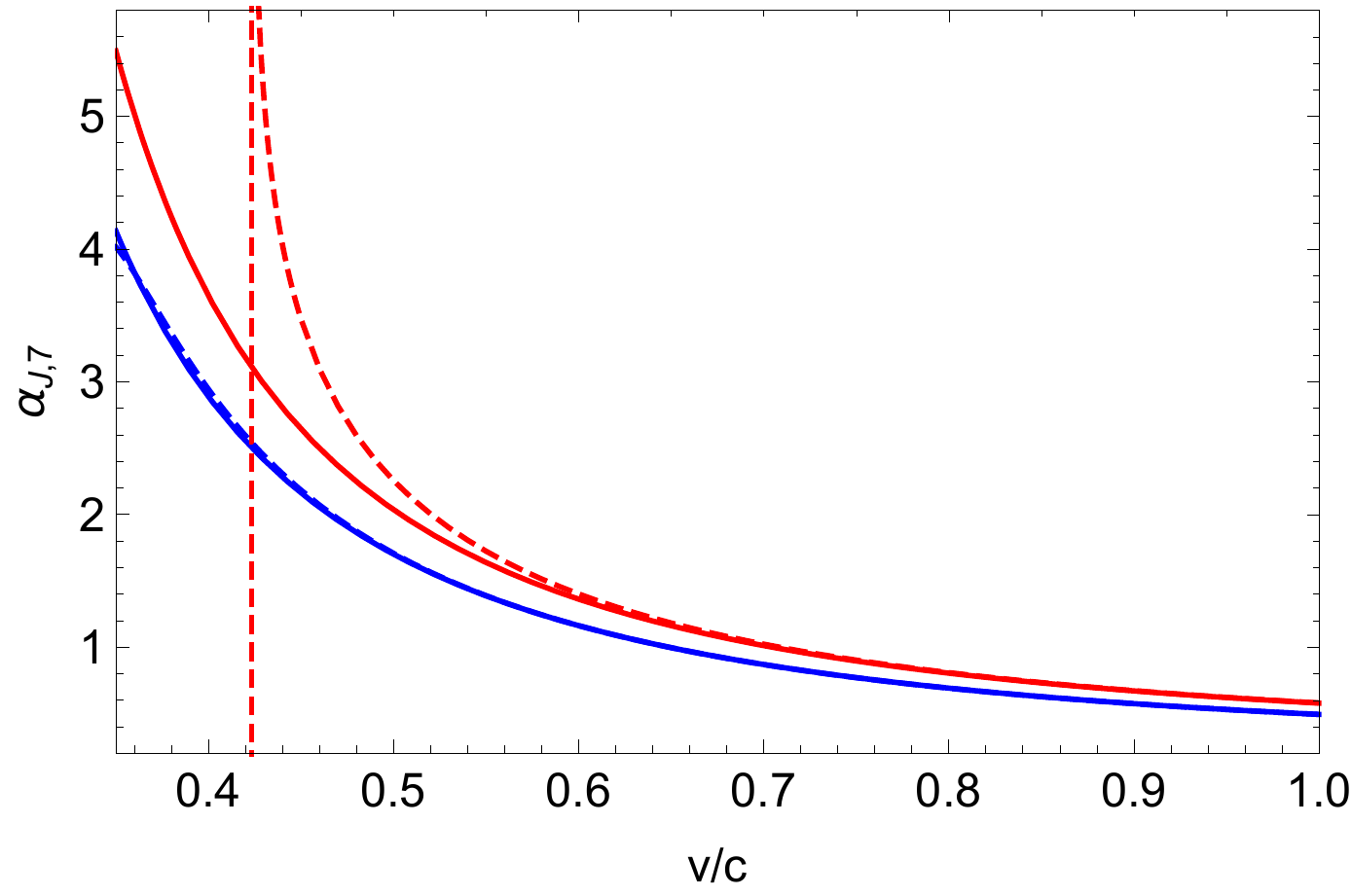}
\hspace{0.5cm}\includegraphics[width=0.45\textwidth]{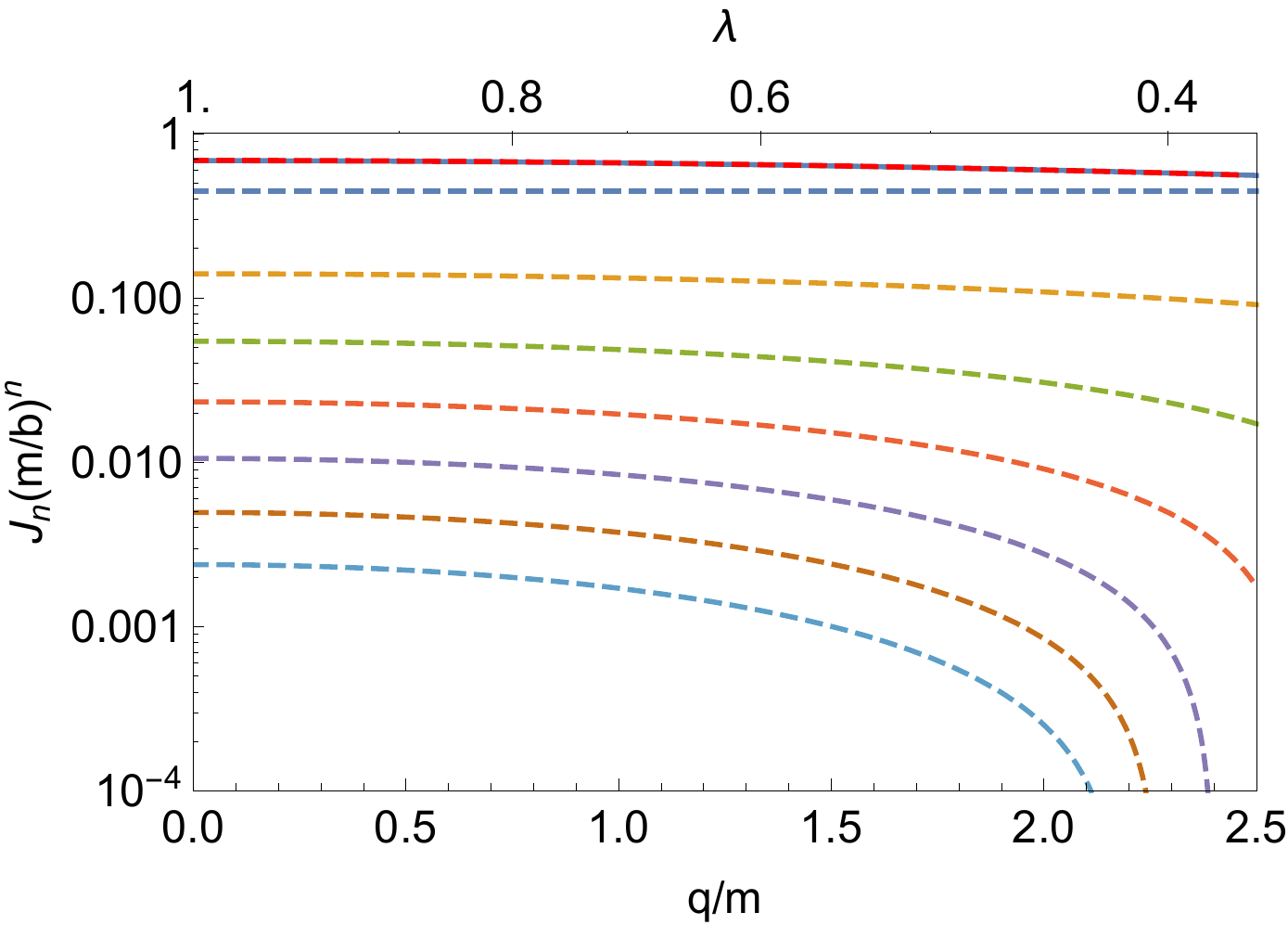}\\
(c)\hspace{8cm}(d)
\caption{Deflection angles in the JNW spacetime. (a) Partial sums \eqref{eq:jnwalphaps} (solid curves) and the exact deflection angle (dashed red curve) for $b/m$ from 7 to 100 and from 7 to 14 in the inset and $v=c,~\lambda=3/4$. From bottom to top curve in each plot, the maximum index in the partial sum increases from 1 to 7. (b) Contribution from each order (from top to bottom curve the order increases from 1 to 7) for $v=c,~\lambda=3/4$. (c) Partial sums and the true value of the deflection angle for $\lambda=3/4$ and $2/3$ and $b=10m$. (d) Contributions from each order of Eq. \eqref{eq:jnwinbcoeffs}, the partial sum $\alpha_{\mathrm{J},7}$ and the true deflection angles for $b=10m,~v=0.9c$.
\label{fig:jnw1}}
\end{figure}
\end{center}

We plot the deflection angle  $\alpha_\mathrm{J}\equiv I_\mathrm{J}(b, v,\lambda)-\pi$ in Fig. \ref{fig:jnw1}.
In Fig. \ref{fig:jnw1} (a), we fix $\lambda=3/4$ so that a photon sphere is located using Eq. \eqref{eq:jnwps} at $r=\frac{10}{3}m$ and plot the partial sum of the deflection angle as the sum of Eqs. \eqref{eq:jnwinbcoeffs} to the $n$-th order
\begin{align}
    \alpha_{\mathrm{J},n}=\sum_{i=1}^nJ_{n}^\prime(v,\lambda)\left(\frac{\beta}{b}\right)^i, \label{eq:jnwalphaps}
\end{align}
and the true value of the deflection angle obtained by numerically integrating Eq. \eqref{eq:iintinr} for the JNW metric. It is seen that similar to the previous Bardeen and Hayward cases, the deflection angle monotonically decreases as $b$ increases, and the contribution from lower order in Eq. \eqref{eq:jnwinbcoeffs} dominates. If $b$ decreases instead, the contribution from the fourth order, Eq. \eqref{eq:j1inb} becomes more apparent. In the whole range of $b$, the partial sum of the highest order $\alpha_{\mathrm{J,7}}$ overlaps with the true value of the deflection angle, even when $b$ is as small as $7m$. At this point, the deflection angle reaches about 1.04 and the gravitational field is not weak anymore.

In Fig. \ref{fig:jnw1} (b), we plot the contribution to the deflection angle from each order in Eq. \eqref{eq:jnwinbcoeffs} and the corresponding orders in the Schwarzschild spacetime. As expected, the contribution decreases linearly in the logarithmic plot as the order increases, suggesting that the  sum as a series of $\frac{\beta}{b}$ will converge. Furthermore, the difference between the JNW and the Schwarzschild cases appears from the very first order, in contrast to the Bardeen and Hayward cases which differs from the Schwarzschild case from much higher orders. The fundamental reason of this is that the JNW spacetime although can reduce to Schwarzschild one when $q=0$, is not a perturbation of the Schwarzschild spacetime in the usual sense \cite{Patil:2011aa}.
Nevertheless, a comparison between the JNW and Schwarzschild results shows that a nonzero scalar charge $q$ typically lowers the contribution from each order.

In Fig. \ref{fig:jnw1} (c), the effect of velocity $v$ is studied for two value of $\lambda$, $\lambda_1=0.8>\lambda_c$ and $\lambda_2=0.4<\lambda_c$. Again, similar to the case of Bardeen and Hayward, smaller velocity will increase the deflection angle. For $\lambda_1=0.8$, it is known that there exists a photon sphere and therefore for a fixed $b$, the true deflection angle is expected to diverge when $v$ approaches the critical value from above. From the plot, it is clear that for $\lambda_1$, the critical value  $v_c=0.42c$,  around and below which clearly the deflection angle $\alpha_{\mathrm{J},7}$ obtained perturbatively becomes invalid. For $\lambda=\lambda_2$ however, since this spacetime does not have a photon sphere, clearly for the entire range of $v$ considered, the  partial summation $\alpha_{\mathrm{J},7}$ agrees very well with the true deflection angle.

In Fig. \ref{fig:jnw1} (d), the dependence of the deflection angle on the parameter $\hat{q}$ or equivalently $\lambda$ is shown. Clearly at the critical $\lambda=\lambda_c=\frac{1}{2}$, i.e., $\frac{q}{m}=\sqrt{3}$, the deflection angle at all order, or its partial sum $\alpha_{\mathrm{J},7}$ and true value, are continuous and smooth. This suggests basically that similar to the Bardeen and Hayward cases, the deflection angle should depend asymptotically on the metric function and therefore insensitive to the appearance/disappearance of the photon sphere at small $r$.

\subsection{Deflection angle in the EBI spacetime\label{subsec:EBI}}
The EBI BH metric describing a nonlinear electrodynamics takes the form \cite{Breton:2002td,Eiroa:2005ag}
\begin{align}
A(r)=B(r)^{-1}=1-\frac{2m}{r}+\frac{2}{\beta^2r}\int_r^{\infty}(\sqrt{\rho^4+\beta^2q^2}-\rho^2)\dd \rho,~C(r)=r^2, \label{eq:ebiadef}
\end{align}
where $m$ is the mass and $q=\sqrt{q_E^2+q_M^2}$ is the charge of the spacetime. In the limit $\beta\to 0$, this reduces exactly to the RN spacetime. The asymptotic behavior of the EBI solution also resemble that of the RN spacetime at large $r$.  The integral in Eq. \eqref{eq:ebiadef} can be re-written using an elliptical function $F$ of the first kind
\be
A(r)=1-\frac{2m}{r}+\frac{2}{3\beta^2}\left\{
r^2-\sqrt{r^4+\beta^2q^2}+\frac{\sqrt{|\beta q|^3}}{r} F\left[ \arccos\lb \frac{r^2-|\beta q|}{r^2+|\beta q|}\rb,\frac{\sqrt{2}}{2}\right]
\right\} \label{eq:ebiadef2}
\ee
Expanding this function at large $r$, one can see that $A(r)$ deviate from the metric function of the RN spacetime starting from the sixth order
\be
A(r\to\infty)= 1-\frac{2 m}{r}+\frac{ q^2}{r^2}-\frac{\beta ^2  q^4}{20 r^6}
+\mathcal{O}\left(\frac{1}{r}\right)^{10}.
\label{eq:ebiarexp}
\ee

\begin{center}
\begin{figure}[htp!]
\includegraphics[width=0.45\textwidth]{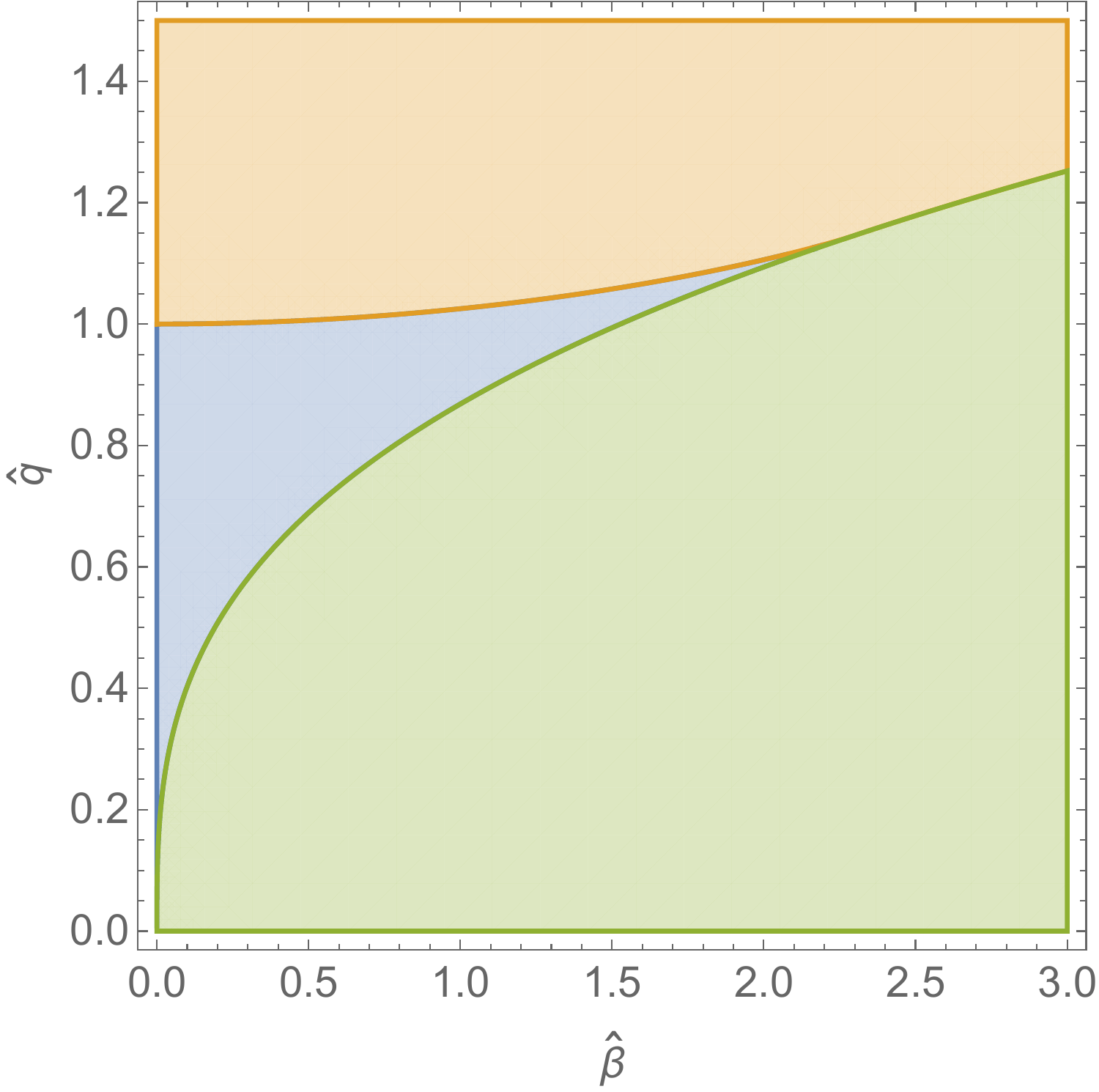}
\caption{The partition of the parameter space spanned by $\hat{\beta}$ and $\hat{q}$ for numbers of event horizons. Green, blue and brow regions allow one, two and zero event horizons. \label{fig:ebi2}}
\end{figure}
\end{center}

Depending on the values of the dimensionless parameters $\hat{\beta}\equiv\frac{\beta}{m}$ and $\hat{q}\equiv \frac{|q|}{m}$, there can be zero, one or two nonzero solutions for $A(r)=0$ (see Fig. \ref{fig:ebi2} for the partition of this parameter space). When the dimensionless parameter $\hat{q}$ is small, i.e., $0\leq \hat{q}\leq \hat{q}_{c1}(\beta)$ (the green region), there is one solution which labels the horizon of a regular BH. When the charge is at a medium value, $\hat{q}_{c1}(\beta)< \hat{q} < \hat{q}_{c2}(\beta)$ (the blue region), similar to the RN BH spacetime, there will be two event horizons. As $\hat{q}$ further increases to beyond $\hat{q} > \hat{q}_{c2}(\beta)$ (the brown region), the two horizons merge and disappear so that there is only a naked singularity at $r=0$ left. One of these critical lines, $\hat{q}_{c1}(\beta)$ can be solved analytically from the metric function \eqref{eq:ebiadef2} as
\begin{align}
\hat{q}_{c1}=\lsb \frac{1024 \pi  \hat{\beta}}{9 \Gamma \left(-\frac{3}{4}\right)^4} \rsb^{\frac{1}{3}}
.\end{align}
The other critical line $\hat{q}_{c2}(\beta)$ can only be solved numerically from the metric function. These two curves meet at the value of $\hat{\beta}=\frac{64\sqrt{2\pi}}{3\Gamma\lb -\frac{3}{4}\rb^2}$, beyond which the $\hat{q}_{c1}(\beta)$ becomes the single boundary separating the one from the zero event horizon regions in this parameter space.

Using this metric, and going through the procedure from Eq. \eqref{eqdxdl} to Eq. \eqref{eq:iinb} for the EBI metric, one can find the change of angular coordinate as a series of $r_0$ up to the eleventh order,
\be
I_{\mathrm{E}}(r_0,v,\hat{\beta},\hat{q})=\sum_{n=0}^{11}E_{n}(v,\hat{\beta},\hat{q})\lb \frac{m}{r_0}\rb ^n+ \mathcal{O}\lb \frac{m}{r_0}\rb ^{12}
\label{eq:iinx0ebi}
\ee
with the coefficients given by
\begin{subequations}
\begin{align}
E_{0}=
    &
    R_{0},\\
E_{1}=
    &
    R_{1},\\
E_{2}=
    &
    R_{2},\\
E_{3}=
    &
    R_{3},\\
E_{4}=
    &
    R_{4},\\
E_{5}=
    &
    R_{5},\\
E_{6}=
    &
    R_{6}+\left(\frac{\pi }{128}+\frac{3 \pi }{64v^2}\right)
   \hat{q}^4\hat{\beta} ^2,\\
E_{7}=
    &
    R_{7}+\left[\frac{12}{175}+\left(\frac{91}{100}-\frac{3 \pi
   }{64}\right) \frac1{v^2}+\left(\frac{3}{4}-\frac{9 \pi }{32}\right)
   \frac1{v^4}\right] \hat{q}^4 \hat{\beta} ^2,\\
E_{8}=
    &
    R_{8}+\left\{\left[\frac{105 \pi }{2048}+\left(\frac{1407 \pi
   }{1280}-\frac{17}{25}\right) \frac1{v^2}+\left(\frac{1239 \pi
   }{640}-\frac{677}{100}+\right) \frac1{v^4}+\left(\frac{57 \pi }{32}-\frac{109}{20}\right)
   \frac1{v^6}\right] \hat{q}^4\right.\nn\\
   &
   \left.+\left[-\frac{21 \pi }{2048}-\frac{41 \pi
   }{320v^2}+\frac{\pi}{640v^4}\right] \hat{q}^6\right\}
   \hat{\beta} ^2,\\
E_{9}=
    &
    R_{9}+\left\{\left[\frac{16}{45}+\left(\frac{1099}{100}-\frac{957 \pi
   }{1280}\right) \frac1{v^2}+\left(\frac{3157}{100}-\frac{1707 \pi
   }{160}\right) \frac1{v^4}+\left(\frac{29209}{600}-\frac{1217 \pi
   }{80}\right) \frac1{v^6} +\left(\frac{1123}{40}-9 \pi \right)
   \frac1{v^8}\right]
   \hat{q}^4\right.\nn\\
   &\left.
   +\left[-\frac{16}{105}+\left(-\frac{2537}{700}+\frac{289 \pi
   }{1280}\right) \frac1{v^2}+\left(-\frac{1063}{200}+\frac{139 \pi
   }{80}\right) \frac1{v^4}+\left(-\frac{13}{40}+\frac{7 \pi }{80}\right)
   \frac1{v^6}\right] \hat{q}^6\right\} \hat{\beta} ^2,\\
E_{10}=
    &R_{10}+\left\{\left[\frac{3969 \pi }{16384}+\left(-\frac{193}{25}+\frac{422013
   \pi }{40960}\right) \frac1{v^2}+\left(-\frac{13919}{100}+\frac{111201 \pi
   }{2560}\right) \frac1{v^4}\right.\right.\nn\\
   &
   \left.+\left(\frac{124977 \pi
   }{1280}-\frac{30371}{100}\right) \frac1{v^6}+\left(\frac{56103 \pi
   }{640}-\frac{55411}{200}\right) \frac1{v^8}+\left(\frac{2463 \pi }{64}-\frac{4823}{40}\right)
   \frac1{v^{10}}\right] \hat{q}^4\nn\\
   &
   +\left[-\frac{1323 \pi
   }{8192}+\left(\frac{618}{175}-\frac{27873 \pi }{5120}\right)
   \frac1{v^2}+\left(\frac{35601}{700}-\frac{5193 \pi }{320}\right)
   \frac1{v^4}+\left(\frac{12587}{200}-\frac{1293 \pi }{64}\right)
   \frac1{v^6}+\left(\frac{129}{40}-\frac{309 \pi }{320}\right)
   \frac1{v^8}\right] \hat{q}^6\nn\\
    &
    \left.+\left[\frac{189 \pi }{16384}+\frac{1987 \pi
    }{8192v^2}+\frac{181 \pi}{5120v^4}+\frac{17 \pi}{2560v^6}\right] \hat{q}^8\right\} \hat{\beta}^2
    -\left(\frac{7 \pi }{4096}+\frac{35 \pi}{2048v^2}\right) \hat{q}^6\hat{\beta} ^4,\\
E_{11}=
    &R_{11}+\left\{\left[\frac{8}{5}+\left(\frac{8731}{100}-\frac{273177 \pi
   }{40960}\right) \frac1{v^2}+\left(\frac{51361}{100}-\frac{3307761 \pi
   }{20480}\right) \frac1{v^4}+\left(\frac{314957}{200}-\frac{638751 \pi
   }{1280}\right) \frac1{v^6}\right.\right.\nn\\
   &
   \left.+\left(\frac{420367}{200}-\frac{429933 \pi
   }{640}\right) \frac1{v^8}+ \left(\frac{1060213}{800}-\frac{134619 \pi
   }{320}\right) \frac1{v^{10}}+\left(\frac{14693}{32}-\frac{4683 \pi
   }{32}\right) \frac1{v^{12}}\right]
   \hat{q}^4\nn\\
   &
   +\left[-\frac{16}{11}+\left(\frac{100883 \pi}{20480}-\frac{35128}{525}\right) \frac1{v^2}+\left(\frac{47233 \pi
   }{512}-\frac{1251673}{4200}\right) \frac1{v^4}+\left(\frac{127821 \pi
   }{640}-\frac{110218}{175}\right) \frac1{v^6}\right.\nn\\
   &
   \left.+\left(-\frac{589141}{1200}+\frac{25133 \pi
   }{160}\right) \frac1{v^8} +\left(-\frac{4831}{240}+\frac{997 \pi
   }{160}\right) \frac1{v^{10}}\right]
   \hat{q}^6+\left[\frac{8}{33}+\left(\frac{9493}{1050}-\frac{25653 \pi
   }{40960}\right) \frac1{v^2}\right.\nn\\
   &
   \left.\left.+\left(\frac{3774}{175}-\frac{129483 \pi
   }{20480}\right) \frac1{v^4}+\left(\frac{3753}{800}-\frac{3609 \pi
   }{2560}\right) \frac1{v^6}+\left(-\frac{11}{160}-\frac{21 \pi }{1280}\right)
   \frac1{v^8}\right] \hat{q}^8\right\} \hat{\beta}
   ^2\nn\\
   &
   +\left[-\frac{32}{2079}+\left(-\frac{491}{1512}+\frac{35 \pi
   }{2048}\right) \frac1{v^2}+\left(-\frac{83}{168}+\frac{175 \pi
   }{1024}\right) \frac1{v^4}\right] \hat{q}^6\hat{\beta}^4,
\end{align}
\end{subequations}
where the $R_n$ are the coefficient of order $n$ in the change of the angular coordinate in RN spacetime. Their values are given in Ref. \cite{Jia:2020dap} and also listed in Eq. \eqref{eq:angrninx0}.

Using Eq. \eqref{eq:xinb}, the expansion parameter $\frac{m}{r_0}$ can be expressed as a power series of $\frac{m}{b}$, given by
\be
\frac{m}{r_0}=\sum_{n=1}^{11}C_{\mathrm{E},n}\lb \frac{m}{b}\rb^n + \mathcal{O}\lb \frac{m}{b}\rb^{12}
\ee
with the coefficients given by
\begin{subequations}\label{eq:x0inbebi}
\begin{align}
C_{\mathrm{E},1}=
    &C_{\mathrm{R},1},\\
C_{\mathrm{E},2}=
    &C_{\mathrm{R},2},\\
C_{\mathrm{E},3}=
    &C_{\mathrm{R},3},\\
C_{\mathrm{E},4}=
    &C_{\mathrm{R},4},\\
C_{\mathrm{E},5}=
    &C_{\mathrm{R},5},\\
C_{\mathrm{E},6}=
    &C_{\mathrm{R},6},\\
C_{\mathrm{E},7}=
    &C_{\mathrm{R},7}-\frac{\hat{q}^4\hat{\beta}^2}{40v^2},\\
C_{\mathrm{E},8}=
    &C_{\mathrm{R},8}+\left(-\frac{1}{10v^2}+\frac{3 }{40v^4}\right)
   \hat{q}^4\hat{\beta} ^2,\\
C_{\mathrm{E},9}=
    &C_{\mathrm{R},9}+\left[\left(-\frac{3}{10 v^2}+\frac{9}{20 v^4}-\frac{3
   }{16v^6}\right) \hat{q}^4+\left(\frac{1}{20v^2}-\frac{3}{80v^4}\right) \hat{q}^6\right] \hat{\beta} ^2,\\
C_{\mathrm{E},10}=
    &C_{\mathrm{R},10}+\left[\left(-\frac{4 }{5v^2}+\frac{9}{5 v^4}-\frac{3}{2 v^6}+\frac{7}{16 v^8}\right) \hat{q}^4+\left(\frac{3}{10 v^2}-\frac{9}{20 v^4}+\frac{3}{16 v^6}\right) \hat{q}^6\right]\hat{\beta} ^2,\\
C_{\mathrm{E},11}=
    &C_{\mathrm{R},11}+\left[\left(-\frac2{v^2}+\frac6 {v^4}-\frac{15}{2 v^6}+\frac{35 }{8v^8}-\frac{63}{64 v^{10}}\right) \hat{q}^4+\left(\frac{6}{5 v^2}-\frac{27}{10v^4}+\frac{9}{4 v^6}-\frac{21}{32 v^8}\right)
   \hat{q}^6\right.\nn\\
   &
   \left.+\left(-\frac{3}{40 v^2}+\frac{9}{80 v^4}-\frac{3}{64 v^6}\right) \hat{q}^8\right] \hat{\beta}
   ^2+\frac{\hat{q}^6\hat{\beta} ^4}{144v^2}.
\end{align}
\end{subequations}
where the $C_{\mathrm{R},i}$ are the corresponding coefficients for the RN spacetime listed in Eq. \eqref{x0inbschsupp}.
Substituting this expansion into Eq. \eqref{eq:iinx0ebi}, we finally obtain the change of the angular coordinate in the EBI spacetime for general velocity, as
\be
I_{\mathrm{E}}(b,v,\hat{\beta},\hat{q})=\sum_{n=0}^{11}E^\prime_{n}(v,\hat{\beta},\hat{q})\lb \frac{m}{b}\rb ^n+ \mathcal{O}\lb \frac{m}{b}\rb ^{12} \label{eq:ebiiinbcoeffs}
\ee
with coefficients
\begin{subequations}
\begin{align}
E^\prime_{0}=
    &
    R^\prime_{0},\\
E^\prime_{1}=
    &
    R^\prime_{1},\\
E^\prime_{2}=
    &
    R^\prime_{2},\\
E^\prime_{3}=
    &
    R^\prime_{3},\\
E^\prime_{4}=
    &
    R^\prime_{4},\\
E^\prime_{5}=
    &
    R^\prime_{5},\\
E^\prime_{6}=
    &
    R^\prime_{6}+\frac{\pi}{128} \left(1+\frac{6}{v^2}\right) \hat{q}^4
   \hat{\beta} ^2,\\
E^\prime_{7}=
    &
    R^\prime_{7}+\frac{4}{175} \left(3+\frac{42 }{v^2}+\frac{35 }{v^4}\right) \hat{q}^4
   \hat{\beta} ^2,\\
E^\prime_{8}=
    &
    R^\prime_{8}+\frac{21 \pi }{2048} \left[\left(5+\frac{120 }{v^2}+ \frac{240 }{v^4}+\frac{64 }{v^6}\right)
   \hat{q}^4-\left(1+\frac{16 }{v^2}+\frac{16 }{v^4}\right)
   \hat{q}^6\right] \hat{\beta} ^2,\\
E^\prime_{9}=
    &
    R^\prime_{9}+\left[\frac{16}{45} \left(1+\frac{36 }{v^2}+\frac{126 }{v^4}+ \frac{84 }{v^6}+\frac{9 }{v^8}\right)
   \hat{q}^4-\frac{16}{105} \left(1+\frac{27 }{v^2}+\frac{63 }{v^4}+\frac{21 }{v^6}\right)
   \hat{q}^6\right] \hat{\beta} ^2,\\
E^\prime_{10}=
    &
    R^\prime_{10}+\frac{63 \pi}{16384}\left[\left(63+\frac{3150 }{v^2}+ \frac{16800 }{v^4}+\frac{20160 }{v^6}+\frac{5760 }{v^8}+\frac{256}{v^{10}}\right) \hat{q}^4\right.\nn\\
   &
   \left.
   -6\left(7+ \frac{280}{v^2}+
   \frac{1120}{v^4}+\frac{896 }{v^6}+\frac{128 }{v^8}\right) \hat{q}^6+3\left(1+\frac{30 }{v^2}+\frac{80 }{v^4}+ \frac{32}{v^6}\right) \hat{q}^8\right]
   \hat{\beta} ^2
   -\frac{7\pi}{4096}\left(1+\frac{10}{v^2}\right)\hat{q}^6\hat{\beta} ^4,\\
E^\prime_{10}=
    &
    R^\prime_{11}+\left[\frac{8}{5} \left(1+\frac{66 }{v^2}+ \frac{495}{v^4}+ \frac{924 }{v^6}+ \frac{495}{v^8}+
   \frac{66}{v^{10}}+\frac{1}{v^{12}}\right) \hat{q}^4\right.\nn\\
   &
   \left.-\frac{16}{11} \left(1+ \frac{55 }{v^2}+\frac{330 }{v^4}+
   \frac{462}{v^6}+\frac{165 }{v^8}+\frac{11 }{v^{10}}\right) \hat{q}^6+\left(\frac{8}{33}+\frac{32}{3v^2}+\frac{48 }{v^4}+\frac{224}{5v^6}+ \frac{8}{v^8}\right) \hat{q}^8\right]
   \hat{\beta} ^2\nn\\
   &
   -\frac{32 }{2079}\left(1+\frac{22 }{v^2}+\frac{33 }{v^4}\right)
   \hat{q}^6\hat{\beta} ^4.
\end{align}\label{eq:einbcoeffs}
\end{subequations}
The $R_n^\prime$ are the corresponding expansion coefficients in the RN spacetime listed in Eq. \eqref{angrninb7to15}. It is seen from these equations that the change of the angular coordinate in the EBI spacetime shows difference from the RN spacetime from the sixth order. Similar to the Bardeen and Hayward vs. the Schwarzschild case, this difference in the deflection angle at the sixth order also should imply that the metric lapse functions of the EBI and RN spacetimes are asymptotically different only from the sixth order. This is indeed verified from the expansion \eqref{eq:ebiarexp}.

\begin{center}
\begin{figure}[htp!]
\includegraphics[width=0.45\textwidth]{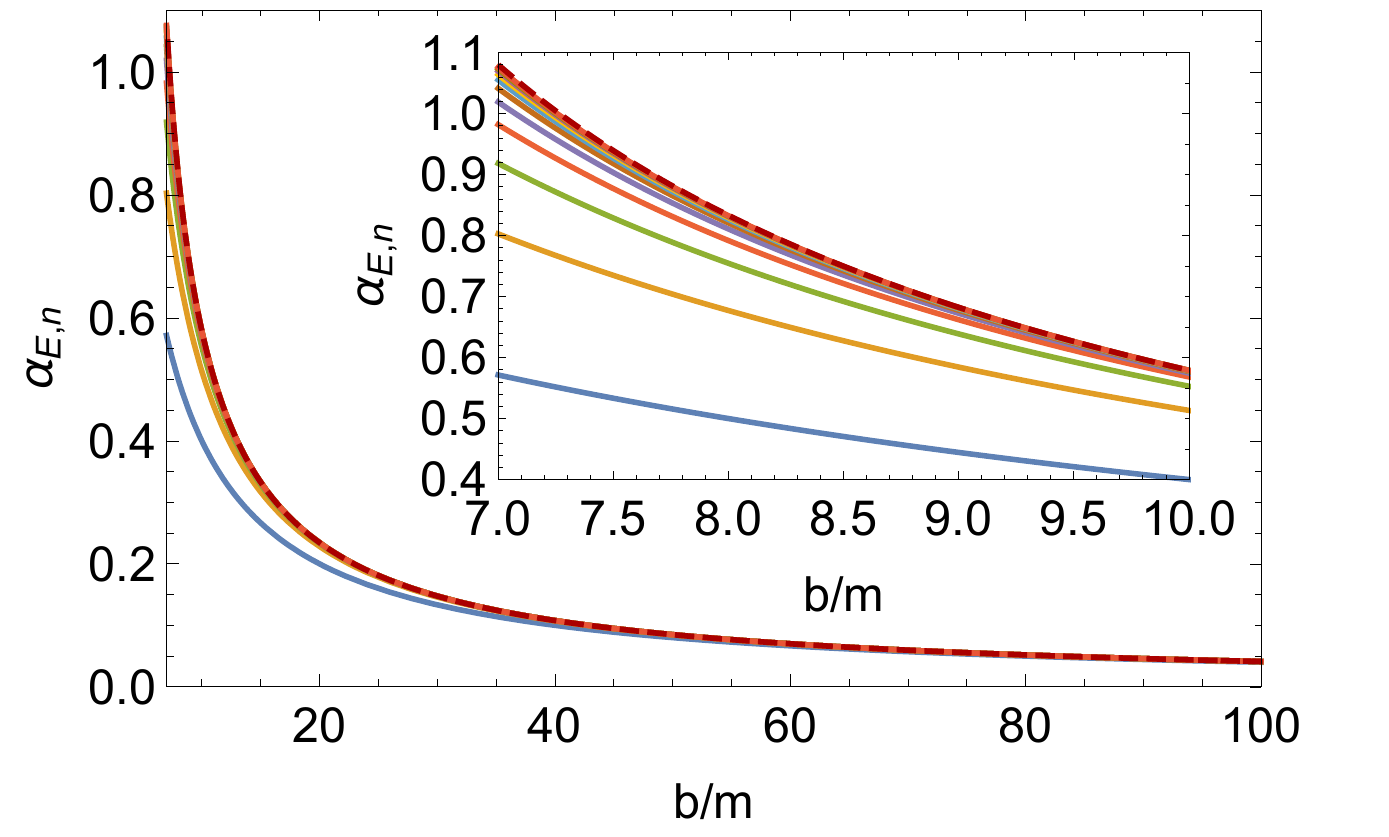}
\hspace{0.5cm}\includegraphics[width=0.45\textwidth]{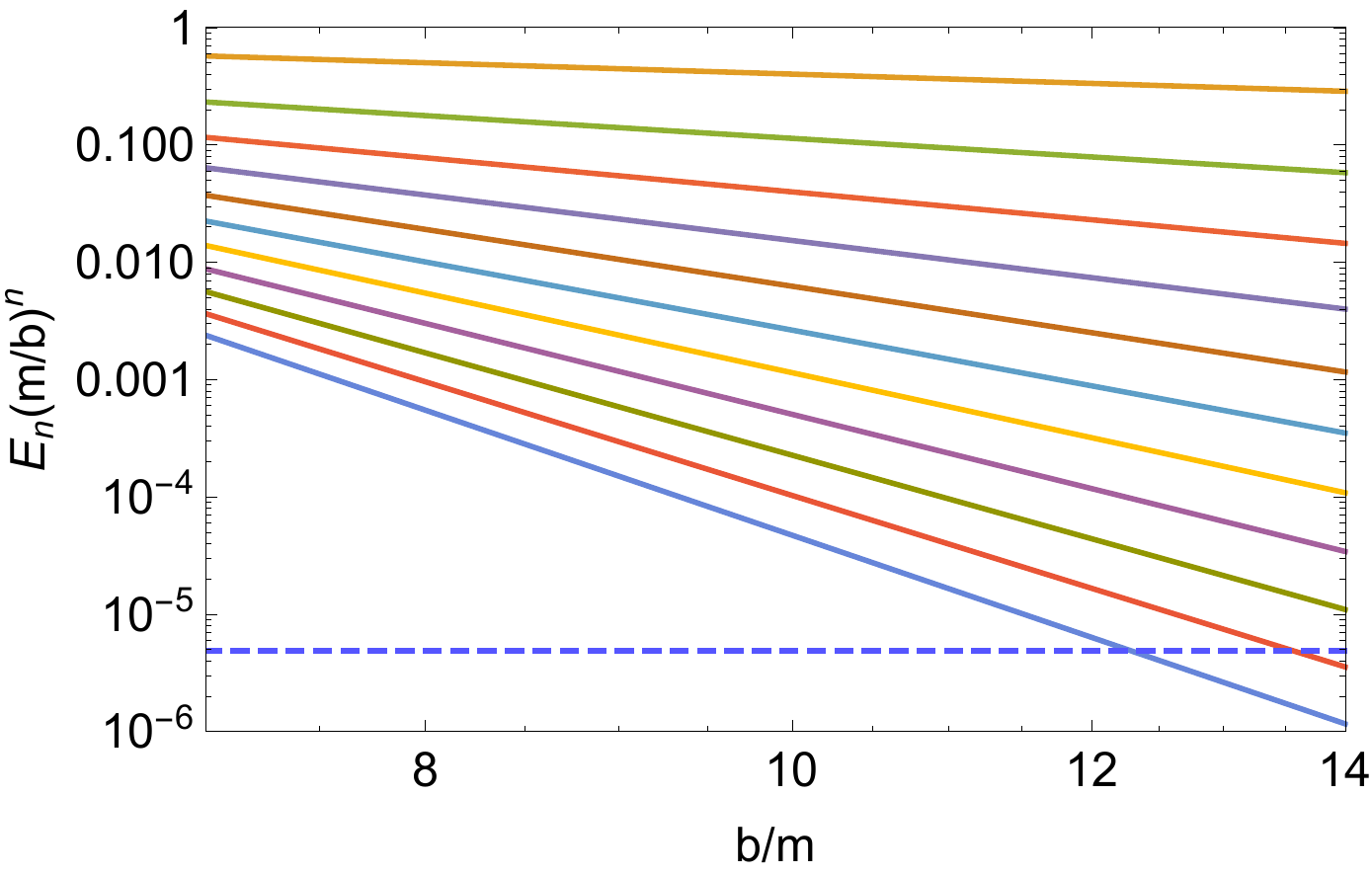}\\
(a)\hspace{8cm}(b)\\
\includegraphics[width=0.45\textwidth]{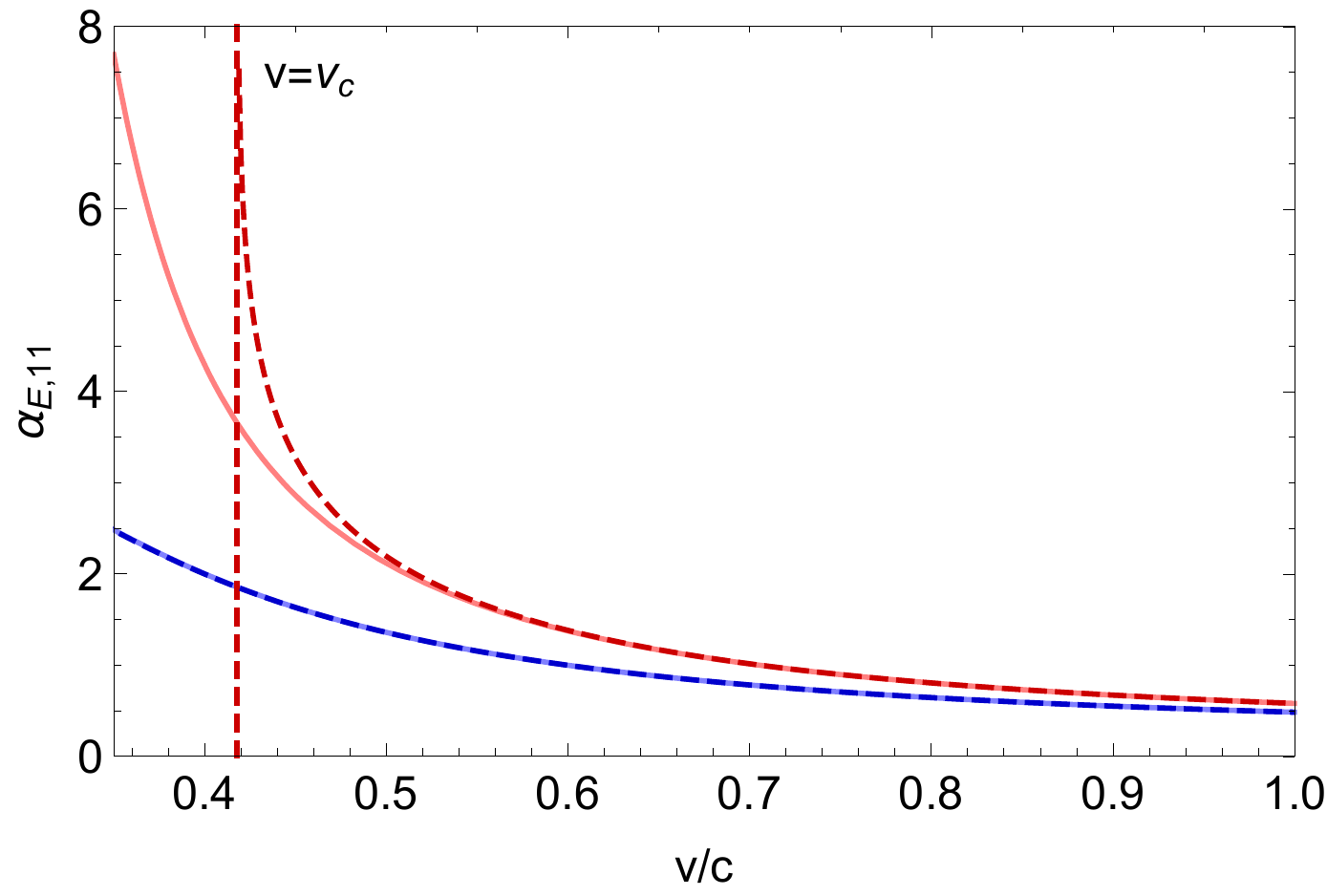}
\hspace{0.5cm}\includegraphics[width=0.45\textwidth]{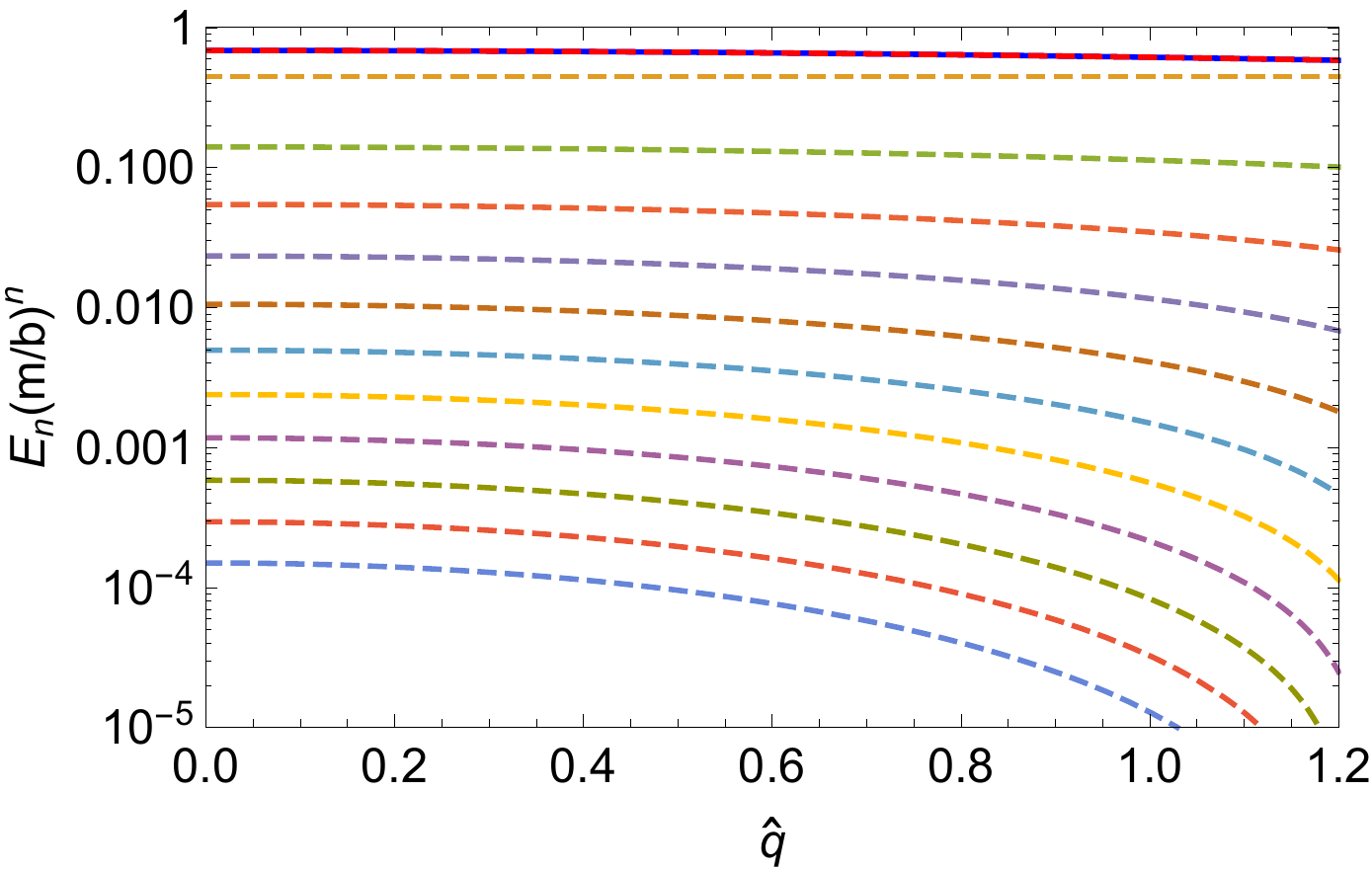}\\
(c)\hspace{8cm}(d)\\
\includegraphics[width=0.45\textwidth]{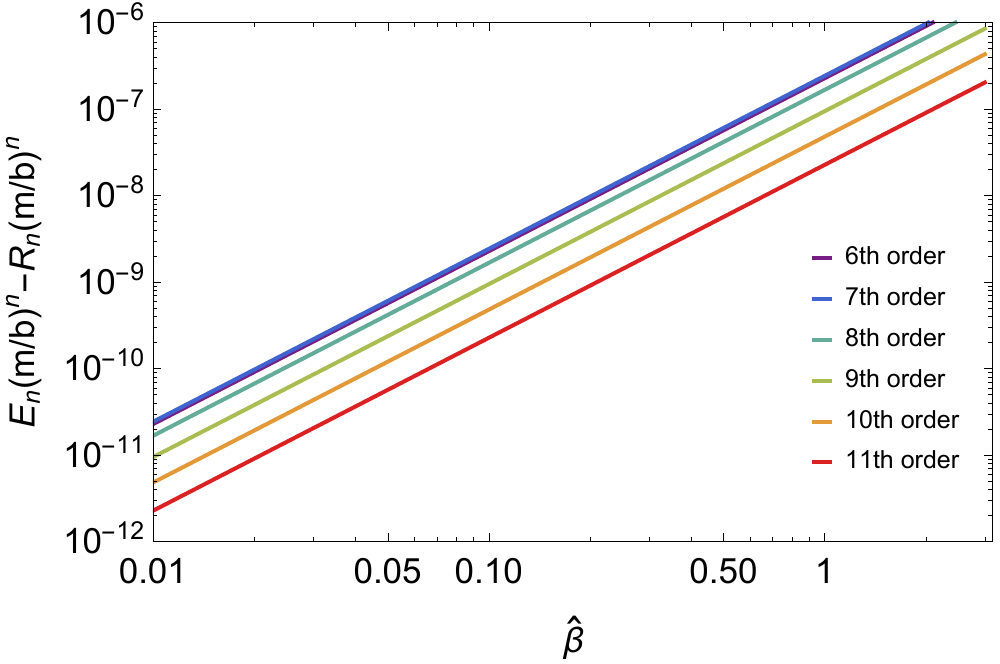}\\
(e)
\caption{Deflection angles in the EBI spacetime. (a) Partial sums \eqref{eq:ebips} (from bottom to top curve in both plots, the maximum index of the partial sum increases from 1 to 11) and the exact deflection angle (dashed red curve) for $v=c,~\hat{\beta}=1$ and $\hat{q}=0.43$. (b) Contribution from each order (from top to bottom curve $n$ increases from 1 to 11) for $v=c,~\hat{\beta}=1$ and $\hat{q}=0.43$. (c) Partial sum $\alpha_{\mathrm{E},11}$ (solid curves) and the true value (dashed curves) of the deflection angles for $\hat{q}=0.43$ (red curves) and $\hat{q}=1.4$ (blue curve) respectively and $b=10m$ and $\hat{\beta}=1$. (d) Contributions from each order (dashed curves), the partial sum $\alpha_{\mathrm{E},11}$ (solid line) and the true value (dashed curve overlapping the solid line) for $b=10m,~v=0.9c$ and $\hat{\beta}=1$. (e) The difference between contribution from each order in the EBI and RN spacetimes for $b=10m,~v=0.9c$ and $\hat{\beta}=1$. \label{fig:ebi1}}
\end{figure}
\end{center}

To study how the deflection angle depends on various kinematic and spacetime parameters, we plot $\alpha_\mathrm{E}$  in Fig. \ref{fig:ebi1}. In Fig. \ref{fig:ebi1} (a), we plot the partial sums $\alpha_{\mathrm{E},n}$ defined using Eqs. \eqref{eq:einbcoeffs} as
\begin{align}
    \alpha_{\mathrm{E},n}=\sum_{i=1}^n E^\prime_i(v,\hat{\beta},\hat{q})\lb \frac{m}{b}\rb ^i. \label{eq:ebips}
\end{align}
It is seen that similar to the cases in the Bardeen, Hayward and JNW spacetime, the partial sum $\alpha_{\mathrm{E},n}$ approaches the true value obtained numerically as $n$ increases. The highest order partial sum $\alpha_{\mathrm{E},11}$ agrees with the true value for $b$ as small as $7m$, at which the gravitational field is not weak anymore and deflection angle is not small.

The Fig. \ref{fig:ebi1} (b) shows the contribution from each order of Eq. \eqref{eq:einbcoeffs}. It is clear that for a fixed $b$, as the order increases, the contribution from each order decreases by a constant small factor, therefore the summation in Eq. \eqref{eq:ebiiinbcoeffs} will converge.

In Fig. \ref{fig:ebi1} (c), the deflection angle is plotted against the velocity $v$ for fixed $b=10m$, $\hat{\beta}=1$ and $\hat{q}=0.43$. As $v$ decreases, the partial sum $\alpha_{\mathrm{E},11}$ and exact deflection angle overlap and increase until about $v\approx 0.55c$ from which the true deflection angle starts to diverge due to the existence of one event horizon in the spacetime for the given choice of $\hat{\beta}$ and $\hat{q}$ and the consequent existence of an critical $v_c\approx 0.42c$. This shows again that the partial sum to high order as an asymptotic expansion although can work when the gravitational field is not weak,  is still not accurate near event horizon.

In Fig. \ref{fig:ebi1} (d), the effect of charge $\hat{q}$ is plotted for two representative values of $\hat{\beta}$, $\hat{\beta}_1=1$ and $\hat{\beta}_2=3$. For the first, it is seen from Fig. \ref{fig:ebi2} that there might exist zero, one or two event horizons as $\hat{q}$ increases, while for the second $\hat{\beta}$, there can only be zero or one event horizon. As seen from the plot, the deflection angles decreases monotonically and smoothly as $\hat{q}$ increases, regardless whether it is near the critical values or not. This shows that our perturbative deflection angle works for all the cases whether there is one, two or no event horizons.

Comparing to the Bardeen, Hayword and JNW spacetimes, the EBI spacetime contains one more parameter, $\hat{\beta}$.
However comparing the same order contribution from Eqs. \eqref{eq:ebiiinbcoeffs} and Eqs. \eqref{eq:angrninx0}, we see that the effect of $\hat{\beta}$ to deflection angle is much smaller than the effect of charge $\hat{q}$. For this reason, in Fig. \ref{fig:ebi1} (e), we are not plotting the contributions from each order in Eq. \eqref{eq:ebiiinbcoeffs}, but the difference between each order of the deflection angles of the RN and EBI spacetimes, against parameter $\hat{\beta}$. It is clear that this difference only starts to appear from the sixth order. Moreover, although differences of order $\hat{\beta}^4$ or above are present in order $E_{10}^\prime$ and above, all the lines in this log-log plot has a slope of value 2, which suggests that the $\hat{\beta}^2$ terms dominate higher orders.

\section{Discussions\label{sec:discussion}}

In this paper we used the perturbative expansion of the integrand in the change of the angular coordinate, \eqref{eq:iintinr}, and integrated it to obtain the
deflection angles in powers of the closest radial coordinate $r_0$ and impact parameter $b$ for signals with general velocity $v$ in the Bardeen, Hayward, JNW and EBI spacetimes to very high orders (indeed they can reach any desired order).

There are a few findings that apply to all general static and spherically symmetric spacetimes worth mentioning here.
The first is that the deflection angles in the weak field limit is determined by the asymptotic behavior of the metric functions solely. Our finding shows that a deviation of the metric function at the $n$-th power of $\frac{m}{b}$ from a given base metric, will result in a deviation of the deflection angle at the same order from that of the base metric. One important consequence of this is the continuous dependence of the deflection angle on (charge) parameters of the spacetime. This is even so when these parameters pass their critical values at which a BH spacetime becomes non-BH one or other critical transit happens, because these dramatic changes usually happen at small $r$ but not asymptotic $r$. In this sense, the deflection angle found perturbatively as a power series of $\frac{m}{b}$ will qualitatively be the same for spacetimes that are drastically different due to the variation of the metric parameter(s) \cite{Linares:2014nda}. This is in clear contrast with the deflection angle in the strong field limit, which experiences drastic change when the parameters pass their critical values for the existence/non-existence of the event horizon or photon sphere \cite{Virbhadra:2007kw}.

The second finding, obtained by comparing with the exact deflection angles calculated numerically, is that the deflection angles found perturbatively in the weak field limit, i.e., large $r_0$ or $b$ limit, can work for surprisingly small $b$ as long as the order is high enough. The deflection angle for the smallest valid $b$ can also be large, at least $0.3\pi$ for the four spacetime we tested.

There are a few directions along which the results in this work are usable or can be extended. The first is to substitute these deflection angles in the lensing equation to use the corresponding GL effects to examine the properties of the central spacetime. Although the qualitative effects of the spacetime (charge) parameters to the deflection angle are not large, quantitatively the deflection angles and consequently the images in the corresponding GL effects are still influenced by these parameters. Therefore they can be put to constrain these parameters. We point out that there are already many works along this line. The GL effect in the Bardeen spacetime has been considered in Refs. \cite{Eiroa:2010wm,Ghaffarnejad:2014zva,Schee:2017hof,Stuchlik:2019uvf}
and in the Hayward spacetime in Refs. \cite{Wei:2015qca, Zhao:2017cwk}, in JNW spacetime in Refs. \cite{Virbhadra:1998dy,bozza:2002,Virbhadra:2002ju,Virbhadra:2007kw} and in the EBI spacetime in Ref. \cite{Eiroa:2005ag}. Since these works are all for null rays, the results here can still be applied to timelike particles such as neutrino and massive GWs.

The second application is to use the GL effect in these spacetime to constrain the velocity of the messenger. For neutrinos, the velocity can be simply connected to their neutrino absolute mass and mass hierarchy. For GW, its velocity itself plays a key role in distinguishing some modified gravity models. For the particular four spacetimes studied in this work, because the velocity appears in the very first $\frac{m}{b}$ order of the deflection angle rather than the third or higher orders in which the spacetime (charge) parameters appear, the application in this direction is expected to be more practical.

\begin{acknowledgments}

This work is supported in part by the NNSF China 11504276 and MOST China 2014GB109004. 
\end{acknowledgments}

\appendix

\section{Formula for the deflection angles in the Schwarzschild and RN spacetimes \label{appd:snexp}}

In Ref. \cite{Jia:2020dap}, the change of the angular coordinate for signal with velocity $v$ and impact parameter $b$ in the Schwarzschild spacetime with mass $m$ were found. In terms of the closest coordinate $r_0$, this takes the form
\be
I_\mathrm{S}(r_0,v)=\sum_{n=0}^\infty S_n(v)\lb \frac{m}{r_0}\rb^n,
\ee
where the coefficients to the thirteenth order are
\begin{subequations}\label{eq:angschinx0}
\begin{align}
S_0=&\pi,\\
S_1=&2\lb 1+\frac{1}{v^2}\rb,\\
S_2=&\frac{3\pi}{4}+(3\pi-2)\frac{1}{v^2}-2\frac{1}{v^4},\\
S_3=&\frac{10}{3}+\lb26-\frac{3\pi}{2}\rb\frac{1}{v^2}-3(2\pi-3)\frac{1}{v^4} +\frac{7}{3}\frac{1}{v^6},\\
S_4=&\frac{105\pi}{64}+\lb\frac{93\pi}{4}-18\rb\frac{1}{v^2}+\lb\frac{69\pi}{4}-86\rb\frac{1}{v^4} +(12\pi-23)\frac{1}{v^6}-3\frac{1}{v^8},\\
S_5=&\frac{42}{5}-\lb\frac{201\pi}{16}-174\rb\frac{1}{v^2}-(117\pi-285)\frac{1}{v^4} -(63\pi-255)\frac{1}{v^6}-\lb24\pi-\frac{207}{4}\rb\frac{1}{v^8} +\frac{83}{20}\frac{1}{v^{10}},\\
S_6=&\frac{1155\pi}{256}+\lb\frac{8787\pi}{64}-114\rb\frac{1}{v^2}+ \lb\frac{10851\pi}{32}-1246\rb\frac{1}{v^4}+\lb \frac{2007\pi}{4}-\frac{4207}{3}\rb\frac{1}{v^6}\nn\\
&+(183\pi-687)\frac{1}{v^8}+\lb48\pi-\frac{443}{4}\rb\frac{1}{v^{10}} -\frac{73}{12}\frac{1}{v^{12}},\\
S_7=&
\frac{858}{35}+\lb \frac{4866}{5}-\frac{9897 \pi }{128}\rb
   \frac{1}{v^2}+\lb 3809-\frac{42105 \pi }{32}\rb  \frac{1}{v^4}+\lb 7431-\frac{35535 \pi }{16}\rb  \frac{1}{v^6}\nn\\
&+\lb \frac{20861}{4}-\frac{3585 \pi }{2}\rb
   \frac{1}{v^8}+\lb \frac{34731}{20}-480 \pi \rb  \frac{1}{v^{10}}+\lb \frac{9253}{40}-96 \pi \rb
   \frac{1}{v^{12}}+\frac{523}{56}\frac{1}{v^{14}},\\
S_8=&
\frac{225225 \pi }{16384}+\lb \frac{185637 \pi }{256}-\frac{3138}{5}\rb\frac{1}{v^2}+\lb\frac{963063 \pi }{256}-\frac{61062}{5}\rb\frac{1}{v^4}\nonumber\\
   &+\lb\frac{331515 \pi }{32}-31487\rb\frac{1}{v^6}+\lb \frac{670665 \pi
   }{64}-34265\rb\frac{1}{v^8}+\lb 5670 \pi -\frac{67199}{4}\rb\frac{1}{v^{10}}\nn\\
   &+\lb 1188 \pi -\frac{84097}{20}\rb\frac{1}{v^{12}}+\lb 192 \pi -\frac{19071}{40}\rb\frac{1}{v^{14}}-\frac{119}{8}\frac{1}{v^{16}},\\
S_9=&
\frac{4862}{63}+\lb\frac{174858}{35}-\frac{858633 \pi }{2048}\rb\frac{1}{v^2}+\lb\frac{179357}{5}-\frac{363561 \pi }{32}\rb\frac{1}{v^4}+\lb\frac{1873517}{15}-\frac{1248837 \pi
   }{32}\rb\frac{1}{v^6}\nonumber\\
   &+\lb\frac{733511}{4}-\frac{119295 \pi }{2}\rb\frac{1}{v^8}+\lb\frac{533009}{4}-\frac{329445 \pi }{8}\rb\frac{1}{v^{10}}+\lb\frac{1979077}{40}-16524 \pi \rb\frac{1}{v^{12}}\nonumber\\
   &+\lb \frac{2765857}{280}-2832 \pi
   \rb\frac{1}{v^{14}}+\lb\frac{2183697}{2240}-384 \pi \rb\frac{1}{v^{16}}+\frac{14051}{576}\frac{1}{ v^{18}},\\
  S_{10}=
       &\frac{2909907 \pi }{65536}+\left(\frac{59009547 \pi }{16384}-\frac{112174}{35}\right)\frac{1}{v^2}+\left(\frac{131274411 \pi }{4096}-\frac{3405538}{35}\right)\frac{1}{v^4}+\left(\frac{37060149 \pi }{256}-\frac{2249681}{5}\right)\frac{1}{v^6}\nonumber\\
       &+\left(\frac{71222277 \pi }{256}-\frac{4427723}{5}\right)\frac{1}{v^8}+\left(\frac{17880387 \pi }{64}-\frac{17282663}{20}\right)\frac{1}{v^{10}}+\left(\frac{1148721 \pi }{8}-\frac{9235633}{20}\right)\frac{1}{v^{12}}\nonumber\\
       &+\left(45444 \pi -\frac{5480777}{40}\right) \frac{1}{v^{14}}+\left(6576 \pi -\frac{6354781}{280}\right)\frac{1}{v^{16}}+\left(768 \pi -\frac{4442329}{2240}\right)\frac{1}{v^{18}}-\frac{13103}{320} \frac{1}{v^{20}},\\
S_{11}=
       &\frac{8398}{33}+\left(\frac{2560186}{105}-\frac{69502047 \pi
   }{32768}\right)
   \frac{1}{v^2}+\left(\frac{9762087}{35}-\frac{687200823
   \pi }{8192}\right)
   \frac{1}{v^4}+\left(\frac{53657057}{35}-\frac{992311497
   \pi }{2048}\right)
   \frac{1}{v^6}\nn\\
   &+\left(\frac{76784969}{20}-\frac{157756977
   \pi }{128}\right)
   \frac{1}{v^8}+\left(\frac{98956459}{20}-\frac{199770417
   \pi }{128}\right)
   \frac{1}{v^{10}}+\left(\frac{141049413}{40}-\frac{36346
   527 \pi }{32}\right)
   \frac{1}{v^{12}}\nn\\
   &+\left(\frac{58958723}{40}-\frac{920367
   \pi }{2}\right)
   \frac{1}{v^{14}}+\left(\frac{812709823}{2240}-119736
   \pi \right)
   \frac{1}{v^{16}}+\left(\frac{344536399}{6720}-14976 \pi
   \right)
   \frac{1}{v^{18}}\nn\\
   &+\left(\frac{54036263}{13440}-1536 \pi
   \right)\frac{1}{v^{20}}+\frac{98601
   }{1408}\frac{1}{v^{22}},\\
S_{12}=
   &\frac{156165009 \pi }{1048576}+\left(\frac{1127476893 \pi
   }{65536}-\frac{1640018}{105}\right)
   \frac{1}{v^2}+\left(\frac{15337199865 \pi
   }{65536}-\frac{14309566}{21}\right)
   \frac{1}{v^4}\nn\\
   &+\left(\frac{811056015 \pi
   }{512}-4948511\right)
   \frac{1}{v^6}+\left(\frac{40234680075 \pi
   }{8192}-\frac{108659839}{7}\right)
   \frac{1}{v^8}+\left(\frac{2069467071 \pi
   }{256}-\frac{505097751}{20}\right)
   \frac{1}{v^{10}}\nn\\
   &+\left(\frac{1897384125 \pi
   }{256}-\frac{1406267327}{60}\right)
   \frac{1}{v^{12}}+\left(\frac{66690645 \pi
   }{16}-\frac{103772803}{8}\right)
   \frac{1}{v^{14}}+\left(\frac{5541675 \pi
   }{4}-\frac{247796467}{56}\right)
   \frac{1}{v^{16}}\nn\\
   &+\left(305280 \pi
   -\frac{1247909311}{1344}\right)
   \frac{1}{v^{18}}+\left(33600 \pi
   -\frac{109722263}{960}\right)
   \frac{1}{v^{20}}+\left(3072 \pi
   -\frac{109282609}{13440}\right)
   \frac{1}{v^{22}}-\frac{15565}{128}
   \frac{1}{v^{24}},\\
S_{13}=
   &\frac{371450}{429}+\left(\frac{132929354}{1155}-\frac{2692560531 \pi
   }{262144}\right)
   \frac{1}{v^{2}}+\left(\frac{201718541}{105}-\frac{9148875
   279 \pi }{16384}\right)
   \frac{1}{v^{4}}\nn\\
   &+\left(\frac{322639187}{21}-\frac{79680420
   705 \pi }{16384}\right)
   \frac{1}{v^{6}}+\left(\frac{1640587857}{28}-\frac{3836959
   9155 \pi }{2048}\right)
   \frac{1}{v^{8}}\nn\\
   &+\left(\frac{3320156743}{28}-\frac{7701893
   7885 \pi }{2048}\right)
   \frac{1}{v^{10}}+\left(\frac{5524434529}{40}-\frac{2824
   294095 \pi }{64}\right)
   \frac{1}{v^{12}}\nn\\
   &+\left(\frac{3935582091}{40}-\frac{1994
   399505 \pi }{64}\right)
   \frac{1}{v^{14}}+\left(\frac{19772726781}{448}-\frac{11
   3269545 \pi }{8}\right)
   \frac{1}{v^{16}}\nn\\
   &+\left(\frac{17027542177}{1344}-3974940
   \pi \right)
   \frac{1}{v^{18}}+\left(\frac{6218405311}{2688}-758400
   \pi \right)
   \frac{1}{v^{20}}+\left(\frac{37267872227}{147840}-74496
   \pi \right)
   \frac{1}{v^{22}}\nn\\
   &+\left(\frac{9707157937}{591360}-6144
   \pi \right) \frac{1}{v^{24}}+\frac{1423159}{6656}\frac{1}{v^{26}}.
\end{align}
\end{subequations}

When expressing $1/x_0$ in powers of $1/b$ in the Schwarzschild spacetime, Eq. \eqref{eq:xinb} can be used, and the result is
\be \frac{m}{r_0}=\sum_{n=1}^\infty C_{\mathrm{S},n} \lb \frac{m}{b}\rb^n,
\ee
where the coefficients are
\begin{subequations}\label{x0inbschsupp}
\begin{align}
C_{\mathrm{S},1}=&1,\\
C_{\mathrm{S},2}=&\frac{1}{v^2},\\
C_{\mathrm{S},3}=&\frac{2}{v^2}+\frac{1}{2v^4},\\
C_{\mathrm{S},4}=&4\lb\frac{1}{v^2}+\frac{1}{v^4}\rb,\\
C_{\mathrm{S},5}=&\lb\frac{8}{v^2}+\frac{18}{v^4}+\frac{3}{v^6}-\frac{1}{8v^8}\rb,\\
C_{\mathrm{S},6}=&16\lb\frac{1}{v^2}+\frac{4}{v^4}+\frac{2}{v^6}\rb,\\
C_{\mathrm{S},7}=&\lb \frac{32}{v^2}+\frac{200}{v^4}+\frac{200}{v^6}+\frac{25}{v^8} -\frac{5}{4v^{10}}+\frac{1}{16v^{12}}\rb,\\
C_{\mathrm{S},8}=&64\lb \frac{1}{v^2}+\frac{9}{v^4}+\frac{15}{v^6}+\frac{5}{v^8}\rb,\\
C_{\mathrm{S},9}=& \lb\frac{128}{v^2}+\frac{1568}{v^4}+\frac{3920}{v^6}+\frac{2450}{v^8}+\frac{245}{v^{10}}-\frac{49}{4 v^{12}}+\frac{7}{8 v^{14}}-\frac{5}{128 v^{16}}\rb ,\\
C_{\mathrm{S},10}=&256\lb
\frac{1}{v^2}+\frac{16}{v^4}+\frac{56}{v^6}+\frac{56}{v^8}
  +\frac{14}{v^{10}}\rb,\\
C_{\mathrm{S},11}=&\frac{512}{v^2}+\frac{10368}{v^4}
  +\frac{48384}{v^6}+\frac{70560}{v^8}
  +\frac{31752}{v^{10}}+\frac{2646}{v^{12}}-\frac{126}{v^{14}}+\frac{81}{8 v^{16}}
  -\frac{45}{64 v^{18}}+\frac{7}{256 v^{20}},\\
C_{\mathrm{S},12}=&1024\lb \frac{1}{v^2}+\frac{25}{v^4}+\frac{150
  }{v^6}+\frac{300}{v^8}+\frac{210}{v^{10}}+\frac{42}{v^{12}}\rb,\\
C_{\mathrm{S},13}=&\frac{2048 }{v^2}+\frac{61952 }{v^4}+\frac{464640
   }{v^6}+\frac{1219680}{v^8}+\frac{1219680
   }{v^{10}}+\frac{426888}{v^{12}}+\frac{30492}{v^{14}}-\frac{5445}{4 v^{16}}+\frac{1815}{16
   v^{18}}-\frac{605}{64 v^{20}}\nn\\
   &+\frac{77}{128
  v^{22}}-\frac{21}{1024 v^{24}}.
\end{align}
\end{subequations}

If we express the change of angular coordinate in terms of impact parameter, it becomes
\be
I_\mathrm{S}(b,v)=\sum_{n=0}^\infty S_n^\prime(v)\lb \frac{m}{b}\rb^n,
\ee
where the coefficients to the ninth order are
\begin{subequations}\label{eq:angschinb}
\begin{align}
S^\prime_0=&\pi ,\\
S^\prime_1=&2\lb 1  +\frac{1}{v^2} \rb  ,\\
S^\prime_2=&\frac{3\pi}{4}\lb 1+\frac{4}{v^2}\rb,\\
S^\prime_3=&2\lb \frac{5}{3}+\frac{15}{v^2}+\frac{5}{v^4}-\frac{1}{3 v^6}\rb,\\
S^\prime_4=&\frac{105\pi}{64}\lb 1 +\frac{16}{v^2} +\frac{16}{v^4}\rb ,\\
S^\prime_5=&2\lb \frac{21}{5}+\frac{105}{v^2}+\frac{210}{v^4}+\frac{42}{v^6}-\frac{3}{v^8}+\frac{1}{5 v^{10}} \rb ,\\
S^\prime_6=&\frac{1155\pi}{256}\lb 1+\frac{36}{v^2} +\frac{30}{v^4} +\frac{4}{v^6}\rb,\\
S^\prime_7=&2\lb \frac{429}{35}+\frac{3003}{5v^2}+\frac{3003}{v^4}+\frac{3003}{ v^6}+\frac{429}{v^8}-\frac{143}{5v^{10}}+\frac{13}{5 v^{12}}-\frac{1}{7v^{14}}\rb,\\
S^\prime_8=&\frac{45045\pi}{16384}\lb 5+\frac{320}{v^2}+\frac{2240}{v^4}+\frac{3584}{v^6}+\frac{1280}{v^8}\rb ,\\
S^\prime_9=&2 \lb \frac{2431}{63}+\frac{21879}{7 v^2}+\frac{29172}{v^4}+\frac{68068}{v^6}+\frac{43758}{v^8}+\frac{4862}{v^{10}}-\frac{884}{3 v^{12}}+\frac{204}{7 v^{14}}-\frac{17}{7 v^{16}}+\frac{1}{9  v^{18}}\rb,\\
S^\prime_{10}=&\frac{2909907 \pi}{65536}\lb 1+\frac{100}{v^{2}}+\frac{1200}{v^{4}}
   +\frac{3840}{v^{6}}+\frac{3840}{v^{8}}+\frac{1024}{v^{10}}\rb,\\
S^\prime_{11}=&
   2\lb\frac{4199}{33}+\frac{46189}{3 v^2}+\frac{230945}{v^4}+\frac{969969 }{v^6}+\frac{1385670
   }{v^8}+\frac{646646}{v^{10}}+\frac{58786
   }{v^{12}}-\frac{3230 }{v^{14}}+\frac{323
   }{v^{16}}-\frac{95}{3 v^{18}}\right.\nn\\
   &
   \left.+\frac{7 }{3
   v^{20}}-\frac{1}{11 v^{22}}\rb,\\
 S^\prime_{12}=&\frac{22309287 \pi}{1048576}\lb 7+\frac{1008}{v^{2}}+\frac{18480}{v^{4}}
    +\frac{98560}{v^{6}}+\frac{190080}{v^{8}}+\frac{135168}{v^{10}}+\frac{28672}{v^{12}}\rb,\\
S^\prime_{13}=&2\lb\frac{185725}{429}+\frac{2414425}{33 v^2}+\frac{4828850}{3
   v^4}+\frac{10623470}{v^6}+\frac{26558675
   }{v^8}+\frac{26558675}{v^{10}}+\frac{9657700
   }{v^{12}}+\frac{742900}{v^{14}}\right.\nn\\
   &
   \left.-\frac{37145
   }{v^{16}}+\frac{10925}{3 v^{18}}-\frac{1150}{3
   v^{20}}+\frac{1150}{33 v^{22}}-\frac{25}{11
   v^{24}}+\frac{1}{13 v^{26}}\rb.
\end{align}
\end{subequations}

The change of the angular coordinate for signal with velocity $v$ and impact parameter $b$ in the RN spacetime with mass $m$ and $Q$ was also found in Ref. \cite{Jia:2020dap}. In terms of the closest coordinate $r_0$ and to the eleventh order, it takes the form
\be
I_\mathrm{R}(r_0,v,\hat{q})=\sum_{n=0}^\infty R_n(v,\hat{q})\lb \frac{m}{r_0}\rb^n,
\ee
where the coefficients are
\begin{subequations}\label{eq:angrninx0}
\begin{align}
R_0=
    &
    S_{0},\\
R_1=
    &
    S_{1},\\
R_2=
    &
    S_{2}-\left(\frac{\pi }{4}+\frac{\pi }{2v^2}\right)\hat{q}^2, \label{rndetailssec}\\
R_3=
    &S_{3}-\lsb 2-\lb\frac{\pi }{2} -11\rb\frac{1}{v^2}
     -\lb\pi-1\rb\frac{1}{v^4}\rsb \hat{q}^2,\\
R_4=
    &
    S_{4}-\lsb\frac{45 \pi }{32}+\lb \frac{121 \pi}{8}-10\rb\frac{1}{v^2}
    -\lb 37-\frac{29 \pi }{4}\rb\frac{1}{v^4}+\lb 2\pi- 3\rb\frac{1}{v^6}\rsb \hat{q}^2
    +\left(\frac{9 \pi }{64}+\frac{7 \pi }{8v^2}-\frac{\pi }{8v^4}\right)\hat{q}^4,\\
R_5=
    &
    S_{5}-\left[\frac{28}{3}+\lb \frac{473}{3}-\frac{85 \pi}{8} \rb\frac{1}{v^2}
    -\lb \frac{153 \pi}{2}-184 \rb\frac{1}{v^4}
    +\lb \frac{665}{6}-27 \pi \rb\frac{1}{v^6}
    -\lb 4 \pi-\frac{43}{6} \rb\frac{1}{v^8}\right]\hat{q}^2\nn\\
    &
    +\left[2+\lb 26-\frac{25 \pi}{16}\rb\frac{1}{v^2}
    -\lb \frac{11 \pi}{2}-\frac{47}{4}\rb\frac{1}{v^4}
    +\lb \frac{\pi}{2}-\frac{1}{4} \rb\frac{1}{v^6}\right]\hat{q}^4,\\
R_6=
    &
    S_{6}-\left[\frac{1575 \pi}{256}+
    \lb \frac{20101 \pi}{128}-\frac{368}{3} \rb\frac{1}{v^2}
    -\lb \frac{3397}{3}-\frac{4921 \pi }{16}\rb\frac{1}{v^4}
    +\lb \frac{2655 \pi}{8}-922 \rb\frac{1}{v^6}\right.\nn\\
    &
    \left.-\lb \frac{1801}{6}-79 \pi \rb\frac{1}{v^8}+\lb 8 \pi-\frac{95}{6} \rb\frac{1}{v^{10}}\right]\hat{q}^2
    +\left[\frac{525 \pi }{256}
    +\lb \frac{1329\pi}{32}-28 \rb\frac{1}{v^2}
    -\lb 204-\frac{441 \pi}{8} \rb\frac{1}{v^4}\right.\nn\\
    &
    \left.+\lb \frac{105 \pi}{4}-\frac{267}{4} \rb\frac{1}{v^6}-\lb\frac{3 \pi}{2}- \frac{5}{4} \rb\frac{1}{v^8}\right]\hat{q}^4
    -\left(\frac{25 \pi }{256}+\frac{157 \pi }{128v^2}
    -\frac{5 \pi }{32 v^4}+\frac{\pi }{16v^6}\right)\hat{q}^6,\\
R_7=
    &
    S_7+\left[-\frac{198}{5}
    +\lb \frac{13381 \pi}{128}-\frac{20413}{15} \rb\frac{1}{v^2}
    +\lb \frac{96527 \pi}{64}-\frac{13091}{3} \rb\frac{1}{v^4}
    +\lb \frac{16249 \pi}{8}-\frac{40831}{6} \rb\frac{1}{v^6}\right.\nn\\
    &
    \left.+\lb \frac{4777 \pi}{4}-\frac{41507}{12} \rb\frac{1}{v^8}
    +\lb 208 \pi-\frac{91307}{120} \rb\frac{1}{v^{10}}
    +\lb 16 \pi-\frac{4049}{120} \rb\frac{1}{v^{12}}\right]\hat{q}^2\nn\\
    &
    +\left[18+\lb \frac{1565}{3}-\frac{4903 \pi}{128} \rb\frac{1}{v^2}
    +\lb \frac{14803}{12}-\frac{6811 \pi}{16} \rb\frac{1}{v^4}
    +\lb \frac{5147}{4}-\frac{3033 \pi}{8} \rb\frac{1}{v^6}\right.\nn\\
    &
    \left.+\lb \frac{6349}{24}-\frac{197 \pi}{2}\rb\frac{1}{v^8}
    +\lb 4 \pi-\frac{103}{24} \rb\frac{1}{v^{10}}\right]\hat{q}^4
    +\left[-2+\lb \frac{411 \pi}{128}-47 \rb\frac{1}{v^2}
    +\lb \frac{1143 \pi}{64}-\frac{205}{4} \rb\frac{1}{v^4}\right.\nn\\
    &
    \left.-\lb \frac{11}{8}+\frac{15 \pi}{16} \rb\frac{1}{v^6}
    +\lb \frac{3 \pi}{8}-\frac{1}{8} \rb\frac{1}{v^8}\right] \hat{q}^6,\\
R_8=
    &
    S_8+\left[-\frac{105105 \pi }{4096}
    +\lb \frac{14938}{15}-\frac{608893 \pi}{512} \rb\frac{1}{v^2}
    +\lb \frac{85417}{5}-\frac{1348113 \pi}{256} \rb\frac{1}{v^4}
    +\lb \frac{109123}{3}-\frac{766485 \pi}{64} \rb\frac{1}{v^6}\right.\nn\\
    &
    \left.+\lb \frac{189181}{6}-\frac{308185 \pi}{32} \rb\frac{1}{v^8}
    +\lb \frac{44805}{4}-3795 \pi \rb\frac{1}{v^{10}}
    +\lb \frac{221449}{120}-516 \pi \rb\frac{1}{v^{12}}
    +\lb \frac{2821}{40}-32 \pi \rb\frac{1}{v^{14}}\right]\hat{q}^2\nn\\
    &
    +\left[\frac{121275 \pi}{8192}
    +\lb \frac{301055 \pi}{512}-\frac{1396}{3} \rb\frac{1}{v^2}
    +\lb \frac{1081745 \pi}{512}-\frac{20473}{3} \rb\frac{1}{v^4}
    +\lb \frac{113345 \pi}{32}-\frac{128557}{12} \rb\frac{1}{v^6}\right.\nn\\
    &
    \left.+\lb \frac{117695 \pi}{64}-\frac{73183}{12} \rb\frac{1}{v^8}
    +\lb 320 \pi-\frac{21215}{24} \rb\frac{1}{v^{10}}
    +\lb \frac{301}{24}-10 \pi \rb\frac{1}{v^{12}}\right]\hat{q}^4\nn\\
    &
    +\left[-\frac{11025 \pi }{4096}
    +\lb 60-\frac{44751 \pi}{512} \rb\frac{1}{v^2}
    +\lb 710-\frac{28461 \pi}{128} \rb\frac{1}{v^4}
    +\lb \frac{2105}{4}-\frac{5757 \pi}{32} \rb\frac{1}{v^6}
    +\lb \frac{81}{8}+\frac{105 \pi}{32} \rb\frac{1}{v^8}\right.\nn\\
    &
    \left.+\lb \frac{7}{8}-\frac{3 \pi}{2} \rb\frac{1}{v^{10}}\right]\hat{q}^6
    +\left(\frac{1225 \pi }{16384}+\frac{803 \pi }{512 v^2}+\frac{39\pi }{512 v^4}+\frac{11 \pi }{64 v^6}-\frac{5 \pi }{128v^8}\right) \hat{q}^8,\\
R_9=
    &
    S_9+\left[-\frac{1144}{7}
    +\lb \frac{398377 \pi}{512}-\frac{993439}{105} \rb\frac{1}{v^2}
    +\lb \frac{1193677 \pi}{64}-\frac{884186}{15} \rb\frac{1}{v^4}
    +\lb \frac{1756347 \pi}{32}-\frac{1757143}{10} \rb\frac{1}{v^6}\right.\nn\\
    &
    +\lb \frac{277265 \pi}{4}-\frac{638909}{3} \rb\frac{1}{v^8}
    +\lb \frac{151885 \pi}{4}-\frac{2952011}{24} \rb\frac{1}{v^{10}}
    +\lb 11094 \pi-\frac{662131}{20} \rb\frac{1}{v^{12}}\nn\\
    &
    \left.+\lb 1232 \pi-\frac{2430129}{560} \rb\frac{1}{v^{14}}
    +\lb 64 \pi-\frac{81603}{560} \rb\frac{1}{v^{16}}\right]\hat{q}^2
    +\left[\frac{572}{5}
    +\lb \frac{29178}{5}-\frac{479403 \pi}{1024} \rb\frac{1}{v^2}\right.\nn\\
    &
    +\lb \frac{605559}{20}-\frac{610641 \pi}{64} \rb\frac{1}{v^4}
    +\lb \frac{290011}{4}-\frac{1447125 \pi}{64} \rb\frac{1}{v^6}
    +\lb \frac{513933}{8}-\frac{83985 \pi}{4} \rb\frac{1}{v^8}\nn\\
    &
    \left.+\lb \frac{963711}{40}-\frac{58815 \pi}{8} \rb\frac{1}{v^{10}}
    +\lb \frac{426581}{160}-948 \pi \rb\frac{1}{v^{12}}
    +\lb 24 \pi-\frac{5363}{160} \rb\frac{1}{v^{14}}\right]\hat{q}^4\nn\\
    &
    +\left[-\frac{88}{3}
    +\lb \frac{51385 \pi}{512}-\frac{3878}{3} \rb\frac{1}{v^2}
    +\lb \frac{101435 \pi}{64}-\frac{15251}{3} \rb\frac{1}{v^4}
    +\lb \frac{20495 \pi}{8}-\frac{66211}{8} \rb\frac{1}{v^6}\right.\nn\\
    &
    \left.+\lb \frac{9275 \pi}{8}-\frac{10331}{3} \rb\frac{1}{v^8}
    -\lb \frac{2321}{48}+\frac{35 \pi}{4} \rb\frac{1}{v^{10}}
    +\lb 5 \pi-\frac{187}{48} \rb\frac{1}{v^{12}}\right]\hat{q}^6\nn\\
    &
    +\left[2+\lb 74-\frac{11177 \pi}{2048} \rb\frac{1}{v^2}
    +\lb \frac{297}{2}-\frac{2837 \pi}{64} \rb\frac{1}{v^4}
    +\lb \frac{97}{4}-\frac{303 \pi}{64} \rb\frac{1}{v^6}
    +\lb \frac{11}{64}-\frac{13 \pi}{8} \rb\frac{1}{v^8}
    +\lb \frac{5 \pi}{16}-\frac{5}{64} \rb\frac{1}{v^{10}}\right] \hat{q}^8,\\
R_{10}=
    &
    S_{10}+\left[-\frac{6891885 \pi}{65536}
    +\lb \frac{140900}{21}-\frac{252273505 \pi}{32768} \rb\frac{1}{v^2}
    +\lb \frac{19342783}{105}-\frac{62200733 \pi}{1024} \rb\frac{1}{v^4}\right.\nn\\
    &
    +\lb \frac{11143912}{15}-\frac{122414497 \pi}{512} \rb\frac{1}{v^6}
    +\lb \frac{37511813}{30}-\frac{100535597 \pi}{256} \rb\frac{1}{v^8}
    +\lb \frac{6043237}{6}-\frac{41702245 \pi}{128} \rb\frac{1}{v^{10}}\nn\\
    &
    \left.+\lb \frac{10252363}{24}-\frac{530825 \pi}{4} \rb\frac{1}{v^{12}}
    +\lb \frac{459593}{5}-30578 \pi \rb\frac{1}{v^{14}}
    +\lb \frac{5586893}{560}-2864 \pi \rb\frac{1}{v^{16}}
    +\lb \frac{167371}{560}-128 \pi \rb\frac{1}{v^{18}}\right]\hat{q}^2\nn\\
    &
    +\left[\frac{2837835 \pi}{32768}
    +\lb \frac{11553985 \pi}{2048}-\frac{14306}{3} \rb\frac{1}{v^2}
    +\lb \frac{39435901 \pi}{1024}-\frac{1745708}{15} \rb\frac{1}{v^4}
    +\lb \frac{16127133 \pi}{128}-\frac{7822791}{20} \rb\frac{1}{v^6}\right.\nn\\
    &
    +\lb \frac{21081145 \pi}{128}-\frac{6302849}{12} \rb\frac{1}{v^8}
    +\lb \frac{1601545 \pi}{16}-\frac{7400429}{24} \rb\frac{1}{v^{10}}
    +\lb \frac{207435 \pi}{8}-\frac{674701}{8} \rb\frac{1}{v^{12}}\nn\\
    &
    \left.+\lb 2636 \pi-\frac{1201277}{160} \rb\frac{1}{v^{14}}
    +\lb \frac{13547}{160}-56 \pi \rb\frac{1}{v^{16}}\right]\hat{q}^4
    +\left[-\frac{945945 \pi }{32768}
    +\lb 1312-\frac{26840595 \pi}{16384} \rb\frac{1}{v^2}\right.\nn\\
    &
    +\lb 27705-\frac{18934095 \pi}{2048} \rb\frac{1}{v^4}
    +\lb \frac{140245}{2}-\frac{23203875 \pi}{1024} \rb\frac{1}{v^6}
    +\lb \frac{502879}{8}-\frac{5010525 \pi}{256} \rb\frac{1}{v^8}\nn\\
    &
    \left.+\lb \frac{69555}{4}-\frac{740715 \pi}{128} \rb\frac{1}{v^{10}}
    +\lb \frac{3007}{16}+\frac{75 \pi}{4} \rb\frac{1}{v^{12}}
    +\lb \frac{225}{16}-15 \pi \rb\frac{1}{v^{14}}\right]\hat{q}^6
    +\left[\frac{218295 \pi }{65536}
    +\lb \frac{2587225 \pi}{16384}-110 \rb\frac{1}{v^2}\right.\nn\\
    &
    +\lb \frac{2671225 \pi}{4096}-1910 \rb\frac{1}{v^4}
    +\lb \frac{425675 \pi}{512}-\frac{10105}{4} \rb\frac{1}{v^6}
    +\lb \frac{37975 \pi}{512}-\frac{613}{2} \rb\frac{1}{v^8}
    +\lb \frac{1195 \pi}{128}-\frac{131}{64} \rb\frac{1}{v^{10}}\nn\\
    &
    \left.+\lb \frac{45}{64}-\frac{25 \pi}{16} \rb\frac{1}{v^{12}}\right]\hat{q}^8
    -\left(\frac{3969 \pi }{65536}
    +\frac{62417 \pi }{32768v^2}
    +\frac{1507 \pi }{2048 v^4}+\frac{381 \pi }{1024 v^6}
    -\frac{77\pi }{512 v^8}+\frac{7 \pi }{256v^{10}}\right) \hat{q}^{10},\\
R_{11}=
    &
    S_{11}+\left[-\frac{41990}{63}
    +\lb \frac{163920193 \pi}{32768}-\frac{18362717}{315} \rb\frac{1}{v^2}
    +\lb \frac{2939299557 \pi}{16384}-\frac{20900769}{35} \rb\frac{1}{v^4}\right.\nn\\
    &
    +\lb \frac{471598515 \pi}{512}-\frac{122425265}{42} \rb\frac{1}{v^6}
    +\lb \frac{523008717 \pi}{256}-\frac{381772261}{60} \rb\frac{1}{v^8}
    +\lb \frac{282754857 \pi}{128}-\frac{280188473}{40} \rb\frac{1}{v^{10}}\nn\\
    &
    +\lb \frac{84979881 \pi}{64}-\frac{164824687}{40} \rb\frac{1}{v^{12}}
    +\lb 426027 \pi-\frac{2294376137}{1680} \rb\frac{1}{v^{14}}
    +\lb 80700-\frac{109249577}{448} \pi \rb\frac{1}{v^{16}}\nn\\
    &
    \left.+\lb 6528 \pi-\frac{909081107}{40320} \rb\frac{1}{v^{18}}
    +\lb 256 \pi-\frac{24593209}{40320} \rb\frac{1}{v^{20}}\right]\hat{q}^2
    +\left[\frac{4420}{7}
    +\lb \frac{748997}{15}-\frac{69100279 \pi}{16384} \rb\frac{1}{v^2}\right.\nn\\
    &
    +\lb \frac{188316941}{420}-\frac{137496109 \pi}{1024} \rb\frac{1}{v^4}
    +\lb \frac{22599985}{12}-\frac{76127315 \pi}{128} \rb\frac{1}{v^6}
    +\lb \frac{408829123}{120}-\frac{70065547 \pi}{64} \rb\frac{1}{v^8}\nn\\
    &
    +\lb \frac{35713057}{12}-\frac{59988655 \pi}{64} \rb\frac{1}{v^{10}}
    +\lb \frac{611485309}{480}-\frac{3294017 \pi}{8} \rb\frac{1}{v^{12}}
    +\lb \frac{303602051}{1120}-\frac{167543 \pi}{2} \rb\frac{1}{v^{14}}\nn\\
    &
    \left.+\lb \frac{1288199}{64}-7000 \pi \rb\frac{1}{v^{16}}
    +\lb -\frac{460919}{2240}+128 \pi\rb\frac{1}{v^{18}}\right]\hat{q}^4
    +\left[-260
    +\lb \frac{24935133 \pi}{16384}-\frac{91848}{5} \rb\frac{1}{v^2}\right.\nn\\
    &
    +\lb \frac{339175455 \pi}{8192}-\frac{556915}{4} \rb\frac{1}{v^4}
    +\lb \frac{153001887 \pi}{1024}-\frac{18959671}{40} \rb\frac{1}{v^6}
    +\lb \frac{105099615 \pi}{512}-\frac{5096833}{8} \rb\frac{1}{v^8}\nn\\
    &
    +\lb \frac{14695425 \pi}{128}-\frac{5865715}{16} \rb\frac{1}{v^{10}}
    +\lb \frac{1567767\pi}{64}-\frac{11871201}{160} \rb\frac{1}{v^{12}}
    -\lb \frac{41269}{64}+30 \pi \rb\frac{1}{v^{14}}\nn\\
    &
    \left.+\lb 42 \pi-\frac{14363}{320} \rb\frac{1}{v^{16}}\right]\hat{q}^6
    +\left[\frac{130}{3}
    +\lb \frac{8080}{3}-\frac{7128915 \pi}{32768} \rb\frac{1}{v^2}
    +\lb \frac{31615}{2}-\frac{37868925 \pi}{8192} \rb\frac{1}{v^4}\right.\nn\\
    &
    +\lb \frac{896225}{24}-\frac{23959375 \pi}{2048} \rb\frac{1}{v^6}
    +\lb \frac{4992479}{192}-\frac{2175535 \pi}{256} \rb\frac{1}{v^8}
    +\lb \frac{147215}{64}-\frac{156075 \pi}{256} \rb\frac{1}{v^{10}}\nn\\
    &
    \left.+\lb \frac{5305}{384}-\frac{2695 \pi}{64} \rb\frac{1}{v^{12}}
    +\lb \frac{25 \pi}{4}-\frac{1475}{384} \rb\frac{1}{v^{14}}\right]\hat{q}^8
    +\left[-2
    +\lb \frac{272293 \pi}{32768}-107 \rb\frac{1}{v^2}
    +\lb \frac{1524629 \pi}{16384}-\frac{685}{2} \rb\frac{1}{v^4}\right.\nn\\
    &
    \left.+\lb \frac{36479 \pi}{1024}-134 \rb\frac{1}{v^6}
    +\lb \frac{2897 \pi}{512}-\frac{191}{64} \rb\frac{1}{v^8}
    +\lb \frac{35}{128}-\frac{449 \pi}{256} \rb\frac{1}{v^{10}}
    +\lb \frac{35 \pi}{128}-\frac{7}{128} \rb\frac{1}{v^{12}}\right]\hat{q}^{10},
\end{align}
\end{subequations}

The expansion of $1/r_0$ in power series of $1/b$ in the RN spacetime is
\begin{subequations}\label{x0inbrn7to15}
\begin{align}
C_{\mathrm{R},1}=
    &
    C_{\mathrm{S},1},\\
C_{\mathrm{R},2}=
    &
    C_{\mathrm{S},2},\\
C_{\mathrm{R},3}=
    &
    C_{\mathrm{S},3} -\frac{\hat{q}^2}{2v^2},\\
C_{\mathrm{R},4}=
    &
    C_{\mathrm{S},4} -\lb\frac{2}{v^2}+\frac{1}{v^4}\rb \hat{q}^2,\\
C_{\mathrm{R},5}=
    &
C_{\mathrm{S},5}- 3\left(\frac{2}{v^2}+\frac{3}{v^4}+\frac{1}{4v^6}\right)\hat{q}^2 + \frac{1}{2}\left(\frac{1}{v^2}+\frac{3}{4 v^4}\right)\hat{q}^4,\\
C_{\mathrm{R},6}=
    &C_{\mathrm{S},6}-16
   \left(\frac{1}{v^2}+\frac{3}{v^4}+\frac{1}{v^6}\right)\hat{q}^2+
   \left(\frac{3}{v^2}+\frac{6}{v^4}+\frac{1}{v^6}\right)\hat{q}^4,\\
C_{\mathrm{R},7}=
    &
    C_{\mathrm{S},7}
   -\left(\frac{40}{v^2}+\frac{200}{v^4}+\frac{150}{v^6}+\frac{25}{2 v^8}-\frac{5}{16 v^{10}}\right)
   \hat{q}^2\nn\\
   &
   +\left(\frac{12}{v^2}+\frac{45}{v^4}+\frac{45}{2 v^6}+\frac{15}{16
   v^8}\right) \hat{q}^4
   -\left(+\frac{1}{2 v^2}+\frac{5}{4
   v^4}+\frac{5}{16 v^6}\right) \hat{q}^6,\\
C_{\mathrm{R},8}=
    &C_{\mathrm{S},8}-48 \left(\frac{2}{v^2}+\frac{15}{v^{4}}+\frac{20} {v^{6}}+\frac{5}{
   v^8}\right) \hat{q}^2+40 \left(\frac{1}{v^{2}}+\frac{6 }{v^{4}}+\frac{6} {v^{6}}+\frac{1}{v^8}\right) \hat{q}^4
   -\left(\frac{4}{v^2}+\frac{18} {v^{4}}+\frac{12}
   {v^{6}}+\frac{1}{v^8}\right) \hat{q}^6,\\
C_{\mathrm{R},9}=
    &C_{\mathrm{S},9}
   -\left(\frac{224}{v^2}+\frac{2352}{v^4}
   +\frac{4900}{v^6}+\frac{2450}{v^8}+\frac{735}{4 v^{10}}-\frac{49}{8 v^{12}}+\frac{7}{32
   v^{14}}\right)
   \hat{q}^2\nn\\
   &+\left(\frac{120}{v^2}+\frac{1050}{v^4}+\frac{1750}{v^6}+\frac{2625}{4
   v^8}+\frac{525}{16 v^{10}}-\frac{35}{64 v^{12}}\right)
   \hat{q}^4\nn\\
   &
   -\left(\frac{20}{v^2}+\frac{140}{v^4}+\frac{175}{v^6}+\frac{175}{4 v^8}+\frac{35}{32
   v^{10}}\right) \hat{q}^6
   +\left(\frac{1}{2 v^2}+\frac{21}{8
   v^4}+\frac{35}{16 v^6}+\frac{35}{128 v^8}\right)
   \hat{q}^8,\\
C_{\mathrm{R},10}=
    &C_{\mathrm{S},10}-512 \left(\frac{1}{v^{2}}+\frac{14 }{v^{4}}+\frac{42 }{v^{6}}+\frac{35 }{v^{8}}+\frac{7
   }{v^{10}}\right) \hat{q}^2+336 \left(\frac{1}{v^{2}}+\frac{12 }{v^{4}}+\frac{30 }{v^{6}}+\frac{20 }{v^{8}}+\frac{3 }{v^{10}}\right) \hat{q}^4\nn\\
   &
   -80 \left(\frac{1}{v^{2}}+\frac{10
   }{v^{4}}+\frac{20 }{v^{6}}+\frac{10 }{v^{8}}+\frac{1}{v^{10}}\right) \hat{q}^6+\left(\frac{5}{v^2}+\frac{40}{v^{4}}+\frac{60 }{v^{6}}+\frac{20 }{v^{8}}+\frac{1}{v^{10}}\right)
   \hat{q}^8,\\
C_{\mathrm{R},11}=
    &C_{\mathrm{S},11}
   -\left(\frac{1152}{v^2}+\frac{20736}{v^4}+
   \frac{84672}{v^6}+\frac{105840}{v^8}+\frac{39690}{v^{10}}+\frac{2646}{v^{12}}-\frac{189}
   {2 v^{14}}+\frac{81}{16 v^{16}}-\frac{45}{256
   v^{18}}\right)
   \hat{q}^2\nn\\
   &
   +\left(\frac{896}{v^2}+\frac{14112}{v^4}+\frac{49392}{v^6}+\frac{51450}{v^8}+\frac{15
   435}{v^{10}}+\frac{3087}{4 v^{12}}-\frac{147}{8 v^{14}}+\frac{63}{128
   v^{16}}\right)
   \hat{q}^4\nn\\
   &
   -\left(\frac{280}{v^2}+\frac{3780}{v^4}+\frac{11025}{v^6}+\frac{18375}{2
   v^8}+\frac{33075}{16 v^{10}}+\frac{2205}{32 v^{12}}-\frac{105}{128
   v^{14}}\right) \hat{q}^6\nn\\
   &
   +\left(\frac{30}{v^2}+\frac{675}{2
   v^4}+\frac{1575}{2 v^6}+\frac{7875}{16 v^8}+\frac{4725}{64 v^{10}}+\frac{315}{256
   v^{12}}\right) \hat{q}^8
   -\left(\frac{1}{2 v^2}+\frac{9}{2
   v^4}+\frac{63}{8 v^6}+\frac{105}{32 v^8}+\frac{63}{256
   v^{10}}\right) \hat{q}^{10},\\
\end{align}
\end{subequations}

Finally the change of the angular coordinate in terms of the impact parameter in the RN spacetime takes the form
\begin{subequations}\label{angrninb7to15}
\begin{align}
R^\prime_0=
    &
    S^\prime_0,\\
R^\prime_1=
    &
    S^\prime_1,\\
R^\prime_2=
    &
    S^\prime_2-\frac{\pi}{4}\left(1+\frac{2}{v^2}\right) \hat{q}^2,\\
R^\prime_3=
    &
    S^\prime_3-2\lb 1+\frac{6}{v^2}+\frac{1}{v^4}\rb \hat{q}^2,\\
R^\prime_4=
    &
    S^\prime_4-\frac{45\pi}{32}\left(1+\frac{12}{v^2}+\frac{8}{v^4}\right) \hat{q}^2+\frac{3\pi}{64}\left(3+\frac{24}{v^2}+\frac{8}{v^4}\right) \hat{q}^4
    ,\\
R^\prime_5=
    &
    S^\prime_5- 4\left(\frac{7}{3}+\frac{140}{3 v^2}+\frac{70}{v^4}+\frac{28}{3 v^6}-\frac{1}{3 v^8}\right)\hat{q}^2 +
   2\left(1+\frac{15}{v^2}+\frac{15}{v^4}+\frac{1}{v^6}\right)\hat{q}^4 ,\\
R^\prime_6=
    &
    S^\prime_6-\frac{1575\pi}{256}\left(1+\frac{30}{v^2}+\frac{80}{v^4}+\frac{32}{v^6}\right)\hat{q}^2
    +\frac{105\pi}{256}\left(5+\frac{120}{v^2}+\frac{240}{v^4}+\frac{64}{v^6}\right)\hat{q}^4
    -\frac{5\pi}{256} \left(5+\frac{90}{v^2}+\frac{120}{v^4}+\frac{16}{v^6}\right) \hat{q}^6,\\
R^\prime_{7}=
    &
    S^\prime_{7}-\frac{6}{5} \left(33+\frac{1386}{v^2}+\frac{5775}{v^4}+\frac{4620}{v^6}+\frac{495}{v^8}
    -\frac{22}{v^{10}}+\frac{1}{v^{12}}\right)\hat{q}^2+2\left(9+\frac{315}{v^2}+\frac{1050}{v^4}
    +\frac{630}{v^6}+\frac{45}{v^8}-\frac{1}{v^{10}}\right)\hat{q}^4\nn\\
    &
    -2 \left(1+\frac{28}{v^2}+\frac{70}{v^4}+\frac{28}{v^6}+\frac{1}{v^8}\right) \hat{q}^6,\\
R^\prime_{8}=
    &
    S^\prime_{8}-\frac{105105 \pi }{4096} \left(1+\frac{56}{v^2}+\frac{336}{v^4}+\frac{448}{v^6}+\frac{128}{v^8}\right)\hat{q}^2
    +\frac{17325 \pi }{8192} \left(7+\frac{336}{v^2}+\frac{1680}{v^4}+\frac{1792}{v^6}+\frac{384}{v^8}\right)\hat{q}^4\nn\\
    &
    -\frac{1575\pi}{4096}\left(7+\frac{280}{v^2}+\frac{1120}{v^4}+\frac{896}{v^6}
    +\frac{128}{v^8}\right)\hat{q}^6
    +\frac{35 \pi}{16384}\left(35+\frac{1120}{v^2}+\frac{3360}{v^4}
    +\frac{1792}{v^6}+\frac{128}{v^8}\right)\hat{q}^8,\\
R^\prime_{9}=
    &
    S^\prime_{9}-\frac{8}{7} \left(143+\frac{10296}{v^2}+\frac{84084}{v^4}+\frac{168168}{v^6}+\frac{90090}{v^8}
    +\frac{8008}{v^{10}}-\frac{364}{v^{12}}+\frac{24}{v^{14}}-\frac{1}{v^{16}}\right) \hat{q}^2\nn\\
    &
    +\frac{4}{5} \left(143+\frac{9009}{v^2}+\frac{63063}{v^4}+\frac{105105}{v^6}+\frac{45045}{v^8}
    +\frac{3003}{v^{10}}-\frac{91}{v^{12}}+\frac{3}{v^{14}}\right)\hat{q}^4\nn\\
    &
    -\frac{8}{3} \left(11+\frac{594}{v^2}+\frac{3465}{v^4}+\frac{4620}{v^6}+\frac{1485}{v^8}
    +\frac{66}{v^{10}}-\frac{1}{v^{12}}\right)\hat{q}^6
    +2 \left(1+\frac{45}{v^2}+\frac{210}{v^4}+\frac{210}{v^6}+\frac{45}{v^8}
    +\frac{1}{v^{10}}\right) \hat{q}^8,\\
R^\prime_{10}=
    &
    S^\prime_{10}-\frac{6891885\pi}{65536}\left(1+\frac{90}{v^2}+\frac{960}{v^4}
    +\frac{2688}{v^6}+\frac{2304}{v^8}+\frac{512}{v^{10}}\right)\hat{q}^2\nn\\
    &
    +\frac{315315\pi}{32768}\left(9+\frac{720}{v^2}
    +\frac{6720}{v^4}+\frac{16128}{v^6}+\frac{11520}{v^8}
    +\frac{2048}{v^{10}}\right) \hat{q}^4
    -\frac{315315\pi}{32768}\left(3+\frac{210}{v^2}+\frac{1680}{v^4}+\frac{3360}{v^6}
    +\frac{1920}{v^8}+\frac{256}{v^{10}}\right)\hat{q}^6\nn\\
    &
    +\frac{10395 \pi}{65536}\left(21+\frac{1260}{v^2}+\frac{8400}{v^4}+\frac{13440}{v^6}
    +\frac{5760}{v^8}+\frac{512}{v^{10}}\right) \hat{q}^8\nn\\
    &
    -\frac{63\pi}{65536}\left(63+\frac{3150}{v^2}+\frac{16800}{v^4}+\frac{20160}{v^6}
    +\frac{5760}{v^8}+\frac{256}{v^{10}}\right)\hat{q}^{10},\\
R^\prime_{11}=
    &
    S^\prime_{11}-\frac{10}{63} \left(4199+\frac{461890}{v^2}+\frac{6235515}{v^4}+\frac{23279256}{v^6}
    +\frac{29099070}{v^8}+\frac{11639628}{v^{10}}+\frac{881790}{v^{12}}-\frac{38760}{v^{14}}
    \right.\nn\\
    &
    \left.+\frac{2907}{v^{16}}-\frac{190}{v^{18}}+\frac{7}{v^{20}}\right) \hat{q}^2\nn\\
    &
    +\frac{4}{7}\left(1105+\frac{109395}{v^2}+\frac{1312740}{v^4}+\frac{4288284}{v^6}
    +\frac{4594590}{v^8}+\frac{1531530}{v^{10}}+\frac{92820}{v^{12}}-\frac{3060}{v^{14}}
    +\frac{153}{v^{16}}-\frac{5}{v^{18}}\right)\hat{q}^4\nn\\
    &
    -4 \left(65+\frac{5720}{v^2}+\frac{60060}{v^4}+\frac{168168}{v^6}+\frac{150150}{v^8}
    +\frac{40040}{v^{10}}+\frac{1820}{v^{12}}-\frac{40}{v^{14}}+\frac{1}{v^{16}}\right)\hat{q}^6\nn\\
    &
    +\frac{10}{3}\left(13+\frac{1001}{v^2}+\frac{9009}{v^4}+\frac{21021}{v^6}
    +\frac{15015}{v^8}+\frac{3003}{v^{10}}+\frac{91}{v^{12}}-\frac{1}{v^{14}}\right)\hat{q}^8\nn\\
    &
    -2\left(1+\frac{66}{v^2}+\frac{495}{v^4}+\frac{924}{v^6}+\frac{495}{v^8}+\frac{66}{v^{10}}
    +\frac{1}{v^{12}}\right) \hat{q}^{10},
\end{align}
\end{subequations}

\end{document}